\documentclass[aps,prb,times,twocolumn,amsmath,amssymb,superscriptaddress,floatfix,showpacs]{revtex4}

\usepackage{color}

\usepackage{graphicx}
\usepackage{subfigure}
\usepackage{bbold}
\usepackage{verbatim}
\usepackage{float}
\usepackage{enumerate}
\usepackage{amsfonts}
\usepackage{multirow}
\usepackage{bbold} 
\usepackage{dsfont}
\usepackage[section]{placeins}
\usepackage{color}
\usepackage{ulem}
\usepackage{tabularx}
\usepackage[colorlinks,bookmarks=false,citecolor=blue,linkcolor=red,urlcolor=blue]{hyperref}

\DeclareMathOperator{\Tr}{Tr}

\begin{document}


\title{
Theory of multiple--phase competition in pyrochlore magnets with 
anisotropic exchange, with application to
Yb$_2$Ti$_2$O$_7$, 
Er$_2$Ti$_2$O$_7$ and
Er$_2$Sn$_2$O$_7$}


\author{Han Yan}
%
\affiliation{Okinawa Institute of Science and Technology Graduate University,
Onna-son, Okinawa 904-0395, Japan}
\affiliation{Clarendon Laboratory, University of Oxford, Parks Rd.,
Oxford OX1 3PU, UK} 

\author{Owen Benton}
%
\affiliation{Okinawa Institute of Science and Technology Graduate University,
Onna-son, Okinawa 904-0395, Japan} 
\affiliation{H.\ H.\ Wills Physics Laboratory,
University of Bristol,  Tyndall Av, Bristol BS8--1TL, UK}

\author{Ludovic Jaubert}
%
\affiliation{Okinawa Institute of Science and Technology Graduate University,
Onna-son, Okinawa 904-0395, Japan}
\affiliation{Rudolf Peierls Centre for Theoretical Physics, University
of Oxford, 1--6 Keeble Rd, Oxford OX1 3NP, UK} 

\author{Nic Shannon}
%
\affiliation{Okinawa Institute of Science and Technology Graduate University,
Onna-son, Okinawa 904-0395, Japan}
\affiliation{Clarendon Laboratory, University of Oxford, Parks Rd.,
Oxford OX1 3PU, UK} 
\affiliation{H.\ H.\ Wills Physics Laboratory,
University of Bristol, Tyndall Av, Bristol BS8--1TL, UK}


\date{\today}


\begin{abstract}
The family of magnetic rare--earth pyrochlore oxides 
R$_2$M$_2$O$_7$
plays host to a diverse array of exotic phenomena, 
driven by the interplay between geometrical frustration and spin--orbit interaction, 
which leads to anisotropy in both magnetic moments and their interactions.  
In this article we establish a general, symmetry--based theory 
of pyrochlore magnets with anisotropic exchange interactions. 
Starting from a very general model of nearest--neighbour exchange
between Kramers ions, 
we find four distinct classical ordered states, all with ${\bf q} = 0$, 
competing with a variety of spin-liquids and unconventional forms 
of magnetic order. 
The finite--temperature phase diagram of this model is determined by
Monte Carlo simulation, supported by classical spin-wave calculations.
We pay particular attention to the region of parameter space
relevant to the widely studied materials Er$_2$Ti$_2$O$_7$, 
Yb$_2$Ti$_2$O$_7$, and Er$_2$Sn$_2$O$_7$. 
We find that many of the most interesting properties of 
these materials can be traced back to the ``accidental'' 
degeneracies where phases with different symmetries meet. 
These include the ordered ground state selection by fluctuations 
in Er$_2$Ti$_2$O$_7$, the dimensional--reduction observed 
in Yb$_2$Ti$_2$O$_7$, and the lack of reported magnetic order in 
Er$_2$Sn$_2$O$_7$.
We also discuss the application of this theory to other pyrochlore oxides 
\end{abstract}




\maketitle

\section{Introduction}
\label{section:introduction}

Like high-energy physics, condensed matter is dominated by the idea of symmetry.
Any physical property which {\it cannot} be traced back to a broken 
symmetry is therefore of enormous fundamental interest. 
In this context, the spin liquid phases found in frustrated magnets 
are a rich source of inspiration~\cite{balents10}.
Perhaps the most widely studied examples are the  ``spin ice'' states in  
Ho$_{2}$Ti$_{2}$O$_{7}$ and Dy$_{2}$Ti$_{2}$O$_{7}$, classical 
spin-liquids famous for their magnetic monopole excitations~\cite{castelnovo12}.
And there is now good reason to believe that a {\it quantum} 
spin-liquid phase, in which the magnetic monopoles are elevated to the role 
of ``elementary'' particles, could exist in spin-ice like materials where quantum 
effects play a larger 
role~\mbox{\cite{hermele04,banerjee08,
savary12-PRL108,shannon12,lee12,
benton12,savary13,gingras14,hao14,pan14,pan15,tokiwa15}}.


\begin{figure}[ht!]
\centering\includegraphics[width=0.9\columnwidth]{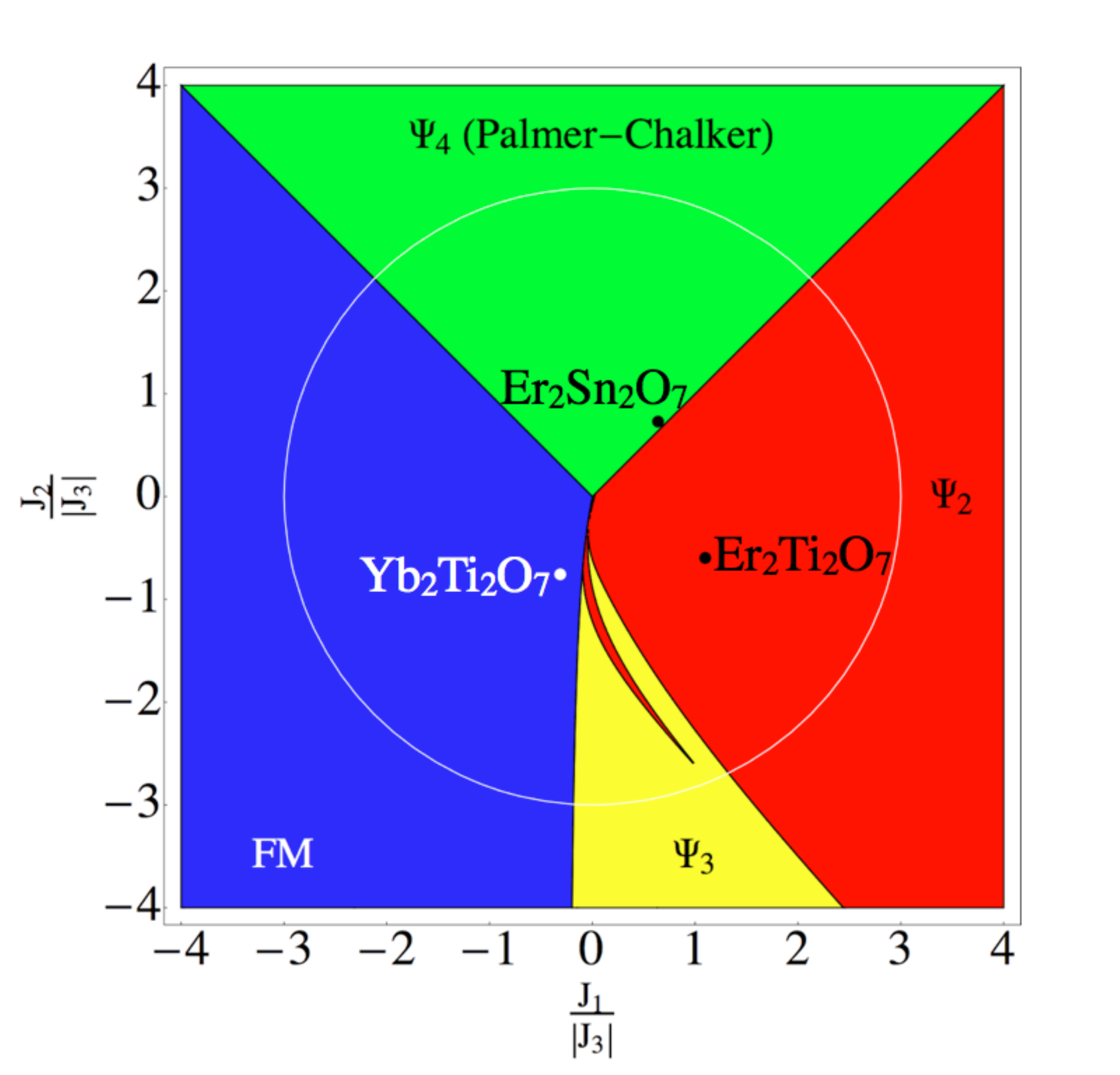} 
\caption{
Classical ground-state phase diagram for a pyrochlore magnet with 
anisotropic exchange interactions.
The model considered is the most general nearest--neighbour 
exchange Hamiltonian on the pyrochlore lattice 
${\mathcal H}_{\sf ex}$~[Eq.~(\ref{eq:Hex1})],  
with symmetric off-diagonal exchange $J_3 < 0$, 
and vanishing Dzyaloshinskii-Moriya interactions ($J_4 = 0$).
There are four distinct ordered phases, illustrated in the insets of
Fig.~\ref{fig:finite-temperature-phase-diagram}.
Points correspond to published estimates of parameters for 
Yb$_2$Ti$_2$O$_7$ [\onlinecite{ross11-PRX1}],
Er$_2$Ti$_2$O$_7$ [\onlinecite{savary12-PRL109}],
and Er$_2$Sn$_2$O$_7$ [\onlinecite{guitteny13}],
setting $J_4 = 0$.
The white circle corresponds to the path through
parameter space shown in Fig.~\ref{fig:finite-temperature-phase-diagram}.
} 
\label{fig:classical-phase-diagram}
\end{figure}


The extraordinary physics of spin ice stems from 
the combination of the geometrical frustration inherent
to the pyrochlore lattice on which the magnetic rare earth ions
R$^{3+}$ reside,
and the strongly anisotropic nature of the interactions between
rare-earth ions \cite{harris97, moessner98-PRB57}.
This mixture of geometrical frustration and strong spin 
anisotropy is common to many pyrochlore materials,
and gives rise to a wide array of interesting
physical behaviors \cite{gardner10}.


The spin ices belong to a wider family
of rare-earth pyrochlore 
oxides R$_2$M$_2$O$_7$ in which the
magnetic ions have a doublet 
ground state, and  highly-anisotropic interactions.
The physical properties of these materials depend on the choice of rare-earth
R$^{3+}$  and transition metal M$^{4+}$, and are fabulously 
diverse.
In addition to spin ices, this family includes a wide range of systems that order
magnetically, spin glasses and systems where local moments couple to itinerant
electrons~\cite{gardner10,bloete69,chern15}.
Materials of current interest include 
Yb$_2$Ti$_2$O$_7$, which exhibits striking ``rod-like'' features in neutron 
scattering~\cite{bonville04,ross09,ross11-PRB84,thompson11}, 
and has been argued to undergo a Higgs transition 
into a ferromagmetically ordered state~\cite{chang12};
Er$_2$Ti$_2$O$_7$,   which appears to offer an elegant worked example 
of (quantum) order by disorder~\cite{champion03,savary12-PRL109,zhitomirsky12,oitmaa13,
rau-arXiv}; 
and Er$_2$Sn$_2$O$_7$, which has yet to been seen to order at {\it any} 
temperature~\cite{matsuhira02,lago05,shirai07,sarte11,guitteny13}.
Alongside continuing investigations into these materials,
the last few years has seen
the synthesis of a steady stream of new rare earth pyrochlore
oxides, exhibiting both ordered \cite{dun13, yaouanc13, li14, dun15, cai16}
and disordered \cite{sibille15} low temperature
states.


\begin{figure}
\centering
\includegraphics[width=0.8\columnwidth]{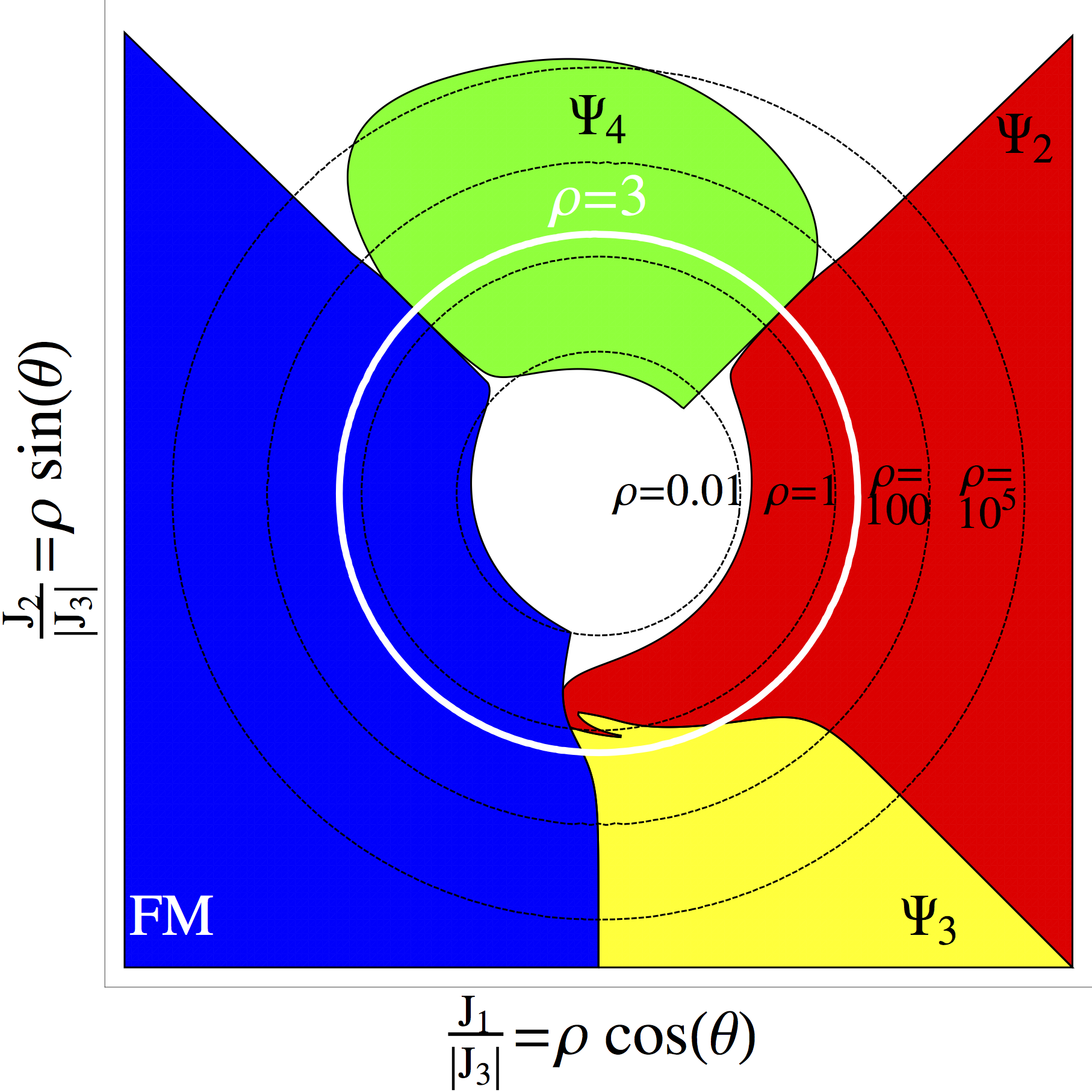}
\caption{ Suppression of classical order by quantum fluctuations 
in pyrochlore magnets with 
anisotropic exchange interactions, as described by 
${\mathcal H}_{\sf ex}$~[Eq.~(\ref{eq:Hex1})], 
with $J_3 < 0$, $J_4=0$.   
Coloured regions show the four ordered phases illustrated in 
Fig.~\ref{fig:finite-temperature-phase-diagram}.  
White regions indicate where quantum fluctuations eliminate 
conventional magnetic order, within a linear spin-wave theory.
Parameters $J_1 /  |J_3| = \rho \cos\theta$, $J_2 /  |J_3| = \rho \sin\theta$
are shown on a log-polar scale with $0 < \rho \lesssim 10^6$.   
The white circle corresponds to the path through
parameter space shown in Fig.~\ref{fig:finite-temperature-phase-diagram}.
}
\label{fig:quantum-phase-diagram}
\end{figure}    


\begin{figure*}[htpb]
\centering
\hspace{3cm}
\centering
\includegraphics[width=0.9\textwidth]{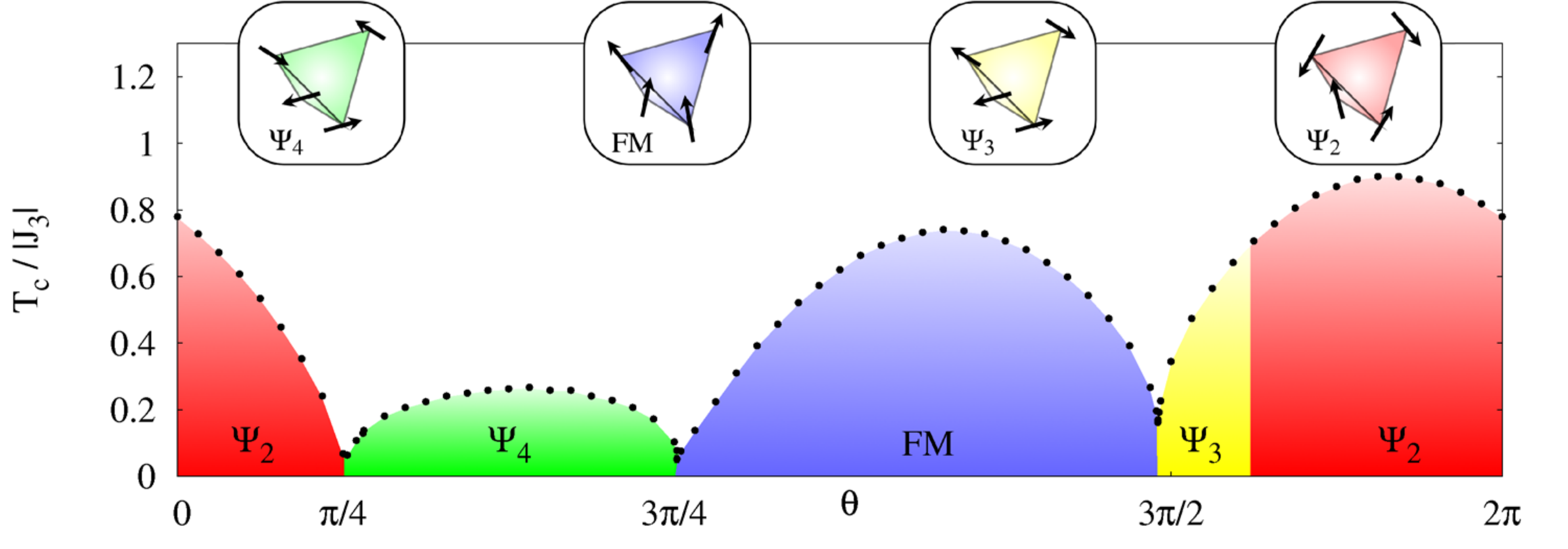}
\caption{
Finite-temperature phase diagram for a pyrochlore magnet with 
anisotropic exchange interactions.
The model considered is ${\mathcal H}_{\sf ex}$ [Eq.~(\ref{eq:Hex1})], 
with $J_1 = 3|J_3| \cos\theta$, $J_2 = 3|J_3| \sin\theta$, $J_3 < 0$, 
and $J_4 \equiv 0$, corresponding to the white circle in 
Fig.~\ref{fig:classical-phase-diagram}.  
Points show finite temperature phase transitions found from 
classical Monte Carlo simulations.  
The four ordered phases, 
Palmer--Chalker ($\Psi_4$), 
non-collinear ferromagnetic (FM), 
coplanar antiferromagnetic ($\Psi_3$) 
and non-coplanar antiferromagnetic ($\Psi_2$), 
are illustrated at the top of the figure.
Each of these phases is six-fold degenerate, with zero crystal momentum, 
and is completely specified by the spin configuration in a single tetrahedron.  
} 
\label{fig:finite-temperature-phase-diagram}
\end{figure*}

Given this ``embarrassment of riches'', it seems reasonable to ask whether 
there is {\it any} common framework which can connect the properties of different 
rare--earth pyrochlore oxides, place new materials in context, and help 
guide the search for novel magnetic states.
In this Article, we enlarge on the results in an earlier preprint~\cite{han-arXiv} 
to develop a broad scenario for these materials, based 
on the concept of multiple--phase competition. 
We go on to show how this approach can be used to explain many of the 
interesting properties of Yb$_2$Ti$_2$O$_7$, Er$_2$Ti$_2$O$_7$ 
and Er$_2$Sn$_2$O$_7$.

Our starting point is the most general model of nearest--neighbour interactions
compatible with the symmetries of the pyrochlore lattice~\cite{curnoe07,mcclarty09,ross11-PRX1}
\begin{eqnarray}
\mathcal{H}_{\sf ex} 
= \sum_{\langle ij \rangle}
   J^{\mu\nu}_{ij} 
   S^\mu_i S^\nu_j \; ,
\label{eq:Hex1}
\end{eqnarray}
where the sum on $\langle ij \rangle$ runs over the nearest--neighbour bonds of the 
pyrochlore lattice, ${\bf S}_i = (S^x_i, S^y_i, S^z_i)$ describes the magnetic moment 
of the rare-earth ion, and the matrix $J^{\mu\nu}_{ij}$ is a function of four independent 
parameters.  
Following the notation of Ross {\it et al.}~[\onlinecite{ross11-PRX1}], we identify 
these as \mbox{``X-Y''} ($J_1$), ``Ising'' ($J_2$), ``symmetric off-diagonal'' ($J_3 $) 
and ``Dzyaloshinskii-Moriya'' ($J_4$) interactions.
This model encompasses an extremely rich variety of different magnetic physics, 
including an exchange-based ``spin-ice'' ($J_1$=$-J_2$=$J_3$=$J_4<0$),
and the Heisenberg antiferromagnet on a pyrochlore lattice ($J_1$=$J_2>0$,  $J_3$=$J_4$=$0$), 
both of which are believed to have spin-liquid ground states~\cite{harris97,moessner98-PRB58}.
Nonetheless, materials such as Er$_2$Ti$_2$O$_7$, which is extremely 
well-described by a nearest--neighbour exchange model~\cite{champion03,savary12-PRL109,zhitomirsky12}, 
 {\it do} order magnetically~\cite{champion03}.

The phase diagram of $\mathcal{H}_{\sf ex}$ [Eq.~(\ref{eq:Hex1})] for a quantum 
spin--1/2 has previously been studied using mean-field and spin--wave approximations, 
with many papers emphasising connections with spin 
ice \cite{savary12-PRL108,lee12,savary13,wong13,hao14}.
In this Article we take a different approach, starting from an analysis of the way in which 
different spin configurations break the point-group symmetries of the pyrochlore lattice.
We show that, for classical spins, the problem of finding the ground state of 
$\mathcal{H}_{\sf ex}$~[Eq.~(\ref{eq:Hex1})]
can be neatly separated into two steps: i) finding the ground state of a single tetrahedron 
and ii) understanding how the spin-configuration on that tetrahedron can be used to tile the 
pyrochlore lattice. 
The first step, in turn, reduces to understanding how the different interactions
in the model transform under the symmetries  ${\sf T}_d$ of a single tetrahedron.
The second step, summarized in a simple set of  ``Lego--brick'' rules, 
enables us to encompass both ordered ground states, which break lattice 
symmetries, and spin-liquids, which do not.

This approach, augmented by spin-wave calculations and extensive classical Monte Carlo 
simulations, makes it possible both to establish a complete phase diagram for 
$\mathcal{H}_{\sf ex}$ [Eq.~(\ref{eq:Hex1})] as a function $(J_1,J_2,J_3,J_4)$, 
and to link ground state properties to predictions for 
neutron-scattering experiments.
In this article, taking our motivation from estimated parameters for 
Yb$_2$Ti$_2$O$_7$ [\onlinecite{ross11-PRX1}], 
Er$_2$Ti$_2$O$_7$ [\onlinecite{savary12-PRL109}]
and Er$_2$Sn$_2$O$_7$ [\onlinecite{guitteny13}],  
we concentrate on ordered phases in the limit \mbox{$J_3 < 0$, $J_4=0$}. 
Here there is a competition between 
four different types of order~:
a Palmer--Chalker \cite{palmer00} phase ($\Psi_4$), 
a non-collinear ferromagnet (FM), 
a coplanar antiferromagnet ($\Psi_3$), 
and a non-coplanar antiferromagnet ($\Psi_2$).
The way in which these phases relate to one another is illustrated in 
Fig.~\ref{fig:classical-phase-diagram}, 
Fig.~\ref{fig:quantum-phase-diagram} and Fig.~\ref{fig:finite-temperature-phase-diagram}.

Crucially, the same symmetry-based approach used to find ordered ground states 
also permits us to explore the way in which these physically distinct states are connected by 
the ``accidental''  degeneracies arising at boundaries between phases with different symmetry.
The enlarged ground-state manifolds at these phase boundaries have far-reaching 
consequences, once quantum and thermal fluctuations are taken into account.   
The common theme which emerges is of systems ``living on the edge'' --- 
the physical properties of materials showing one type of magnetic order
being dictated by the proximity of another, competing, ordered phase.

\begin{figure}
\centering
\includegraphics[width=0.9\columnwidth]{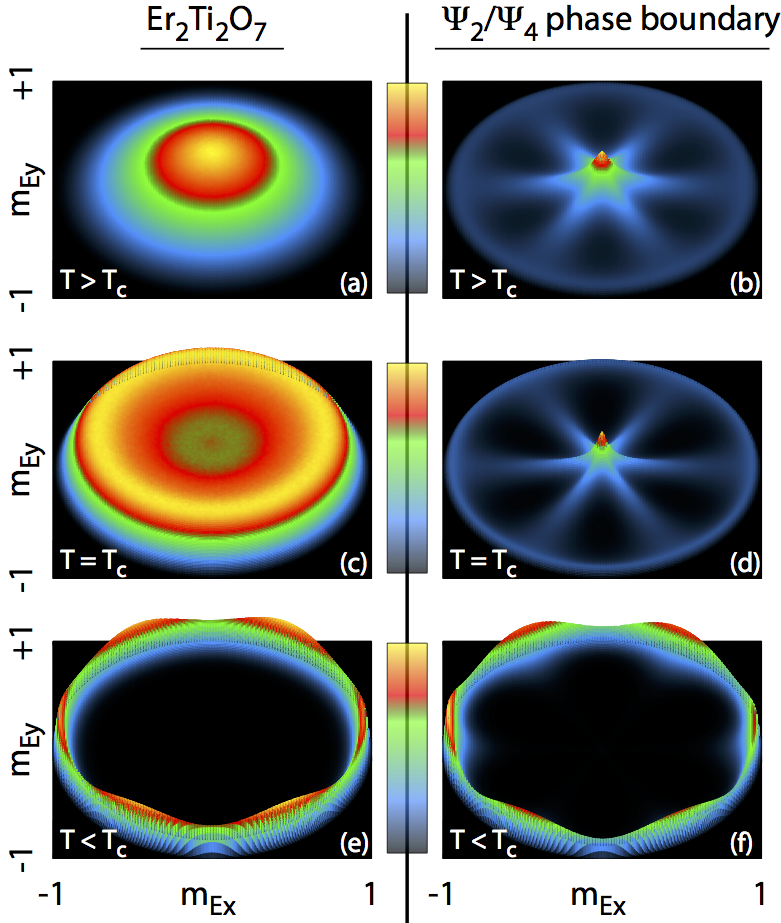}
\caption{
Selection of an ordered ground state by thermal fluctuations in Er$_2$Ti$_2$O$_7$.
For $T< T_c$, fluctuations select six states with non--coplanar antiferromagnetic 
order ($\Psi_2$) from a one--dimensional manifold of degenerate ground states.
The entropic selection of these six states can be traced to 
an enlarged ground--state manifold found on the boundary with the 
Palmer--Chalker phase ($\Psi_4$).
Plots show the probability distribution of the order parameter 
${\bf m}_{\sf E} = (m_{{\sf E}_x}, m_{{\sf E}_y})$, as described in Section~\ref{section:living-on-the-edge}.
}
\label{fig:figure-of-doom-lite}
\end{figure}

Thus, in Yb$_2$Ti$_2$O$_7$, we find ferromagnetic 
order proximate to competing,  ``$\Psi_3$'' and ``$\Psi_2$'' phases, which 
manifest themselves in the ``rods'' seen in neutron scattering.
Meanwhile, in Er$_2$Ti$_2$O$_7$,  we discover that the 
reason fluctuations select the well--established ``$\Psi_2$'' 
ground state~\cite{champion03,savary12-PRL109,zhitomirsky12}, 
is proximity to a neighboring Palmer--Chalker phase, as illustrated 
in Fig.~\ref{fig:figure-of-doom-lite}.  
And in the case of Er$_2$Sn$_2$O$_7$, we find that fluctuations 
of Palmer--Chalker order predominate, but that all forms of magnetic order 
are strongly suppressed by the proximity of a degenerate ground-state manifold 
connected to a neighbouring ``$\Psi_2$'' phase.

We note that the same approach of combining symmetry analysis and the ``Lego--brick'' rules 
can also be used to systematically search for unconventional ordered states
and new (classical) spin-liquid phases on the pyrochlore lattice.
This is a theme which will be developed elsewhere \cite{benton16, benton-thesis}.
The remainder of the present article is structured as follows :


In Section~\ref{section:model} we introduce a general model of nearest--neighbour 
exchange interactions on a pyrochlore lattice and, restricting to classical spins, establish
the conditions under which the model has a magnetically ordered ground state.
We also provide a complete classification of possible ordered states in terms of the 
irreducible representations of the tetrahedral symmetry group $T_d$.


In Section~\ref{section:classical-ground-states} we show that this symmetry analysis can be used to 
determine the classical ground state of $\mathcal{H}_{\sf ex}$~[Eq.~(\ref{eq:Hex1})]
for arbitrary parameters $(J_1,J_2,J_3,J_4)$.
The nature of the ground states in the limit 
$(J_3 < 0, J_4 = 0)$, which is of particular relevance to real materials,
 is explored in some detail, including analysis
of the degenerate manifolds arising at the phase boundaries of the model.


In Section~\ref{section:spin-wave-theory} we explore the spin-wave excitations associated 
with these ordered phases.
This enables us to make predictions for neutron scattering, 
and to develop a ground state phase diagrams for classical and semiclassical spins, 
focusing again on the limit $J_3 < 0$ and  $J_4 = 0$,  
as illustrated in Fig.~\ref{fig:classical-phase-diagram}.  
It also enables us to identify regions of the phase diagram where strong quantum
fluctuations are liable to eliminate classical order entirely, as illustrated in 
Fig.~\ref{fig:quantum-phase-diagram}.


In Section~\ref{section:finite-temperature} we use classical Monte Carlo simulation to explore the 
finite-temperature phase transitions which separate each of the ordered
phases from the high-temperature paramagnet.
The results of this analysis are summarized in Fig.~\ref{fig:finite-temperature-phase-diagram}.


In Section~\ref{section:living-on-the-edge} we study the finite--temperature consequences 
of the enlarged ground--state manifolds arising at the boundary between different 
ordered phases.
This is illustrated in Fig.~\ref{fig:figure-of-doom-lite}.   


In Sections~\ref{section:Er2Ti2O7},~\ref{section:Yb2Ti2O7} 
and~\ref{section:Er2Sn2O7}, we discuss the implications of 
these results for the rare-earth pyrochlore oxides 
Er$_{2}$Ti$_{2}$O$_{7}$, Yb$_2$Ti$_2$O$_7$ and 
Er$_{2}$Sn$_{2}$O$_{7}$, respectively.
Other rare--earth pyrochlore magnets to which the theory might apply are  
discussed briefly in Section~\ref{section:other-pyrochlores}. 


We conclude in Section~\ref{conclusions} with a summary of our results,
and an overview of some of the interesting open issues.


Technical details of calculations are reproduced in a short series of 
Appendices at the end of the Article~:
Appendix~\ref{appendix:local-coordinate-frame} provides details of the local 
coordinate frame throughout the Article, and the associated form of the $g$--tensor.
Appendix~\ref{appendix:linear-spin-wave} provides technical details of the linear 
spin--wave calculations described in Section~\ref{section:spin-wave-theory}.
Appendix~\ref{appendix:classical-MC} provides technical details associated
with the classical Monte Carlo simulations described in 
Section~\ref{section:finite-temperature}.  


As far as possible, we have endeavoured to make 
Sections~\ref{section:Er2Ti2O7}--\ref{section:other-pyrochlores}, 
describing the application of the theory to experiments on rare--earth pyrochlores, 
self--contained.     
Readers chiefly interested in these materials may safely
omit the theoretical development in Sections~\ref{section:classical-ground-states} 
to~\ref{section:living-on-the-edge} of the Article.


\section{Microscopic model of anisotropic exchange}
\label{section:model}

\subsection{Magnetism at the level of a single ion}

Pyrochlore oxides, A$_2$B$_2$O$_7$, are a ubiquitous feature of igneous rocks 
throughout the world.  
This broad family of materials takes its name from the mineral  ``pyrochlore'' 
\mbox{[(Ca, Na)$_2$Nb$_2$O$_6$(OH,F)]},  
which burns with a green ($\chi \lambda \omega \rho \grave{o} \varsigma$) 
fire ($\pi \grave{\upsilon} \rho$), and shares its crystal structure with a great many 
other oxides, halides and chalcogenides.
Here we concentrate on those pyrochlore oxides in which the B-cation 
is a non-magnetic transition metal, such as Ti$^{4+}$ or Sn$^{4+}$, 
while the remaining cation A$^{3+}$ is a rare-earth ion with a magnetic
doublet ground state.
These magnetic ions form a {\it pyrochlore lattice}, built of corner-sharing tetrahedra,
which shares the same cubic symmetry $Fd\overline{3}m$ as the parent material.

Even within this restricted group of rare-earth oxides, the interplay between strong 
spin-orbit coupling, and the crystal electric field (CEF) at the A-cation
site, leads to a huge variation in the magnetic properties of the rare-earth ion.
For example, Dy$^{3+}$ provides the strong Ising moment in the spin-ice 
Dy$_2$Ti$_2$O$_7$, while Er$^{3+}$ forms a moment with 
XY-like character in Er$_2$Ti$_2$O$_7$~[\onlinecite{cao09-PRL103}].

The goal of this article is not to explore the intricate CEF ground states 
of rare-earth ions (see \textit{e.g.} Ref.~[\onlinecite{onoda11,rau15-PRB}] 
for a discussion on this topic), but rather to understand the way in which the anisotropic exchange 
interactions between them shape the magnetism of rare-earth pyrochlore oxides.
We therefore concentrate on materials in which the ground state of the rare-earth ion 
is a Kramers doublet, with an odd number of electrons, 
like Yb$^{3+}$ ([Xe]4f$^{13}$) or Er$^{3+}$ ([Xe]4f$^{11}$).

In this case, as long as the temperature is small compared with the lowest-lying CEF 
excitation, the magnetic ion can be described by a pseudospin-1/2 degree of freedom
\begin{eqnarray}
[S^\mu ,  S^\nu] = i \epsilon_{\mu\nu\xi} S^\xi.
\end{eqnarray}
It is important to note that, even with the restriction to Kramers doublets, 
there is more than one possibility for how $S^{\mu}$ will transform under 
space--group operations \cite{huang14}. 
In this article we will focus on the case where $S^{\mu}$
transforms like a magnetic dipole, which is the case 
appropriate to Yb$^{3+}$ and Er$^{3+}$ based
pyrochlores.
We note that an alternative ``dipolar-octupolar" case
may be realized in Dy$^{3+}$, Nd$^{3+}$ and Ce$^{3+}$ based
pyrochlores \cite{huang14,sibille15,li17}.


Where $S^{\mu}$ transforms like a magnetic dipole, 
it will be associated with an effective magnetic moment 
\begin{eqnarray}
m_i^\mu &=& \sum_{\nu=1}^3 g_i^{\mu\nu}  S_i^{\nu} 
\label{eq:g-tensor-global}
\end{eqnarray}
where $\mu$, $\nu = \{x, y, z\}$.  
Since the magnetic anisotropy of rare-earth ion is determined by
the local CEF, the g-tensor $g_i^{\mu \nu}$ is site-dependent, as described in 
Appendix~\ref{appendix:local-coordinate-frame}. 
This has important consequences for the magnetic correlations measured in neutron 
scattering experiments, discussed below.

\subsection{Anisotropy in exchange interactions}


The interplay between spin-orbit coupling and CEF leads to anisotropy in the 
interactions between rare-earth ions, just as it leads to anisotropy in the magnetic 
ground state of an individual ion~\cite{rau15-PRB}. 
It is possible to make estimates of exchange interactions in a pyrochlore 
oxides from knowledge of the CEF ground state and low-lying excitations~\cite{molavian07,onoda11}.  
However for the purposes of this article it is sufficient to consider the constraints
on these interactions imposed by the symmetry of the lattice.


In the case of Kramers ions on a pyrochlore lattice, the most general form of 
nearest--neighbour exchange can be broken down into a sum over tetrahedra $t$
\begin{eqnarray}
\mathcal{H}_{\sf ex}
   &=& \sum_{\langle ij \rangle} J^{\mu\nu}_{ij} S^\mu_i S^\nu_j 
   =  \sum_t \mathcal{H}_{\sf ex}^{\sf tet} [t]  \; , 
   \protect\label{eq:Hex}
\end{eqnarray}
where 
\begin{eqnarray}   
\mathcal{H}_{\sf ex}^{\sf tet}  [t]
     =  \sum_{i,j \in t} {\mathbf S}_i {\bf J}_{ij}^{[t]} {\mathbf S}_j \; .
\end{eqnarray}
Here, ${\mathbf S}_i = (S^x_i, S^y_i, S^z_i)$, and ${\bf J}_{ij}^{[t]}$ is a 
$3 \times 3$ matrix specific to the bond $ij$, within tetrahedron $t$.
The  exchange interactions ${\bf J}_{ij}$ do not, in general, possess {\it any} 
continuous spin-rotation invariance.
Nonetheless, the form of exchange ${\bf J}_{ij}$ {\it is} strongly constrained by 
the symmetry of the bond $ij$, and the interactions on different bonds must also 
be related by lattice symmetries.   

\begin{figure}
\centering\includegraphics[width=0.8\columnwidth]{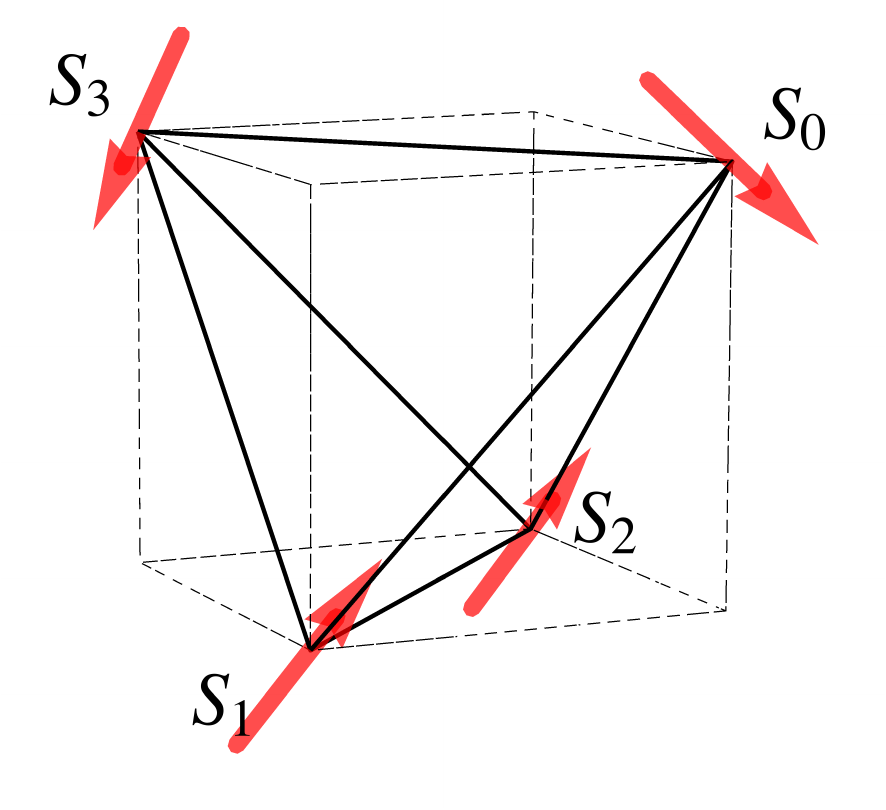}
\caption{
\label{fig:tetrahedron}
A single tetrahedron within the pyrochlore lattice, 
showing the convention used in labelling sites.  
The positions of the magnetic sites relative to the 
centre of the tetrahedron are defined in 
Appendix~\ref{appendix:local-coordinate-frame}. 
}
\end{figure}

Once these constraints are taken into account~\cite{curnoe07}, ${\bf J}_{ij}$ 
is a function of just four independent parameters and, for the six bonds which 
make up the tetrahedron shown in Fig.~\ref{fig:tetrahedron}, can be written 
\begin{eqnarray}
&{\bf J}_{01} 
  = \begin{pmatrix}
    J_2 & J_4 &J_4 \\
   -J_4 & J_1 &J_3 \\
   -J_4 & J_3 &J_1
   \end{pmatrix} 
   \quad
&{\bf J}_{02} 
  = \begin{pmatrix}
    J_1 & -J_4 & J_3 \\
    J_4 & J_2 & J_4 \\
    J_3 & -J_4 & J_1
   \end{pmatrix}    \nonumber\\
&{\bf J}_{03} 
  = \begin{pmatrix}
    J_1 & J_3 & -J_4 \\
    J_3 & J_1 & -J_4 \\
    J_4 & J_4 & J_2
   \end{pmatrix} 
   \quad
&{\bf J}_{12} 
  = \begin{pmatrix}
    J_1 & -J_3 & J_4 \\
    -J_3 & J_1 & -J_4 \\
    -J_4 & J_4 & J_2
   \end{pmatrix}    \nonumber\\
&{\bf J}_{13} 
  = \begin{pmatrix}
     J_1 & J_4 & -J_3 \\
     -J_4 & J_2 & J_4 \\
     -J_3 & -J_4 & J_1
   \end{pmatrix} 
   \quad
&{\bf J}_{23} 
  = \begin{pmatrix}
    J_2 & -J_4 & J_4 \\
    J_4 & J_1 & -J_3 \\
    -J_4 & -J_3 & J_1
   \end{pmatrix} \nonumber\\ 
   \label{eq:Jij}
\end{eqnarray}
where we label lattice sites and interactions following the conventions
of Ross {\it et al.}~[\onlinecite{ross11-PRX1}].  
The structure of these matrices imply that the different contributions to 
the interaction ${\bf J}_{ij}$ [Eq.~(\ref{eq:Jij})] can be approximately identified as 
\begin{itemize}
\item $J_1\rightarrow$ ``XY'' with respect to the local bond 
\item $J_2\rightarrow$ ``Ising'' with respect to the local bond 
\item $J_3\rightarrow$ symmetric off diagonal exchange
\item $J_4\rightarrow$ Dzyaloshinskii-Moriya
\end{itemize}


\begin{table}
\begin{tabular}{ | c | c | c | c |}
\hline
    & 
    Yb$_2$Ti$_2$O$_7$~[\onlinecite{ross11-PRX1}] & 
    Er$_2$Ti$_2$O$_7$~[\onlinecite{savary12-PRL109}] & 
    Er$_2$Sn$_2$O$_7$~[\onlinecite{guitteny13}] \\
\hline
\hline
$J_1$ & $-0.09\ \text{meV}$ & $0.11\ \text{meV}$  & $0.07\ \text{meV}$ \\
\hline
$J_2$ & $-0.22\ \text{meV}$ & $-0.06\ \text{meV}$  & $0.08\ \text{meV}$ \\
\hline
$J_3$  & $-0.29\ \text{meV}$ & $-0.10\ \text{meV}$  & $-0.11\ \text{meV}$ \\
\hline
$J_4$ & $0.01\ \text{meV}$ & $-0.003\ \text{meV}$  & $0.04\ \text{meV}$ \\
\hline
\hline
$J_{zz}$ & $0.17\ \text{meV}$ & $-0.025\ \text{meV}$  & $0$ \\
\hline
$J_{\pm}$ & $0.05\ \text{meV}$ & $0.065\ \text{meV}$  & $0.014\ \text{meV}$ \\
\hline
$J_{\pm \pm}$ & $0.05\ \text{meV}$ & $0.042\ \text{meV}$  & $0.074\ \text{meV}$ \\
\hline
$J_{z \pm}$ & $-0.14\ \text{meV}$ & $-0.009\ \text{meV}$  & $0$ \\
\hline
\end{tabular}
\caption{
Estimates of the parameters for anisotropic near-neighbour exchange,  
taken from experiments on Yb$_2$Ti$_2$O$_7$~[\onlinecite{ross11-PRX1}], 
Er$_2$Ti$_2$O$_7$~[\onlinecite{savary12-PRL109}], 
and Er$_2$Sn$_2$O$_7$~[\onlinecite{guitteny13}].  
Values are quoted for exchange interactions in both the crystal coordinate frame 
$\mathcal{H}_{\sf ex}$~[Eq.~(\ref{eq:Hex1})], and the local coordinate frame, 
${\mathcal H}^{\sf local}_{\sf ex}$~[Eq.~(\ref{eq:Hross})], 
following the notation of \mbox{Ross {\it et al.}~[\onlinecite{ross11-PRX1}]}.  
An alternative set of parameters for Yb$_2$Ti$_2$O$_7$
has recently been proposed by Robert {\it et al.}~[\onlinecite{robert15}].
}
\label{table:J-from-experiment}
\end{table}


The anisotropic nearest--neighbour exchange model, 
$\mathcal{H}_{\sf ex}$~[Eq.~(\ref{eq:Hex1})],  has been applied 
with considerable success to a number of pyrochlore oxides.
In the case of Yb$_2$Ti$_2$O$_7$, $\mathcal{H}_{\sf ex}$~[Eq.~(\ref{eq:Hex1})] 
has been shown to give an excellent description of
spin-wave spectra measured in magnetic field~\cite{ross11-PRX1}.
Thermodynamic quantities, calculated from $\mathcal{H}_{\sf ex}$~[Eq.~(\ref{eq:Hex1})]
using the parameters from [\onlinecite{ross11-PRX1}], also gave a very good
description of experiments~\cite{applegate12,hayre12}.
Parameters for Er$_2$Ti$_2$O$_7$ have been extracted from equivalent 
inelastic neutron scattering experiments\cite{savary12-PRL109}, 
and from measurements of the field-dependence of magnetisation at low 
temperature~\cite{bonville13}.
The model $\mathcal{H}_{\sf ex}$~[Eq.~(\ref{eq:Hex1})], using parameters taken 
from neutron scattering~\cite{savary12-PRL109}, has been shown to give 
good agreement with the observed spin wave spectrum in
Er$_2$Ti$_2$O$_7$,
consistent with quantum order--by--disorder~[\onlinecite{savary12-PRL109}].
Anisotropic nearest--neighbour exchange parameters for Er$_2$Sn$_2$O$_7$
have also been estimated from measurements of the 
magnetization curve~\cite{guitteny13}.


Representative estimates of the exchange parameters 
$(J_1,\, J_2,\, J_3,\, J_4)$ taken from experiment on 
Yb$_2$Ti$_2$O$_7$~[\onlinecite{ross11-PRX1}], 
Er$_2$Ti$_2$O$_7$~[\onlinecite{savary12-PRL109}], 
and Er$_2$Sn$_2$O$_7$~[\onlinecite{guitteny13}], 
are shown in \mbox{Table~\ref{table:J-from-experiment}}. 
The typical scale of interactions is $|J| \sim 0.1\ \text{meV}$ 
\mbox{(i.e. $|J| \sim 1\ \text{K}$)}, with typical uncertainty in estimates of order 
$\delta J \sim 0.02\ \text{meV}$ 
(i.e. $\delta J \sim 0.2\ \text{K}$)~[\onlinecite{ross11-PRX1,savary12-PRL109,guitteny13}].  
In all of these cases, the symmetric off-diagonal exchange interaction \mbox{$J_3$} is negative, 
while the Dzyaloshinskii-Moriya interaction $J_4$ is relatively small~\footnote{
In contrast, it has recently been found that Dzyaloshinskii-Moriya interactions 
in the ``breathing--pyrochlore'' material Ba$_3$Yb$_2$Zn$_5$O$_{11}$ are 
rather large \cite{rau16, haku16}.}.

\subsection{Anisotropic exchange in a local frame}
\label{section:local-frame}


Since both the anisotropy in magnetic ground state of the rare-earth ion, 
and the anisotropy in its interactions, are dictated by the local CEF field, 
it is often convenient to describe them in a local coordinate frame 
$$
\{ {\bf x}_i^{\sf \ local},  {\bf y}_i^{\sf \ local}, {\bf z}_i^{\sf \ local} \}
$$
 such that the axis ${\bf z}_i^{\sf \ local} $aligns with the
$C_3$ symmetry axis of  the local CEF on site $i$, as described in 
Appendix~\ref{appendix:local-coordinate-frame}.  
We introduce a $SU(2)$ (pseudo) spin-1/2 in this local frame
\begin{eqnarray}
[\mathsf{S}_i^\alpha ,  \mathsf{S}_i^\beta] 
   = i \epsilon_{\alpha\beta\gamma} \mathsf{S}_i^\gamma \; ,
\label{eq:Slocal}
\end{eqnarray}
where $\alpha$, $\beta$, $\gamma = \{ {\bf x}_i^{\sf \ local},  {\bf y}_i^{\sf \ local}, {\bf z}_i^{\sf \ local} \}$.
Note that throughout the manuscript ${\sf S}_i^{\alpha}$ will refer to the spin components in
this local frame, while $S_i^{\alpha}$ refers to the spin components in the global, crystal,
coordinate system.

In the local coordinate frame, the most general form of exchange interactions between Kramers ions on the 
pyrochlore lattice can be written~\cite{ross11-PRX1}
\begin{eqnarray}
{\mathcal H}^{\sf local}_{\sf ex} 
    &=& \sum_{\langle ij\rangle} 
               \Big\{ J_{zz} \mathsf{S}_i^z \mathsf{S}_j^z - J_{\pm}
                (\mathsf{S}_i^+ \mathsf{S}_j^- + \mathsf{S}_i^- \mathsf{S}_j^+) 
                \nonumber \\ 
       && + J_{\pm\pm} \left[\gamma_{ij} \mathsf{S}_i^+ \mathsf{S}_j^+ + \gamma_{ij}^*
                 \mathsf{S}_i^-\mathsf{S}_j^-\right] 
                 \nonumber \\
       && + J_{z\pm}\left[ 
                           \mathsf{S}_i^z (\zeta_{ij} \mathsf{S}_j^+ + \zeta^*_{ij} \mathsf{S}_j^-) 
                           + {i\leftrightarrow j}
                 \right]\Big\}
\label{eq:Hross}                 
\end{eqnarray}
where the matrix
\begin{equation}
\zeta 
    = \left(\begin{array}{cccc}
            0 & -1 & e^{i\frac{\pi}{3}} & e^{-i\frac{\pi}{3}}\\
           -1 & 0 & e^{-i\frac{\pi}{3}} & e^{i\frac{\pi}{3}}\\
            e^{i\frac{\pi}{3}} & e^{-i\frac{\pi}{3}} & 0 & -1\\
            e^{-i\frac{\pi}{3}} & e^{i\frac{\pi}{3}} & -1 & 0
      \end{array}\right)
    \quad 
       \gamma=-\zeta^*
\end{equation}
encode the change in coordinate frame between different sublattices.


The relationship between the parameters in this local frame, 
$(J_{zz}, J_{\pm},J_{\pm \pm},J_{z \pm})$, and exchange parameters in the 
global frame of the crystal axes $(J_1,\, J_2,\, J_3,\, J_4)$, is given in 
Table~\ref{table:J-in-local-frame}.
Corresponding estimated parameters from experiment on 
Yb$_2$Ti$_2$O$_7$~[\onlinecite{ross11-PRX1}], 
Er$_2$Ti$_2$O$_7$~[\onlinecite{savary12-PRL109}], 
and Er$_2$Sn$_2$O$_7$~[\onlinecite{guitteny13}], 
are shown in \mbox{Table~\ref{table:J-from-experiment}}. 


\begin{table}
\begin{tabular}{ | c | c |}
\hline
\multirow{2}{*}{}
   interaction in local & exchange parameters   
   \\
   coordinate frame & in global frame
\\
\hline
\multirow{1}{*}{}
$J_{zz}$ & $-\frac{1}{3} (2 J_1 - J_2 + 2J_3 + 4J_4) $\\
\multirow{1}{*}{}
$J_{\pm}$ & $\frac{1}{6} (2 J_1 - J_2 -J_3 - 2 J_4)$ \\
\multirow{1}{*}{}
$J_{\pm \pm}$ & $\frac{1}{6} (J_1 + J_2 - 2 J_3 + 2 J_4)$ \\
\multirow{1}{*}{}
$J_{z \pm}$ & $\frac{1}{3 \sqrt{2}} (J_1 + J_2 + J_3-J_4)$ \\
\hline
\end{tabular}
\caption{
Relationship between the parameters of the anisotropic nearest--neighbour 
exchange model in the local coordinate frame, 
${\mathcal H}^{\sf local}_{\sf ex}$~[Eq.~(\ref{eq:Hross})], and the exchange parameters 
in the crystal coordinate frame $\mathcal{H}_{\sf ex}$~[Eq.~(\ref{eq:Hex1})].
The notation used for the different components of the interaction 
follows Ross {\it et al.}~[\onlinecite{ross11-PRX1}].
}
\label{table:J-in-local-frame}
\end{table}

\subsection{Proof of the existence of a classical ground state with ${\bf q}=0$, 
4-sublattice order}
\label{theorem-on-q=0}

Finding the ground state of $\mathcal{H}_{\sf ex}$~[Eq.~(\ref{eq:Hex1})], for a 
quantum (pseudo)spin-1/2, and arbitrary exchange interactions 
$(J_1, J_2, J_3, J_4)$, is a very difficult problem, in general only tractable
as a mean-field theory~\cite{savary12-PRL108,lee12,savary13,wong13,hao14}.
However, many rare-earth pyrochlores are known to have relatively simple 
ground states, with vanishing crystal momentum ${\bf q}=0$, implying a 4-sublattice 
magnetic order~\cite{gardner10}.  
Here we show that, under the restriction that ${\bf S}_i$ is a classical variable, 
$\mathcal{H}_{\sf ex}$ [Eq.~(\ref{eq:Hex})] {\it always} possesses a ground state of this type. 
In Section~\ref{Lego-brick-rules}, below, we explore the conditions under which this 
classical ground state is unique.


We begin with the simple observation that, since $\mathcal{H}_{\sf ex}$~[Eq.~(\ref{eq:Hex1})] is 
expressed as a sum over individual tetrahedra, any state which minimizes the energy of each 
individual tetrahedron must be a ground state.
It is convenient to split this sum into two pieces 
\begin{eqnarray}
\mathcal{H}_{\sf ex} 
  = \sum_{t \in \sf A} \mathcal{H}_{\sf ex}^{\sf A}[t]
  + \sum_{t' \in \sf B} \mathcal{H}_{\sf ex}^{\sf B}[t'] ;\,
\label{eq:HAandB}
\end{eqnarray}
where $\sf A$ and $\sf B$ refer to the two distinct sublattices of tetrahedra, with
\begin{eqnarray}
  \mathcal{H}_{\sf ex}^{\sf A}[t]&= & \sum_{i,j \in  t} {\mathbf S}_i {\bf J}_{ij}^{\sf [A]} {\mathbf S}_j \\
  \mathcal{H}_{\sf ex}^{\sf B}[t']&= & \sum_{i,j \in  t'} {\mathbf S}_i {\bf J}_{ij}^{\sf [B]} {\mathbf S}_j\;.
\end{eqnarray} 
The interactions ${\bf J}_{ij}^{\sf [A]} $ and ${\bf J}_{ij}^{\sf [B]} $ are related by inversion 
about a single site $\mathcal{I}$ 
\begin{eqnarray}
{\bf J}_{ij}^{\sf [B]}  &=&  \mathcal{I} \cdot {\bf J}_{ij}^{\sf [A]} \cdot \mathcal{I} \; .
\label{eq:inversion}
\end{eqnarray} 
Since $\mathcal{I}^2=1$, we can write
\begin{eqnarray}
{\bf S}_i \cdot {\bf J}_{ij}^{\sf [A]} \cdot {\bf S}_j
   &=& {\bf S}_i \cdot \mathcal{I}^2 \cdot {\bf J}_{ij}^{\sf [A]} \cdot \mathcal{I}^2  \cdot {\bf S}_j \nonumber\\
   &=& {\bf S}_i \cdot \mathcal{I} \cdot {\bf J}_{ij}^{\sf [A]} \cdot \mathcal{I}  \cdot {\bf S}_j 
\end{eqnarray}   
where we have used the fact that the spin ${\bf S}_i$ is invariant under lattice inversion.     
This implies 
\begin{eqnarray}   
   {\bf J}_{ij}^{\sf [A]}     &=&  \mathcal{I} \cdot {\bf J}_{ij}^{\sf [A]} \cdot \mathcal{I} = {\bf J}_{ij}^{\sf [B]} \; .
\end{eqnarray}
It follows that interactions  for {\it any} tetrahedron $t$ is the same, 
regardless of which tetrahedral sublattice it belongs to, and we can safely write 
\begin{eqnarray}
\mathcal{H}_{\sf ex}^{\sf tet}  = \sum_{i,j \in t} {\mathbf S}_i {\bf J}_{ij} {\mathbf S}_j 
\label{eq:Htet}
\end{eqnarray}
where ${\bf J}_{ij} $ are given by Eq.~(\ref{eq:Jij}) and the sum runs over all pairs of  sites
$i,j$ in a given tetrahedron $t$, which may now be of either sublattice.


The proof we are seeking follows directly from this 
result [Eq.~(\ref{eq:Htet})]~:
for classical spins, \mbox{$[\mathcal{H}_{\sf ex}^{\sf A} , 
\mathcal{H}_{\sf ex}^{\sf B}] = 0$},
and we can construct a ground state of $\mathcal{H}_{\sf ex}$ 
by choosing {\it any} state which
minimizes the energy of a single tetrahedron, and repeating it 
across all ${\sf A}$-sublattice
(or ${\sf B}$-sublattice) tetrahedra. 
Since every spin is shared between one ${\sf A}$-- and 
one ${\sf B}$--sublattice tetrahedron, 
and the Hamiltonians for ${\sf A}$-- or ${\sf B}$--sublattices 
are equivalent [Eq.~(\ref{eq:Htet})], 
any such classical spin-configuration which minimizes the 
energy on one tetrahedral sublattice, simultaneously 
minimizes the energy on the other tetrahedral sublattice, 
and is a ground state of $\mathcal{H}_{\sf ex}$~[Eq.~(\ref{eq:Hex1})].


It follows that there {\it always} exists a classical, ${\mathbf q} = 0$ ground state of 
$\mathcal{H}_{\sf ex}$~[Eq.~(\ref{eq:Hex1})], with 4-sublattice long-range magnetic order, 
for arbitrary exchange interactions $(J_1, J_2, J_3, J_4)$.
This is true {\it even} in the presence of finite Dzyaloshinskii-Moriya interaction $J_4$.


Such a ${\mathbf q} = 0$ ground state has a finite, discrete degeneracy associated with the breaking 
of point-group and time-reversal symmetries
(in the case of classical spins, time-reversal corresponds to the inversion of all spins 
${\bf S}_i \to - {\bf S}_i$).   
This degeneracy must be at least $2$ (time-reversal), and is typically $6$ 
($C_3$ rotations $\otimes$ time-reversal), for the ordered phases considered in this article.

\subsection{Conditions for the uniqueness of 4-sublattice order --- the ``Lego--brick'' rules}
\label{Lego-brick-rules}


The existence of a classical ground state of $\mathcal{H}_{\sf ex}$~[Eq.~(\ref{eq:Hex1})] with 
4-sublattice magnetic order, for arbitrary exchange interactions $(J_1, J_2, J_3, J_4)$ 
constitutes a enormous simplification, since it is much easier to determine the 
spin-configuration which minimizes the energy of a single 
tetrahedron (as described in Section~\ref{section:classical-ground-states}, below) 
than to find the ground state of the entire lattice.   
However, as we shall see, many of interesting properties of 
rare-earth pyrochlores follow 
from the fact that while such a classical ground state must 
exist, it need not be unique.


Establishing the uniqueness of a 4--sublattice ground state, up to the discrete degeneracy
of the state itself, amounts to determining the number of ways in which the spin configurations 
which minimize the energy of a single tetrahedron can be used to tile the entire lattice.  
For many purposes, it is convenient to think of these as a set of ``Lego--brick'' rules for 
fitting together spin-configurations on a lattice.
These rules allow us to determine the degeneracy, and nature, of the ground states of the
whole lattice, using the ground states of a single tetrahedron.


The rules can be stated as follows:
\begin{enumerate}
\item{
If the spin on every site of the tetrahedron points in a different direction, 
in each of its classical ground--states, then the 4--sublattice  
ground state of the lattice is unique (up to global symmetry operations).
In this case, the degeneracy of the ground state of the lattice is the same as that of 
a single tetrahedron.}
\item{If, within the set of
  ground states 
for a single tetrahedron,
there are two states in which a single site has the same spin orientation, 
the 4--sublattice ground state of the lattice is not unique.  
In this case, the system undergoes a dimensional reduction into 
independent kagome planes, and the number of classical ground states 
is at least $\mathcal{O}(2^L)$, where $L$ is the linear size of the system.}
\item{
If, within the set of ground states 
for a single tetrahedron, there is a pair of states which have the
same spin orientation
on {\it two} sites, the 4--sublattice ground state is also not unique.   
However in this case, the number of classical ground states must grow 
as {\it at least} $\mathcal{O}(2^{L^2})$, corresponding to dimensional 
reduction into independent chains of spins.
In the special case of spin--ice, 
the corresponding classical ground--state degeneracy is extensive.
}
\end{enumerate}


\begin{figure}
\centering
\includegraphics[width=0.9\columnwidth]{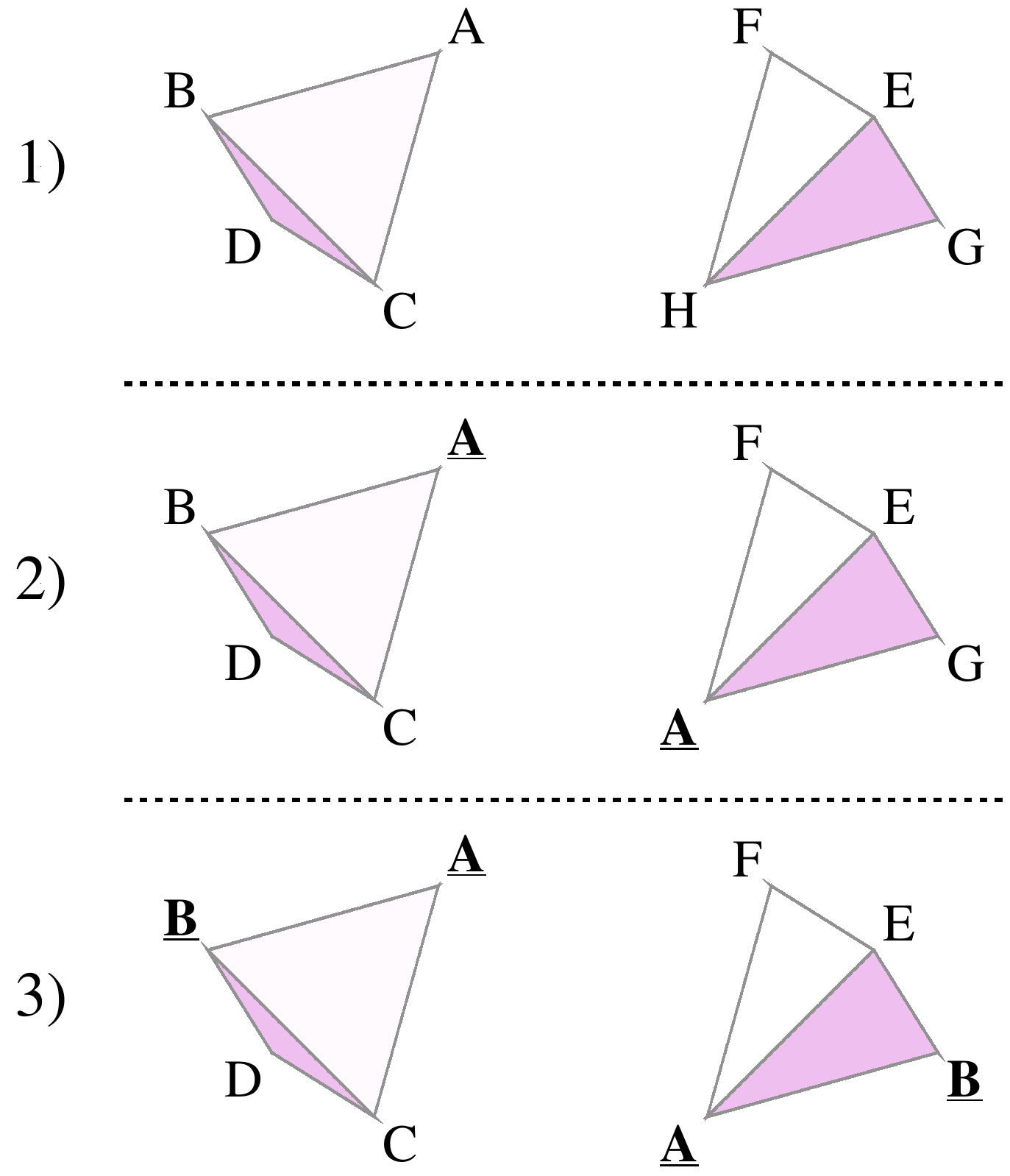}
\caption{
The ``Lego--brick'' rules describing how the ground states of a single 
tetrahedron can be connected to tile the pyrochlore lattice.  
The two tetrahedra in the left and right panels represent a pair
of tetrahedra in distinct ground--state configurations.
Distinct spin orientations on the sites of each tetrahedron 
are denoted by letters {\bf a}---{\bf h}.   
Three cases are shown.
In case (1)~all of the ground states for a single tetrahedron have 
     different spin orientations for any given site. 
This means that
two tetrahedra in distinct ground states cannot be joined together
because they do not share a common spin orientation on any site.
In this case the 4--sublattice ground state of the lattice is unique with $q=0$ order, 
up to global symmetry operations.   
In case (2)~ there are two
ground states configurations for a tetrahedron which 
share a common spin orientation on a single site, 
here denoted {\bf \underline{a}}.  
These tetrahedra can be joined together by sharing the spin in orientation 
{\bf \underline{a}}.
In this case the ground state of the lattice has 
a degeneracy of at least $\mathcal{O}(2^L)$. 
Indeed, successive kagome layers of spins can be independently in BDC or FEG configurations.
In case (3) there is a pair of ground states which
 share common spin orientations on two sites,
here denoted {\bf \underline{a}} and {\bf \underline{b}}.
These tetrahedra can be joined together by sharing 
the spin in orientation  {\bf \underline{a}}
or the spin in orientation {\bf \underline{b}}.
In this case the ground state  of the lattice 
has a degeneracy of at least $\mathcal{O}(2^{L^2})$.   
}
\label{fig:lego.brick.rules}
\end{figure}    


The first rule guarantees the uniqueness of a 4-sublattice ground 
state, where the spin on every site of the tetrahedron points in a different direction 
in each of the ground states of a single tetrahedron.   
Away from phase boundaries, this is true for {\it all} of the 4-sublattice ${\mathbf q} = 0$ 
ordered phases discussed in this article.
However, it is clear from Rules 2 and 3 that 
if two of the ground states of a single tetrahedron share a common spin --- i.e.  
the spin on a given site points in the same direction in more than one ground state --- then 
it is {\it always} possible to construct other ground states, with finite ${\mathbf q}$.


To give a concrete example of how these ``Lego--brick'' rules work, let us assume that two different ground 
states for a single tetrahedron have identical orientation of the spin on site $0$, but different 
orientation of the spins on sites $1$, $2$ and $3$.   
In this case it is possible to divide the pyrochlore lattice into a set of parallel kagome planes,
containing spins associated with sites $1$,$2$ and $3$ of a tetrahedron, separated by 
triangular-lattice planes associated with site $0$.   
Since each successive kagome plane can take on one of two different spin 
configurations, the number of such ground states grows as $2^{\sf N_K}$, 
where ${\sf N_K}$ is the number of kagome planes, 
and encompasses all possible ${\mathbf q} \parallel [111]$.
Dimensional reduction of this type occurs on the classical phase 
boundary between ordered FM and Palmer--Chalker phases discussed in 
Section~\ref{FM-meets-PC} of this article.


An example where Rule~3 applies, and a set of independent chains emerges
in the ground state manifold, is the phase boundary 
between the Palmer--Chalker phase 
and the non--coplanar antiferromagnet 
discussed in Section~\ref{ssec:T2E}.
However the ``Lego--brick'' rules permit even larger ground--state 
degeneracies, as is known from the ``two--in, two--out"  states, 
made famous by the spin--ice problem.
In this case there are a total of $6$ possible ground states for a single 
tetrahedron, but each possible spin orientation, on each site, belongs 
to $3$ different ground states.
According to Rule~3, the 4--sublattice classical ground-state
--- a ferromagnet --- should not be unique, and the total number 
of classical grounds states must grow as {\it at least} $\mathcal{O}(2^{L^2})$.   
In fact, the ground--state degeneracy of spin ice is extensive, scaling 
as \mbox{$\Omega_{\sf ice} \sim (3/2)^{N/2}$}, where $N$ is the total 
number of sites in the lattice~\cite{pauling35}.  
This manifold of spin--ice states includes ground states with 
all possible ${\mathbf q}$.


\begin{table*}
\begin{tabular}{ | c | c |  c | }
\hline
\multirow{2}{*}{}
   order  & 
   definition in terms   & 
   associated
\\
   parameter & 
   of spin components & 
   ordered phases 
\\
\hline
\multirow{1}{*}{}
   $m_{\sf A_2}$ & 
   $\frac{1}{2 \sqrt{3} } 
     \left(S_0^x+S_0^y+S_0^z+S_1^x-S_1^y-S_1^z-S_2^x+S_2^y-S_2^z-S_3^x-S_3^y+S_3^z
     \right)$ & 
    ``all in-all out'' 
\\   
\hline
\multirow{1}{*}{} 
   ${\bf m}_{\sf E}$ & 
   $\begin{pmatrix}
         \frac{1}{2 \sqrt{6} } \left( -2 S_0^x + S_0^y + S_0^z - 2 S_1^x - S_1^y-S_1^z+2 S_2^x + S_2^y-
              S_2^z +2 S_3^x-S_3^y +S_3^z \right) \\
         \frac{1}{2 \sqrt{2}} \left( -S_0^y+S_0^z+S_1^y-S_1^z-S_2^y-S_2^z+S_3^y+S_3^z \right)
     \end{pmatrix}$ &
     $\Psi_2$ and $\Psi_3$ \\ 
\hline
\multirow{1}{*}{}
   ${\bf m}_{\sf T_{1, A}}$  & 
   $\begin{pmatrix}
        \frac{1}{2} (S_0^x+S_1^x+S_2^x+S_3^x) \\
        \frac{1}{2} (S_0^y+S_1^y+S_2^y+S_3^y) \\
        \frac{1}{2} (S_0^z+S_1^z+S_2^z+S_3^z)
   \end{pmatrix} $ &
   collinear FM
\\
\hline
\multirow{1}{*}{}
   ${\bf m}_{\sf T_{1, B}}$  & 
   $\begin{pmatrix}
          \frac{-1}{2\sqrt{2}} (S_0^y+S_0^z-S_1^y-S_1^z-S_2^y+S_2^z+S_3^y-S_3^z)  \\
          \frac{-1}{2\sqrt{2}} (S_0^x+S_0^z-S_1^x+S_1^z-S_2^x-S_2^z+S_3^x-S_3^z)  \\
          \frac{-1}{2\sqrt{2}} ( S_0^x+S_0^y-S_1^x+S_1^y+S_2^x-S_2^y-S_3^x-S_3^y) 
   \end{pmatrix}$ &
    non-collinear FM 
\\
\hline
\multirow{1}{*}{}
   ${\bf m}_{\sf T_2} $ & 
   $\begin{pmatrix}
        \frac{1}{2 \sqrt{2}} 
        \left(
         -S_0^y+S_0^z+S_1^y-S_1^z+S_2^y+S_2^z-S_3^y-S_3^z
        \right) 
        \\
        \frac{1}{2 \sqrt{2}} 
        \left(
        S_0^x-S_0^z-S_1^x-S^z_1-S_2^x+S_2^z+S_3^x+S_3^z
        \right) \\
        \frac{1}{2 \sqrt{2} }
        \left(
        -S_0^x+S_0^y+S_1^x+S_1^y-S_2^x-S_2^y+S_3^x-S_3^y
        \right)
      \end{pmatrix} $  &
       Palmer--Chalker ($\Psi_4$)     
\\ 
\hline
\end{tabular}
\caption{
Order parameters ${\bf m}_\lambda$, describing how the point-group symmetry of a 
single tetrahedron within the pyrochlore lattice is broken by magnetic order.    
Order parameters transform according to irreducible representations of the point-group 
${\sf T}_d$, and are expressed in terms of linear combinations of spin-components 
${\bf S}_i = (S^x_i, S^y_i, S^z_i)$, 
in the global frame of the crystal axes --- cf. $\mathcal{H}_{\sf ex}$~[Eq.~(\ref{eq:Hex1})].   
Labelling of spins within the tetrahedron follows the convention of 
Ross~{\it et al.}~[\onlinecite{ross11-PRX1}] --- cf. Fig.~\ref{fig:tetrahedron}.  
The notation $\Psi_i$ for ordered phases is taken from~[\onlinecite{poole07}].
}
\label{table:m_lambda_global}
\end{table*}

\subsection{Representation theory}
\label{representation-theory} 

Except in very specific limits, such as the Heisenberg model 
($J_1 = J_2 = J, J_3 = J_4 = 0$), Hamiltonian $\mathcal{H}_{\sf ex}$~[Eq.~(\ref{eq:Hex1})] does 
not possess {\it any} continuous spin-rotation symmetry.
The key to unlocking its properties, therefore, is to understand how different 
ordered states, and indeed different spin-fluctuations, break the space-group
symmetries of the pyrochlore lattice.
This task is made easier by the realization that a classical ground state 
with ${\bf q}=0$ exists for all possible $(J_1, J_2, J_3 , J_4)$, 
as discussed in Section~\ref{theorem-on-q=0}.  
It is therefore possible to restrict discussion to the point-group
symmetries of the lattice.   


In what follows, we explore the consequences of applying representation
theory for these point-group operations to a general model 
of anisotropic nearest-neighbor exchange on the pyrochlore lattice, 
$\mathcal{H}_{\sf ex}$~[Eq.~(\ref{eq:Hex1})].  
This  analysis serves a two-fold purpose~: it reduces the Hamiltonian 
for a single tetrahedron $\mathcal{H}_{\sf ex}^{\sf tet}$~[Eq.~(\ref{eq:Htet})] 
to a diagonal form, and it provides a set of order 
parameters with which to characterise the ${\bf q}=0$, 4-sublattice ordered phases 
found in real materials.


The point-group symmetry of the pyrochlore lattice is the cubic symmetry group 
${\sf O}_h = {\sf T}_d \times {\sf I}$.
Here ${\sf T}_d$ is symmetry group of a single tetrahedron, 
and ${\sf I} = \{\epsilon, \mathcal{I} \}$, where $\epsilon$ is the identity
and $\mathcal{I}$ corresponds to the lattice inversion introduced 
in Eq.~(\ref{eq:inversion}). 
For classical spins, lattice inversion plays a benign role 
(cf. Section~\ref{theorem-on-q=0}), and it is sufficient to consider 
${\sf T}_d$ alone.
The group ${\sf T}_d$ has 24 elements~\cite{kovalev93}, corresponding to the symmetries 
of the tetrahedron~:
$8 \times C_3$ --- $\frac{2 \pi}{3}$ rotation around a $[111]$ axis;   
$3 \times C_2$ --- $\pi$ rotation around $[100]$ axis;
$6 \times S_4$ --- $\frac{\pi}{2}$ rotation around a $[100]$ axis followed by reflection 
in the same $[100]$ plane; 
$6 \times \sigma_d$ --- reflection in $[011]$ plane; 
and $\epsilon$ --- the identity \cite{dresselhaus10}.   


The different ways in which classical ground states with ${\bf q}=0$ break the symmetries of a 
tetrahedron can be fully characterised by introducing order parameters ${\bf m}_\lambda$
which transform with the non-trivial irreducible representations 
\mbox{$\lambda = \{\ {\sf A}_2$, ${\sf E}$, ${\sf T}_1$, ${\sf T}_2\ \}$} 
of ${\sf T}_d$.
These order parameters are formed by linear combinations of spin components, and can be 
expressed in either global coordinate frame of $\mathcal{H}_{\sf ex}$~[Eq.~(\ref{eq:Hex1})] 
-- cf. Table~\ref{table:m_lambda_global} -- 
or in the local coordinate frame of  ${\mathcal H}_{\sf ex}^{\sf local}$~[Eq.~(\ref{eq:Hross})] 
-- cf. Table~\ref{table:m_lambda_local}.  


The anisotropic exchange Hamiltonian $\mathcal{H}_{\sf ex}^{\sf tet}$~[Eq.~(\ref{eq:Htet})] 
can be transcribed exactly in terms of same set of irreps.
\begin{eqnarray}
{\mathcal H}_{\sf ex}^{\sf tet}   
  &\equiv&  \frac{1}{2}
    \left[ 
       a_{\sf A_2} \, m_{\sf A_2}^2 
       + a_{\sf E}\, {\bf m}^2_{\sf E} 
       + a_{\sf T_2}\, {\bf m}^2_{\sf T_2} 
       + a_{\sf T_{1, A}}\, {\bf m}^2_{\sf T_{1, A}} 
    \right.
    \nonumber\\
    && \quad \left. 
       + a_{\sf T_{1, B}}\, {\bf m}^2_{\sf T_{1, B}} 
       + a_{\sf T_{1, AB}}\, {\bf m}_{\sf T_{1, A}} \cdot {\bf m}_{\sf T_{1, B}}
    \right], 
\label{eq:HTd}
\end{eqnarray}
where the coefficients 
\begin{eqnarray}
a_{\sf A_2} &=& -2 J_1+J_2-2(J_3+2J_4)\nonumber\\
a_{\sf E} &=& -2 J_1+J_2+J_3+2J_4\nonumber\\ 
a_{\sf T_2} &=& -J_2+J_3-2J_4\nonumber\\
a_{\sf T_{1,A}} &=& 2 J_1+J_2\nonumber\\ 
a_{\sf T_{1, B}} &=& -J_2-J_3+2J_4\nonumber\\ 
a_{\sf T_{1, AB}} &=& -\sqrt{8} J_3
\end{eqnarray}  
are completely determined by the parameters of $\mathcal{H}_{\sf ex}$~[Eq.~(\ref{eq:Hex1})].
Equivalent expressions for $a_\lambda$ can be found in terms of the parameters 
of  ${\mathcal H}_{\sf ex}^{\sf local}$~[Eq.~(\ref{eq:Hross})]. 
We note that a similar analysis, applied to Er$_{2}$Ti$_{2}$O$_{7}$, 
appears in Ref.~[\onlinecite{mcclarty09}], with
a different
choice of basis for the two appearances of the $\sf T_1$ irrep.
Note that this is entirely different from the analysis appearing in
Ref. [\onlinecite{curnoe07}] which classifies wavefunctions for spin-$1/2$ 
on a tetrahedron according to the irreps of $T_d$, not linear combinations of
spin operators.


\begin{table*}
\begin{tabular}{ | c | c |  }
\hline
\multirow{2}{*}{}
   order  & 
   definition in terms  
\\
   parameter & 
   of local spin components 
\\
\hline
\multirow{1}{*}{}
   $m_{\sf A_2}$ & 
   $\frac{1}{2} 
     \left( \mathsf{S}_0^z+\mathsf{S}_1^z+\mathsf{S}_2^z+\mathsf{S}_3^z
     \right)$ 
\\   
\hline
\multirow{1}{*}{} 
   ${\bf m}_{\sf E}$ & 
   $\dfrac{1}{2}
   \begin{pmatrix}
        \mathsf{S}_0^x+\mathsf{S}_1^x+\mathsf{S}_2^x+\mathsf{S}_3^x\\
        \mathsf{S}_0^y+\mathsf{S}_1^y+\mathsf{S}_2^y+\mathsf{S}_3^y 
     \end{pmatrix}$  \\ 
\hline
\multirow{1}{*}{}
   ${\bf m}_{\sf T_{1, A}}$  & 
   $\begin{pmatrix}
        \frac{1}{2\sqrt 3} (-\sqrt 2 \mathsf{S}_0^x + \mathsf{S}_0^z - \sqrt 2 \mathsf{S}_1^x + \mathsf{S}_1^z + \sqrt 2 \mathsf{S}_2^x - \mathsf{S}_2^z + 
 \sqrt 2 \mathsf{S}_3^x - \mathsf{S}_3^z) \\
        \frac{1}{12} (\sqrt 6 \mathsf{S}_0^x - 3 \sqrt 2  \mathsf{S}_0^y + 2 \sqrt 3 \mathsf{S}_0^z - \sqrt 6  \mathsf{S}_1^x + 
  3 \sqrt 2  \mathsf{S}_1^y - 2 \sqrt 3 \mathsf{S}_1^z + \sqrt 6 \mathsf{S}_2^x - 3 \sqrt 2 \mathsf{S}_2^y + 
  2 \sqrt 3 \mathsf{S}_2^z - \sqrt 6  \mathsf{S}_3^x + 3 \sqrt 2 \mathsf{S}_3^y - 2 \sqrt 3 \mathsf{S}_3^z)\\
          \frac{1}{12} (\sqrt 6 \mathsf{S}_0^x + 3 \sqrt 2  \mathsf{S}_0^y + 2 \sqrt 3 \mathsf{S}_0^z - \sqrt 6  \mathsf{S}_1^x - 
  3 \sqrt 2  \mathsf{S}_1^y - 2 \sqrt 3 \mathsf{S}_1^z - \sqrt 6 \mathsf{S}_2^x - 3 \sqrt 2 \mathsf{S}_2^y - 
  2 \sqrt 3 \mathsf{S}_2^z + \sqrt 6  \mathsf{S}_3^x + 3 \sqrt 2 \mathsf{S}_3^y + 2 \sqrt 3 \mathsf{S}_3^z)
   \end{pmatrix} $ 
\\
\hline
\multirow{1}{*}{}
   ${\bf m}_{\sf T_{1, B}}$  & 
   $\begin{pmatrix}
        \frac{1}{2\sqrt 3} (-\sqrt 2 \mathsf{S}_0^x - \mathsf{S}_0^z - \sqrt 2 \mathsf{S}_1^x - \mathsf{S}_1^z + \sqrt 2 \mathsf{S}_2^x + \mathsf{S}_2^z + 
 \sqrt 2 \mathsf{S}_3^x +\mathsf{S}_3^z) \\
        \frac{1}{12} (\sqrt 6 \mathsf{S}_0^x - 3 \sqrt 2  \mathsf{S}_0^y - 2 \sqrt 3 \mathsf{S}_0^z - \sqrt 6  \mathsf{S}_1^x + 
  3 \sqrt 2  \mathsf{S}_1^y + 2 \sqrt 3 \mathsf{S}_1^z + \sqrt 6 \mathsf{S}_2^x - 3 \sqrt 2 \mathsf{S}_2^y - 
  2 \sqrt 3 \mathsf{S}_2^z - \sqrt 6  \mathsf{S}_3^x + 3 \sqrt 2 \mathsf{S}_3^y + 2 \sqrt 3 \mathsf{S}_3^z)\\
          \frac{1}{12} (\sqrt 6 \mathsf{S}_0^x + 3 \sqrt 2  \mathsf{S}_0^y - 2 \sqrt 3 \mathsf{S}_0^z - \sqrt 6  \mathsf{S}_1^x - 
  3 \sqrt 2  \mathsf{S}_1^y + 2 \sqrt 3 \mathsf{S}_1^z - \sqrt 6 \mathsf{S}_2^x - 3 \sqrt 2 \mathsf{S}_2^y+
  2 \sqrt 3 \mathsf{S}_2^z + \sqrt 6  \mathsf{S}_3^x + 3 \sqrt 2 \mathsf{S}_3^y - 2 \sqrt 3 \mathsf{S}_3^z)
   \end{pmatrix} $ 
\\
\hline
\multirow{1}{*}{}
   ${\bf m}_{\sf T_2} $ & 
   $\begin{pmatrix}
        \frac{1}{2} 
        \left(
                -\mathsf{S}_0^y-\mathsf{S}_1^y+\mathsf{S}_2^y+\mathsf{S}_3^y 
        \right) 
        \\
        \frac{1}{4} 
        \left(
        \sqrt 3  \mathsf{S}_0^x + \mathsf{S}_0^y - \sqrt 3 \mathsf{S}_1^x - \mathsf{S}_1^y + \sqrt 3 \mathsf{S}_2^x + \mathsf{S}_2^y - 
 \sqrt 3 \mathsf{S}_3^x - \mathsf{S}_3^y
        \right) \\
        \frac{1}{4} 
        \left(
       - \sqrt 3  \mathsf{S}_0^x + \mathsf{S}_0^y + \sqrt 3 \mathsf{S}_1^x - \mathsf{S}_1^y + \sqrt 3 \mathsf{S}_2^x - \mathsf{S}_2^y - 
 \sqrt 3 \mathsf{S}_3^x + \mathsf{S}_3^y
        \right)
      \end{pmatrix} $    
\\ 
\hline
\end{tabular}
\caption{
Order parameters ${\bf m}_\lambda$, describing how the point-group symmetry of a 
single tetrahedron within the pyrochlore lattice is broken by magnetic order.    
Order parameters are irreducible representations of the point-group 
${\sf T}_d$, and are expressed in terms of linear combinations of spin-components 
$\mathsf{S}_i = (\mathsf{S}^x_i, \mathsf{S}^y_i, \mathsf{S}^z_i)$, 
in the local frame of the magnetic ions --- cf. ${\mathcal H}_{\sf ex}^{\sf local}$~[Eq.~(\ref{eq:Hross})].  
For convenience, in this table, the local axes $(x^{\sf local},y^{\sf local},z^{\sf local})$ are simply written $(x,y,z)$.
Labelling of spins within the tetrahedron follows the convention of 
Ross~{\it et al.}~[\onlinecite{ross11-PRX1}] --- cf. Fig.~\ref{fig:tetrahedron}.   
}
\label{table:m_lambda_local}
\end{table*}

%
Symmetry permits a finite coupling $a_{\sf T_{1, AB}} \ne 0 $ between the 
two distinct ${\sf T_1}$ irreps ${\bf m}_{\sf T_{1, A}}$ and ${\bf m}_{\sf T_{1, B}}$.
This can be eliminated from ${\mathcal H}_{\sf ex}^{\sf tet}$ by a coordinate 
transformation 
\begin{eqnarray}
{\bf m}_{\sf T_{1, A'}} &=& \cos \theta_{\sf T_1} \ {\bf m}_{\sf T_{1, A}} 
        - \sin \theta_{\sf T_1} \ {\bf m}_{\sf T_{1, B}} \nonumber\\
{\bf m}_{\sf T_{1, B'}} &=& \sin \theta_{\sf T_1} \ {\bf m}_{\sf T_{1, A}} 
        + \cos \theta_{\sf T_1}\ {\bf m}_{\sf T_{1, B}} 
 \label{eq:rotatedT1}
\end{eqnarray}
where 
\begin{eqnarray}
\theta_{\sf T_1} = \frac{1}{2}\arctan{ \left( \frac{\sqrt{8} J_3}{2J_1+2J_2+J_3-2J_4} \right)}.
\label{eq:FMangle}
\end{eqnarray}
is the canting angle between spins and the relevant [100] axis in the ferromagnetic ground state. 
The Hamiltonian ${\mathcal H}_{\sf ex}^{\sf tet}$ then becomes
\begin{eqnarray}
{\mathcal H}_{\sf ex}^{[{\sf T_d}]} 
   &=& \frac{1}{2} \big[ a_{\sf A_2} m_{\sf A_2}^2+ a_{\sf E} {\bf m}^2_{\sf E}
           + a_{\sf T_2}{\bf m}^2_{\sf T_2}  \nonumber \\
     && + a_{\sf T_{1, A^{\prime}}} {\bf m}^2_{\sf T_{1, A^{\prime}}}
           + a_{\sf T_{1, B^{\prime}}} {\bf m}^2_{\sf T_{1, B^{\prime}}} \big].
\label{eq:HTd}
\end{eqnarray}
with coefficients  given in Table~\ref{table:coefficients}.
We wish to emphasize that ${\mathcal H}_{\sf ex}^{[{\sf T_d}]}$~[Eq.~(\ref{eq:HTd})] 
is an {\it exact transcription} of ${\mathcal H}_{\sf ex}^{{\sf tet}}$~[Eq.~(\ref{eq:Htet})] 
and {\it not} a phenomenological Landau theory.
As such, ${\mathcal H}_{\sf ex}^{[{\sf T_d}]}$~[Eq.~(\ref{eq:HTd})] 
is subject to the constraint that every spin has fixed length.

For the majority of the discussion in this article, we shall be concerned with classical 
vectors ${\bf S}_i$ representing a (pseudo)spin-1/2, with 
\begin{eqnarray}
S = 1/2 \; ,
\label{eq:S.is.half}
\end{eqnarray}
in which case
\begin{eqnarray}
|{\bf S}_i|^2=1/4 \; .
\label{eq:contraint-on-S}
\end{eqnarray}
For spins belonging to a single tetrahedron, we can use 
symmetry~\footnote{ The first line of Eq.~(\ref{eq:spinconstraints}) is a quadratic invariant 
of $T_d$, while the next three form a basis for the $T_1$ irreducible representation.} 
to express this constraint as
\begin{eqnarray}
{\bf S}_0^2+{\bf S}_1^2+{\bf S}_2^2+{\bf S}_3^2&=&1 \nonumber \\
{\bf S}_0^2+{\bf S}_1^2-{\bf S}_2^2-{\bf S}_3^2&=&0 \nonumber \\
{\bf S}_0^2-{\bf S}_1^2+{\bf S}_2^2-{\bf S}_3^2&=&0 \nonumber \\
{\bf S}_0^2-{\bf S}_1^2-{\bf S}_2^2+{\bf S}_3^2&=&0 \; .
\label{eq:spinconstraints}
\end{eqnarray}
The constraint of fixed spin-length, Eq.~(\ref{eq:spinconstraints}), 
plays a central role in determining the allowed classical ground states, below.


We note in passing that the addition of a  single--ion anisotropy term 
$-D ({\bf S}\cdot{\bf z}_i^{\sf local})^2$ can easily be included in the analysis
by a simple modification of the coefficients
$a_\lambda$ in ${\mathcal H}_{\sf ex}^{[{\sf T_d}]}$~[Eq.~(\ref{eq:HTd})], 
and so do not affect any of the conclusions reached about ground states, below.  
However, since interactions of this form contribute only to a trivial energy constant
for a Kramers doublet, we will not pursue this point further here.

\section{Analysis of classical phase diagram for $T=0$}
\label{section:classical-ground-states} 

\subsection{General considerations}

Given the existence of a classical ground state with ${\bf q}=0$, 4-sublattice order, 
it is easy to determine a ground-state phase diagram directly from the Hamiltonian 
${\mathcal H}_{\sf ex}^{[{\sf T_d}]}$~[Eq.~(\ref{eq:HTd})].
The method developed below is quite general and can be applied for arbitrary  $(J_1,J_2,J_3,J_4)$.    
However, for concreteness, we concentrate on the limit
\begin{eqnarray}
J_3<0 \;  , \; J_4\equiv  0  \; ,
\end{eqnarray}
which is of particular relevance to known pyrochlore materials,
leading to the phases shown in 
Fig.~\ref{fig:classical-phase-diagram}---Fig.~\ref{fig:finite-temperature-phase-diagram}.


\begin{table*}[!t]
\begin{tabular}{ | c | c  | c |}
\hline
coefficient   & 
definition in terms of & 
definition in terms of\\
of $|{\bf m}_\lambda|^2$ &  
parameters of $\mathcal{H}_{\sf ex}$~[Eq.~(\ref{eq:Hex1})] & 
parameters of ${\mathcal H}^{\sf local}_{\sf ex}$~[Eq.~(\ref{eq:Hross})] \\
\hline
$a_{\sf A_2}$ & $-2J_1+J_2-2(J_3+2J_4)$ & $3 J_{zz}$\\
\hline
$a_{\sf E}$ & $-2 J_1+J_2+J_3+2J_4$ & $-6J_\pm $\\
\hline
$a_{\sf T_2}$ & $-J_2+J_3-2J_4$ & $2J_\pm-4J_{\pm\pm}$\\
\hline
 & $(2 J_1+J_2)\cos^2(\theta_{\sf T_1})$ & $\frac{1}{3} (4 J_\pm + 8 J_{\pm\pm} + 8 \sqrt 2 J_{z\pm} - J_{zz})\cos^2(\theta_{\sf T_1})$  \\
$a_{\sf T_{1, A'}}$ & $-(J_2+J_3- 2J_4)\sin^2(\theta_{\sf T_1}) $ & $+\frac{2}{3} (1 J_\pm + 2 J_{\pm\pm} - 4 \sqrt 2 J_{z\pm} - J_{zz})\sin^2(\theta_{\sf T_1}) $\\
& $+\sqrt{2} J_3 \sin(2 \theta_{\sf T_1})$ & 
$+\frac{\sqrt 2 }{3}  (-2 J_\pm - 4 J_{\pm\pm} + 2 \sqrt 2 J_{z\pm} - J_{zz}) \sin(2 \theta_{\sf T_1})$\\
\hline
 &$(2 J_1+J_2)\sin^2(\theta_{\sf T_1})-$ & $\frac{1}{3} (4 J_\pm + 8 J_{\pm\pm} + 8 \sqrt 2 J_{z\pm} - J_{zz})\sin^2(\theta_{\sf T_1})$ \\
$a_{\sf T_{1, B'}}$ & $(J_2+J_3-2J_4)\cos^2(\theta_{\sf T_1}) $ &$-\frac{2}{3} (1 J_\pm + 2 J_{\pm\pm} - 4 \sqrt 2 J_{z\pm} - J_{zz})\cos^2(\theta_{\sf T_1})$\\
&$-\sqrt{2} J_3 \sin(2 \theta_{\sf T_1})$ & 
$-\frac{\sqrt 2 }{3}  (-2 J_\pm - 4 J_{\pm\pm} + 2 \sqrt 2 J_{z\pm} - J_{zz}) \sin(2 \theta_{\sf T_1})$\\
\hline
\end{tabular}
\caption{Coefficients $a_\lambda$ of the scalar invariants $|{\bf m}_\lambda|^2$ 
appearing in ${\mathcal H}_{\sf ex}^{[{\sf T_d}]}$~[Eq.~(\ref{eq:HTd})].
Coefficients are expressed as a function of ($J_1$, $J_2$, $J_3$, $J_4$), 
the parameters of $\mathcal{H}_{\sf ex}$~[Eq.~(\ref{eq:Hex1})]; 
and ($J_{zz}$, $J_\pm$, $J_{\pm\pm}$, $J_{z\pm}$), the 
parameters of ${\mathcal H}^{\sf local}_{\sf ex}$~[Eq.~(\ref{eq:Hross})], 
with the canting angle $\theta_{\sf T_1}$ defined in Eq.~(\ref{eq:FMangle}).
The classical ground states of $\mathcal{H}_{\sf ex}$ [Eq.~(\protect\ref{eq:Hex})] can be found 
by identifying the coefficient(s) $a_\lambda$ with the lowest value, 
and imposing the constraint of fixed spin-length,  
Eq.~(\ref{eq:spinconstraints}), on the associated ${\bf m}_\lambda$.
}
\label{table:coefficients}
\end{table*}


The classical ground state of 
${\mathcal H}_{\sf ex}^{[{\sf T_d}]}$~[Eq.~(\ref{eq:HTd})]  can be found by first 
identifying the irrep $\lambda^*$ for which 
$a_{\lambda^{*}}$ takes on the minimum value, and then imposing the constraint 
on the total length of the spin [Eq.~(\ref{eq:spinconstraints})] on ${\bf m}_{\lambda^*}$,
which implies that
\begin{equation}
 m_{\sf A_2}^2 
       + {\bf m}^2_{\sf E} 
       + {\bf m}^2_{\sf T_2} 
       + {\bf m}^2_{\sf T_{1, A'}} 
       + {\bf m}^2_{\sf T_{1, B'}} 
        \equiv \sum_{\lambda}{\bf m}_{\lambda}^2 =1 \; .
\label{eq:constraint}
\end{equation}
Such an approach is possible because 
each individual
 order parameter ${\bf m}_\lambda$ 
can reach a maximum value of unity 
within physical spin configurations
\begin{eqnarray}
\max {\bf m}_{\lambda}^2 =1 \;.
\end{eqnarray}
This method of determining the classical ground state is completely general
and, once generalized to the lattice, is not restricted to conventionally 
ordered states~\cite{benton-thesis, benton16}.

\begin{figure}
\centering\includegraphics[width=0.9\columnwidth]{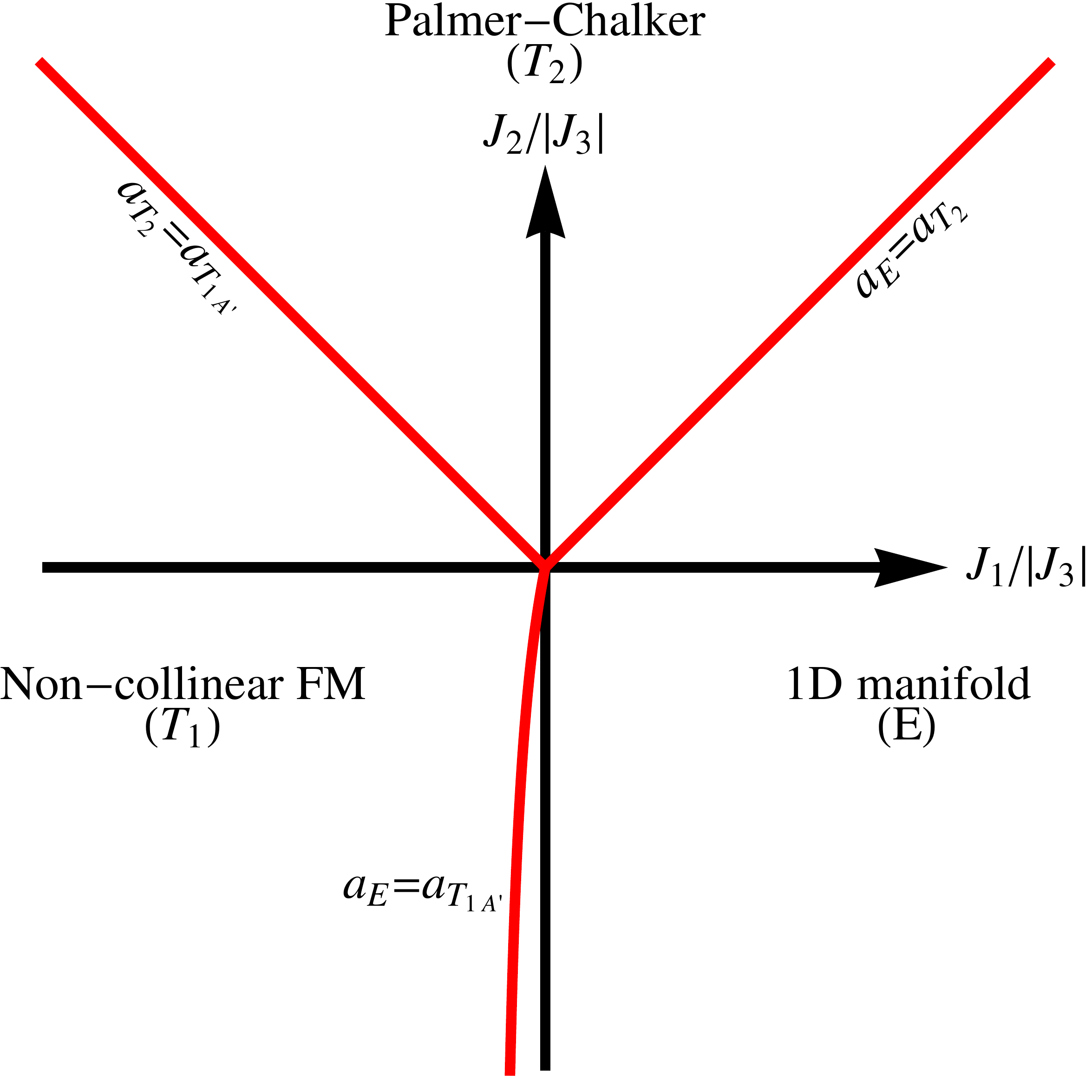}
\caption{
Classical ground state phase diagram of $\mathcal{H}_{\sf ex}$ [Eq.~(\ref{eq:Hex})]
for $J_3<0$, $J_4=0$, as a function of $(J_1,J_2)/|J_3|$.
In the absence of fluctuations, the ground states are 
a non-collinear FM transforming with the ${\sf T_1}$ irrep of $T_d$; 
a one--dimensional manifold of states transforming with the ${\sf E}$ irrep of ${\sf T_d}$; 
and the Palmer--Chalker phase, a coplanar antiferromagnet transforming with 
the ${\sf T_2}$ irrep of $T_d$.
All three phases have long-range 4-sublattice order.
Analytical expressions for the boundaries between phases are given in 
Eq.~(\ref{eq:boundary-T2-E}--\ref{eq:boundary-E-T1}), with coefficients
$a_\lambda$ defined in Table~\ref{table:coefficients}.  
\label{fig:phase}}
\end{figure}

In the
limit $J_3<0 \;  , \; J_4\equiv  0$, 
the coefficients $a_\lambda$ with the  
lowest values are $a_{\sf E}$, $a_{\sf T_{1 A'}}$, or $a_{\sf T_2}$, 
and the corresponding ${\bf q}=0$ ordered ground states found have 
${\sf E}$, ${\sf T}_1$ and ${\sf T}_2$ symmetry.
The boundaries between these phases occur where
\begin{eqnarray}
   a_{\sf T_2} &=& a_{\sf E} < a_{\sf T_2},\;  a_{\sf T_{1 B^{\prime}}},\;  a_{\sf A_2}  \nonumber\\ 
        && \Rightarrow   J_2 = J_1 > 0 \label{eq:boundary-T2-E} \\
   a_{\sf T_2} &=& a_{\sf T_{1, A^{\prime}}}  < a_{\sf E},\; a_{\sf T_{1 B^{\prime}}},\; a_{\sf A_2}  \nonumber\\ 
       &&  \Rightarrow J_2 = -J_1 > 0  \label{eq:boundary-T2-T1} \\    
   a_{\sf E} &=& a_{\sf T_{1, A^{\prime}}} < a_{\sf T_2},\; a_{\sf T_{1 B^{\prime}}},\; a_{\sf A_2}   \nonumber\\ 
        &&  \Rightarrow J_2 = \frac{J_1 (4 J_1 - 5 J_3)}{4 J_1 - J_3} < 0 \label{eq:boundary-E-T1}
\end{eqnarray}
as illustrated in Fig.~\ref{fig:phase}.


In what follows, we explore the classical ground states 
with ${\sf E}$, ${\sf T}_1$ and ${\sf T}_2$ symmetry in some detail, 
paying particular attention to what happens on phase boundaries 
where more than one order parameter can take on a finite value.
We will not consider the ``all--in, all out'' ground state, a simple 
Ising--type order with two degenerate ground states.
``All--in, all out'' order has a finite value of the order parameter 
$m_{\sf A_2}$ [cf~Table~\ref{table:m_lambda_local}], which requires
$a_{\sf A_2}$ to be the lowest coefficient.
This only occurs for $J_4 > 0$ and/or $J_3 > 0$, 
and so falls outside the scope of this Article.

\subsection{Non-collinear FM with $\sf T_1$ symmetry}
\label{ferromagnet}

We begin by considering what happens where interactions are predominantly 
ferromagnetic (i.e. $J_1, J_2<0$), as in Yb$_2$Ti$_2$O$_7$~[\onlinecite{ross11-PRX1}] --- cf.~Table~\ref{table:J-from-experiment}.
For most of this region,  as might be expected, the classical configuration with the lowest energy
is a state with a finite magnetisation.
This is the ground state throughout the
region bounded by
\mbox{$a_{\sf T_{1,A^\prime}} = a_{\sf T_2} $} [Eq.~(\ref{eq:boundary-T2-T1})] ,
and \mbox{$a_{\sf T_{1,A^\prime}} = a_{\sf E}$} [Eq.~(\ref{eq:boundary-E-T1})]
--- cf. Fig.~\ref{fig:phase}.
Here the energy 
is  minimised by setting 
\begin{eqnarray}
   {\mathbf m}^2_{\sf T_{1,A'}} = 1
\end{eqnarray}
and 
\begin{eqnarray}
   m_{\sf A_2} = {\bf m}_{\sf E} = {\bf m}_{\sf T_2} = {\bf m}_{\sf T_{1 B^{\prime}}}=0
\end{eqnarray}
The constraints on the spin lengths [Eq.~(\ref{eq:spinconstraints})]
further imply that 
\begin{eqnarray}
   m_{\sf T_ {1 A^{\prime}}}^y m_{\sf T_ {1 A^{\prime}}}^z&=&0 \nonumber \\
   m_{\sf T_ {1 A^{\prime}}}^x m_{\sf T_ {1 A^{\prime}}}^z&=&0 \nonumber \\
   m_{\sf T_ {1 A^{\prime}}}^x m_{\sf T_ {1 A^{\prime}}}^y&=&0.
\end{eqnarray}
It follows that there are 6 possible ground states
\begin{eqnarray}
   {\bf m}_{\sf T_ {1 A^{\prime}}} =
  \begin{pmatrix}
      \pm1\\
       0\\
       0
   \end{pmatrix},
   \begin{pmatrix}
       0\\
       \pm1\\
       0
    \end{pmatrix},
    \begin{pmatrix}
       0\\
       0\\
       \pm1
     \end{pmatrix}.
\end{eqnarray}
Written in terms of spins, these are 6, non-collinear ferromagnetic (FM) 
ground states, with typical spin configuration
\begin{eqnarray}
\mathbf{S}_0 &=& 
  S \left (\sin \theta_{\sf T_1} /\sqrt{2},
   \sin \theta_{\sf T_1} /\sqrt{2},
   \cos\theta_{\sf T_1} \right)
   \nonumber \\
\mathbf{S}_1 &=& 
 S  \left (-\sin \theta_{\sf T_1} /\sqrt{2},
   \sin \theta_{\sf T_1} /\sqrt{2},
   \cos\theta_{\sf T_1} \right)
   \nonumber \\
\mathbf{S}_2 &=& 
  S \left (\sin \theta_{\sf T_1} /\sqrt{2},
   -\sin \theta_{\sf T_1} /\sqrt{2},
   \cos\theta_{\sf T_1} \right)
   \nonumber \\
\mathbf{S}_3 &=& 
  S \left( -\sin \theta_{\sf T_1} /\sqrt{2},
   -\sin \theta_{\sf T_1} /\sqrt{2},
   \cos\theta_{\sf T_1} \right)
\label{eq:fmgs2}
\end{eqnarray}
where $\theta_{\sf T_1}$ is given by Eq.~(\ref{eq:FMangle}) and, 
following Eq.~(\ref{eq:S.is.half}), $S=1/2$.

\begin{figure}
\centering
\subfigure[]{%
  \includegraphics[width=.4\linewidth]{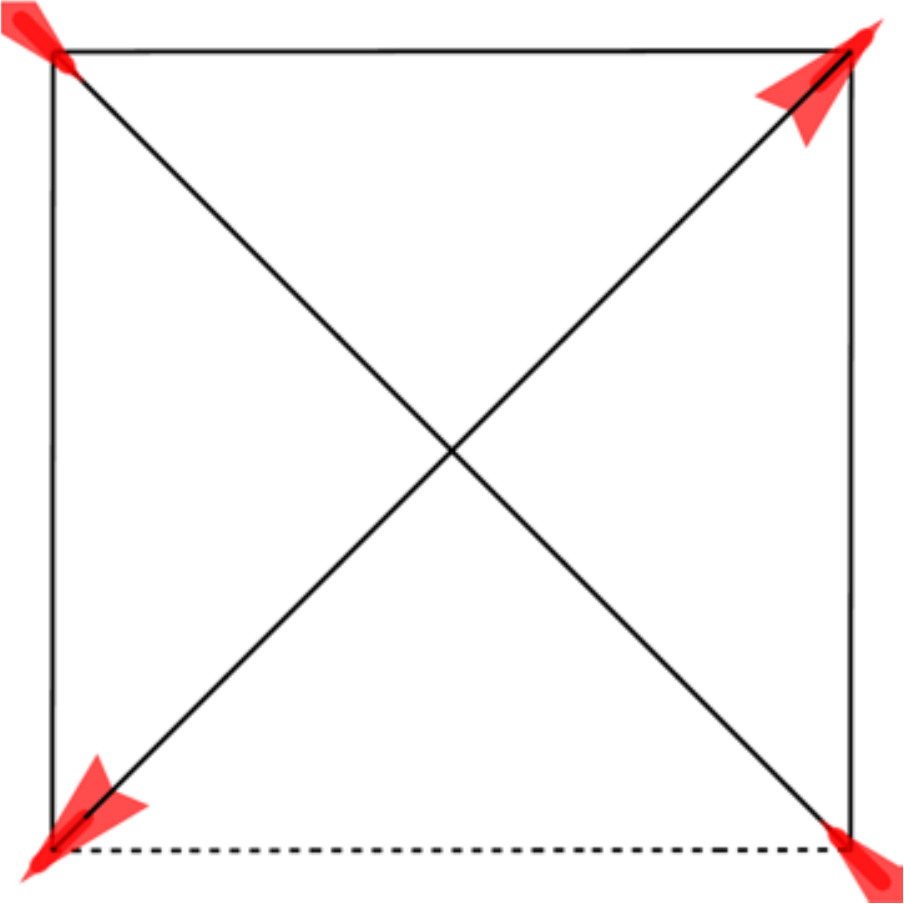}}%
    \qquad
\subfigure[]{%
  \includegraphics[width=.4\linewidth]{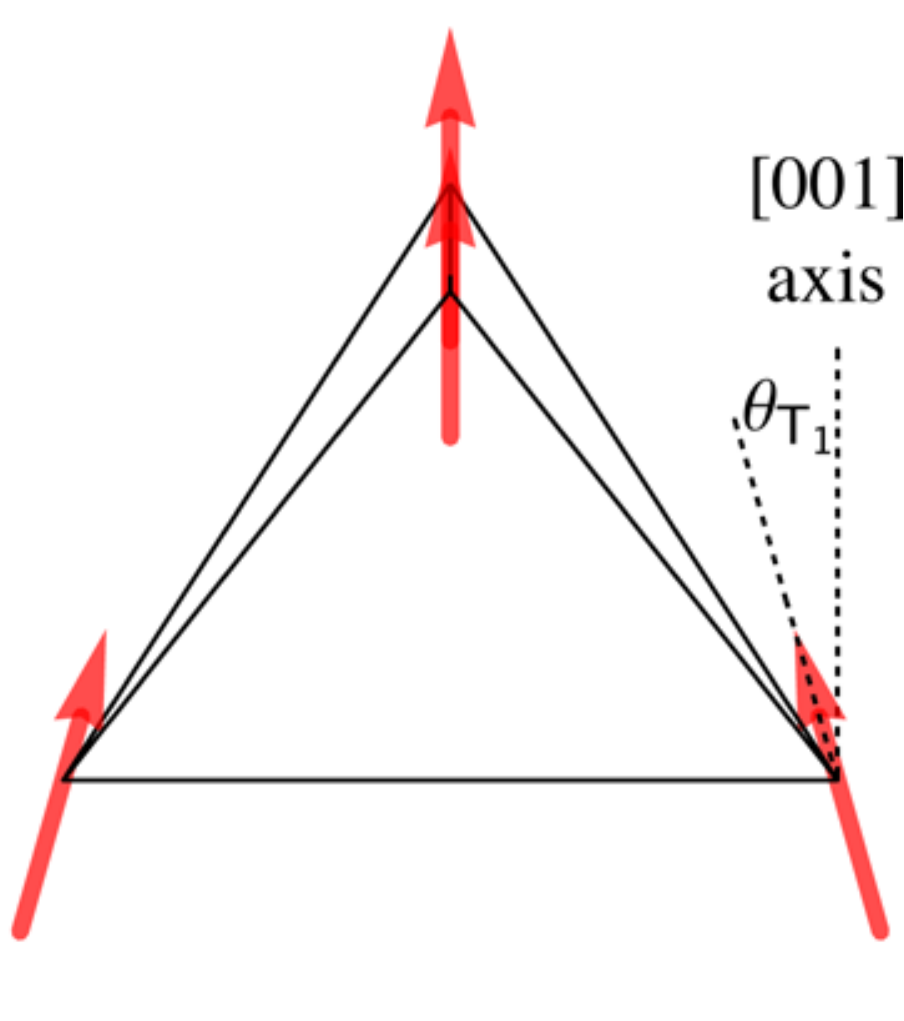}}%
\caption{\label{fig:ferrophase} 
Spin-configuration in the 4-sublattice non-collinear FM phase,
transforming with the $\sf T_1$ irrep of $T_d$ :
(a) viewed along the $[001]$ axis;   
(b) viewed slightly off the $[110]$ axis. 
The magnetisation is aligned with the $[001]$ axis.
Spins are canted into the plane perpendicular to this, 
with canting angle $\theta_{\sf T_1}$.
The canting is of an ``ice-like''  form 
such that the projection of the spin configuration onto a
$[001]$ plane has two spins oriented into the tetrahedron
and two oriented out, as shown in (a).
}
\end{figure}

The magnetisation of this FM ground state, illustrated in Fig.~\ref{fig:ferrophase}, 
is parallel to a $[001]$ axis, with spins canted away from this axis, in an ``ice-like'' 
manner.
This state has been identified as the ground state in Yb$_2$Sn$_2$O$_7$, 
where it was referred to as a ``splayed FM''~[\onlinecite{yaouanc13}], and in most 
samples of Yb$_2$Ti$_2$O$_7$ which order at low temperature~\cite{yasui03,chang12,chang14,lhotel14,jaubert15, robert15, gaudet16, hallas16}
although a different form of canting has recently been claimed in Ref. \onlinecite{yaouanc16}.
It is also the observed ordered state of the Tb based pyrochlore Tb$_2$Sn$_2$O$_7$
[\onlinecite{mirebeau05, petit12}]


\subsection{One--dimensional manifold of states with ${\sf E}$ symmetry}
\label{1Dmanifold}

For a wide range of parameters, predominantly with antiferromagnetic ``XY'' 
interactions $J_1 > 0$, the classical ground state of 
${\mathcal H}_{\sf ex}^{[{\sf T_d}]}$~[Eq.~(\ref{eq:HTd})] is a one--dimensional
manifold of states which transforms with the ${\sf E}$ irrep of ${\sf T}_d$.  
These ground states occur in a region bounded by 
\mbox{$a_{\sf E} = a_{\sf T_{1,A^{\prime}}} $} [Eq.~(\ref{eq:boundary-E-T1})] 
and \mbox{$a_{\sf E} = a_{\sf T_2}$} [Eq.~(\ref{eq:boundary-T2-E})] 
--- cf~Fig.~\ref{fig:phase} --- and is characterised by spins lying in the  
``XY'' plane normal to the local $[111]$ axis on each site 
[cf~Eqs.~(\ref{eq:local-111-axis},\ref{eq:local-easy-plane-x},\ref{eq:local-easy-plane-y})].


For this range of parameters, the classical ground state energy can be minimised by setting
\begin{eqnarray}
   {\bf m}_{\sf E}^2=1
   \label{eq:Enorm}
\end{eqnarray}
and
\begin{eqnarray}
   {m}_{\sf A_2}={\bf m}_{\sf T_2}
   = {\bf m}_{\sf T_{1 A^\prime}}={\bf m}_{\sf T_{1 B^{\prime}}} = 0 \; .
   \label{eq:Ezerofields}
\end{eqnarray}
These solutions {\it automatically} satisfy the constraint on the total length 
of the spin Eq.~(\ref{eq:spinconstraints}) and are conveniently characterised by writing 
\begin{eqnarray}
   {\bf m}_{\sf E} = \ (\cos \theta_{\sf E},\  \sin \theta_{\sf E}) \; . 
   \label{eq:thetaE}
\end{eqnarray}

\begin{figure}
\centering
\subfigure[]{%
  \includegraphics[width=.4\linewidth]{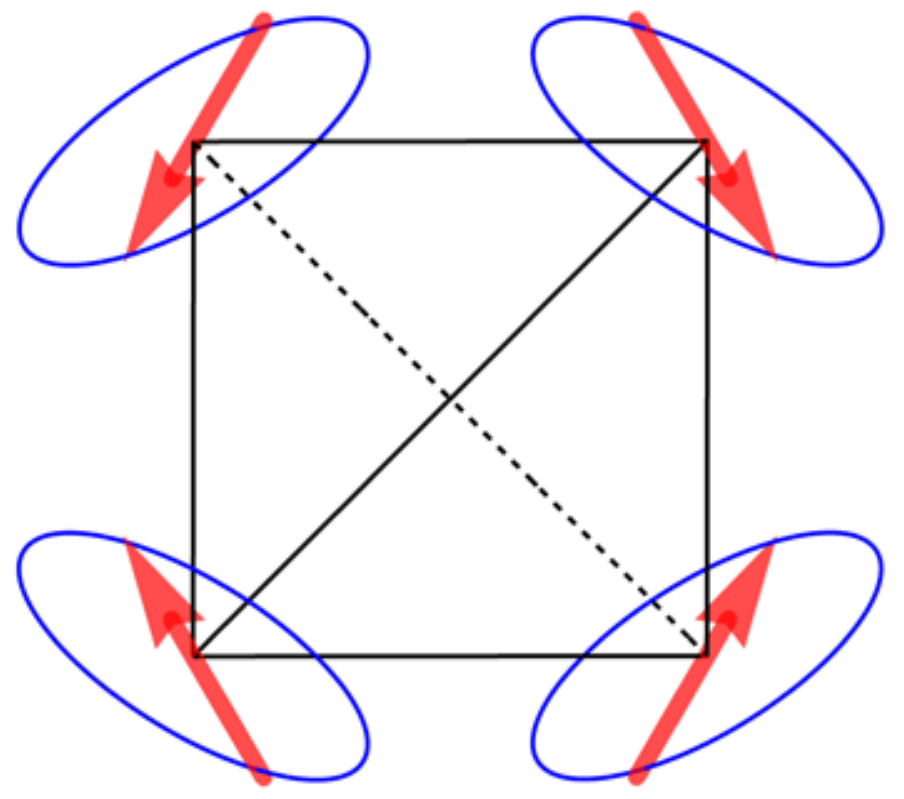}}%
    \qquad
\subfigure[]{%
  \includegraphics[width=.4\linewidth]{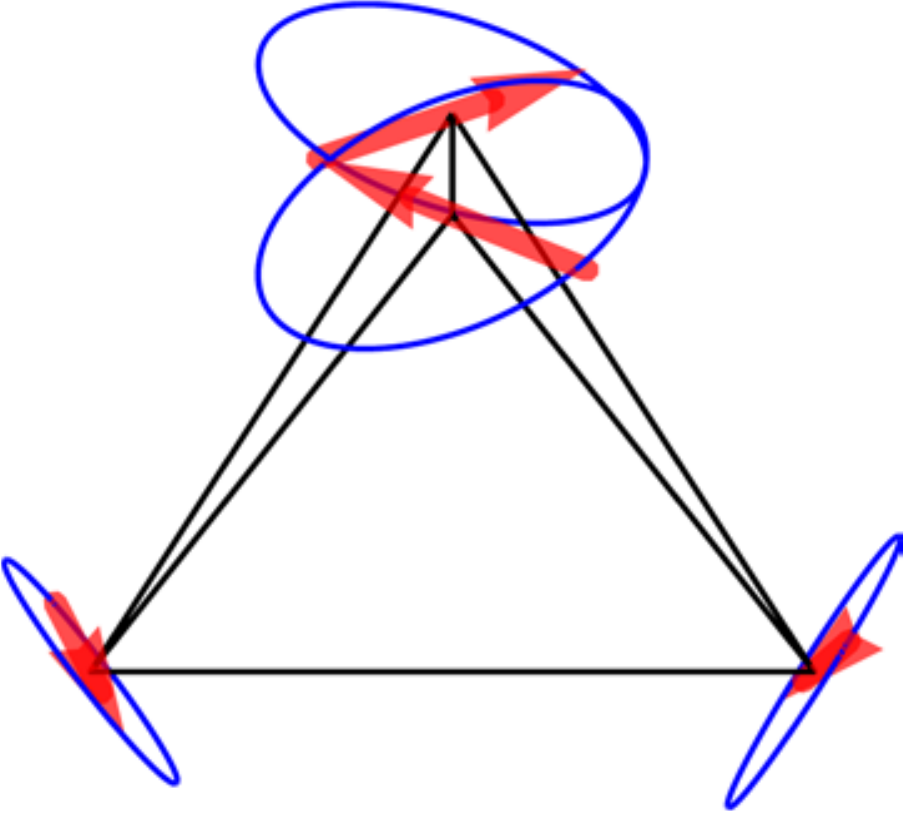}}%
\caption{
Example of a spin configuration within 
the one--dimensional manifold of states 
transforming with the ${\sf E}$ irrep of $T_d$~:
(a) viewed along $[001]$ axis; 
(b) viewed slightly off the $[110]$ axis.
The manifold possesses 4-sublattice long-range order, with spins
lying in the  ``XY'' 
plane perpendicular to the local $[111]$ axis at each site.
The manifold is continuous, and can be parameterised
with a single angle $\theta_{\sf E}$.
The manifold can be generated by a clockwise rotation of all
spins around their respective 
local axes.
\label{fig:u1phase}}
\end{figure}


It follows that the ground state is a continuous, one--dimensional 
manifold of states parameterised by the single angle 
\mbox{$0 \leq \theta_{\sf E} < 2\pi$}.
The spin configuration in this manifold is given by
\begin{eqnarray}
\mathbf{S}_0 & = & S \bigg(
                             \sqrt{\frac{2}{3}}\cos(\theta_{\sf E}),\:
                             \sqrt{\frac{2}{3}}\cos(\theta_{\sf E}+\frac{2 \pi}{3}),\nonumber \\
                             && \qquad \sqrt{\frac{2}{3}}\cos(\theta_{\sf E}-\frac{2\pi}{3})
                             \bigg) \nonumber\\
\mathbf{S}_1 & = & S \bigg(
                             \sqrt{\frac{2}{3}}\cos(\theta_{\sf E}),\:
                             -\sqrt{\frac{2}{3}}\cos(\theta_{\sf E}+\frac{2\pi}{3}), \nonumber \\
                            	&& \qquad -\sqrt{\frac{2}{3}}\cos(\theta_{\sf E}-\frac{2\pi}{3})
                             \bigg) \nonumber\\
\mathbf{S}_2 & = & S \bigg(
                             -\sqrt{\frac{2}{3}}\cos(\theta_{\sf E}),\:
                             \sqrt{\frac{2}{3}}\cos(\theta_{\sf E}+\frac{2\pi}{3}),\nonumber \\
                            && \qquad -\sqrt{\frac{2}{3}}\cos(\theta_{\sf E}-\frac{2\pi}{3})
                             \bigg) \nonumber\\
\mathbf{S}_3 & = & S \bigg(
                             -\sqrt{\frac{2}{3}}\cos(\theta_{\sf E}),\:
                             -\sqrt{\frac{2}{3}}\cos(\theta_{\sf E}+\frac{2\pi}{3}),\: \nonumber \\
                            && \qquad \sqrt{\frac{2}{3}}\cos(\theta_{\sf E}-\frac{2\pi}{3})
                             \bigg).
                             \label{eq:u1gs2} 
\end{eqnarray}
with spins lying in the local ``XY'' plane [cf. Eqs.~(\ref{eq:local-easy-plane-x}) and~(\ref{eq:local-easy-plane-y})].


\begin{figure}
\centering
\subfigure[]{%
  \includegraphics[width=.4\linewidth]{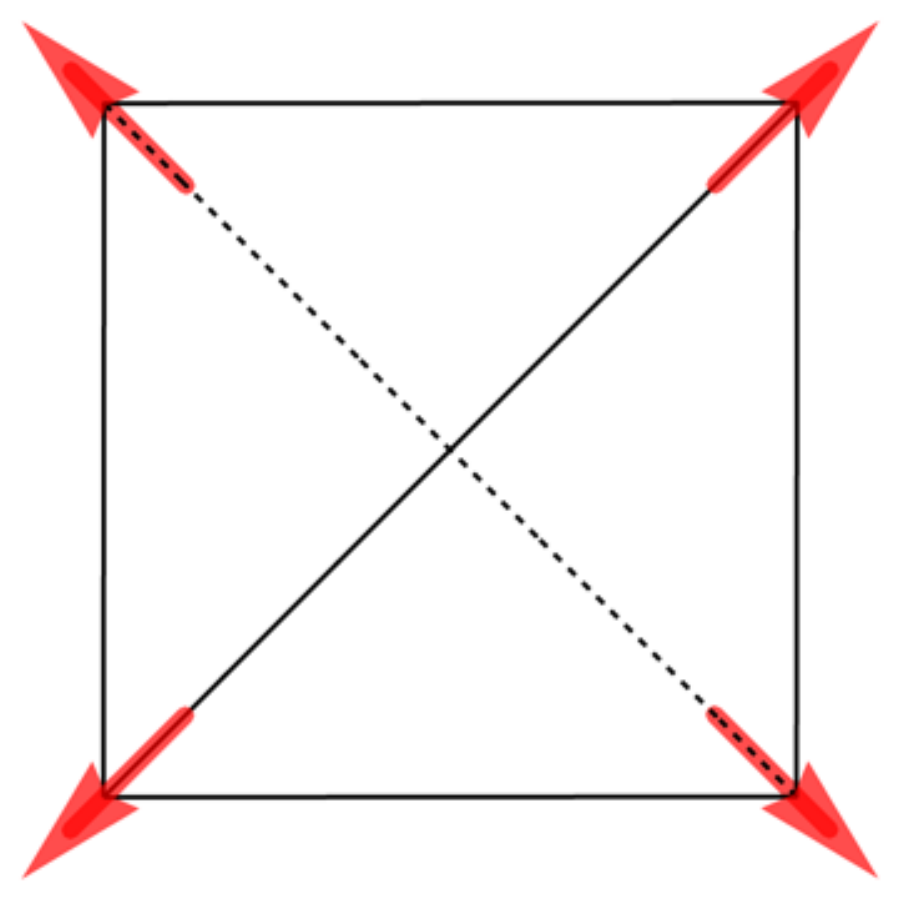}}%
    \qquad
\subfigure[]{%
  \includegraphics[width=.4\linewidth]{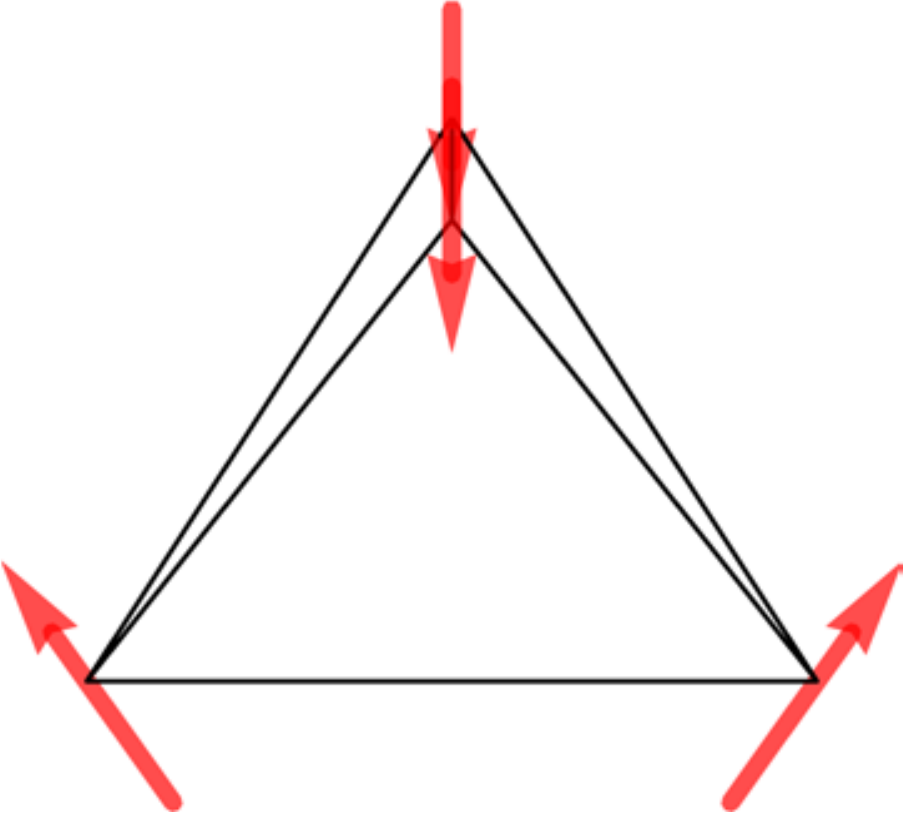}}%
\caption{
Spin configuration in the 4-sublattice non-coplanar antiferromagnet, $\Psi_2$, 
selected by fluctuations from the one--dimensional manifold of states 
transforming with ${\sf E}$~:
(a) viewed along $[001]$ axis; 
(b) viewed slightly off the $[110]$ axis.
At the phase boundary with the Palmer--Chalker phase, 
each of the six $\Psi_2$ ground states can be transformed continuously
into a Palmer--Chalker state without leaving the ground--state manifold.
\label{fig:psi2}}
\end{figure}

\subsection{Non-coplanar antiferromagnet, $\Psi_2$, with ${\sf E}$ symmetry}

%

%
It is now well--established that in Er$_2$Ti$_2$O$_7$ quantum fluctuations
~\cite{savary12-PRL109,zhitomirsky12,wong13}, classical thermal
fluctuations at low temperature \cite{champion03}
and thermal fluctuations near the ordering temperature
~\cite{oitmaa13, zhitomirsky14} fluctuations all act within the one--dimensional manifold 
of classical ground states described in Section~\ref{1Dmanifold}, 
to select a non--coplanar antiferromagnet, $\Psi_2$, illustrated in Fig.~(\ref{fig:psi2}).   
Structural disorder, meanwhile, favours the coplanar antiferromagnet, $\Psi_3$ [\onlinecite{maryasin14,andreanov15}], 
illustrated in Fig.~(\ref{fig:psi3}).   
Together, this pair of states form a basis for the ${\sf E}$ irrep of ${\sf T}_d$ [\onlinecite{poole07}].


The $\Psi_2$ ground state is six-fold degenerate, with spins canted symmetrically 
out of the $[100]$ plane.
The six spin configurations for $\Psi_2$ states are given by Eq.~(\ref{eq:u1gs2}) with 
$\theta_{\sf E} = \frac{n \pi}{3}$, $n=0,1,2\ldots 5$.
The $\Psi_2$ state is characterised by the primary order parameter ${\bf m}_{\sf E}$ 
[cf.~Table~\ref{table:m_lambda_global}], and by $c_{\sf E} > 0$, where (cf.~Refs.[\onlinecite{chern-arXiv,zhitomirsky14}])
\begin{eqnarray}
c_{\sf E} = \langle \cos 6\theta_{\sf E} \rangle 
\label{eq:ctheta}
\end{eqnarray}

Symmetry allows for fluctuations to induce a finite value of $m_{\sf A_2}$
in the $\Psi_2$ state~\cite{javanparast15}, but classically this must vanish as $T\to0$ since
the energy is minimised by $m_{\sf A_2}=0$ within the region of 
phase diagram which favours the $\Psi_2$ state.

\subsection{Coplanar antiferromagnet, $\Psi_3$, with ${\sf E}$ symmetry}
\label{subsec:psi3}

For parameters bordering on the non-collinear FM phase, fluctuations
select a coplanar antiferromagnet, $\Psi_3$, from the one--dimensional 
manifold of states transforming with ${\sf E}$.
The $\Psi_3$ ground state is six-fold degenerate, with spins lying in 
a common $[100]$ plane.

\begin{figure}
\centering
\subfigure[]{%
  \includegraphics[width=.4\linewidth]{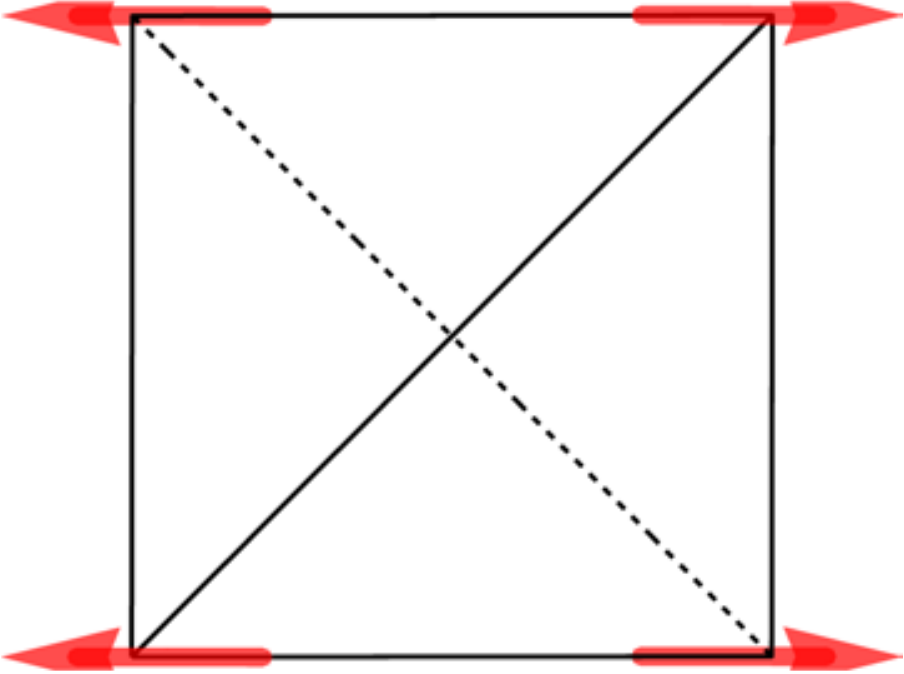}}%
  \qquad
\subfigure[]{%
  \includegraphics[width=.4\linewidth]{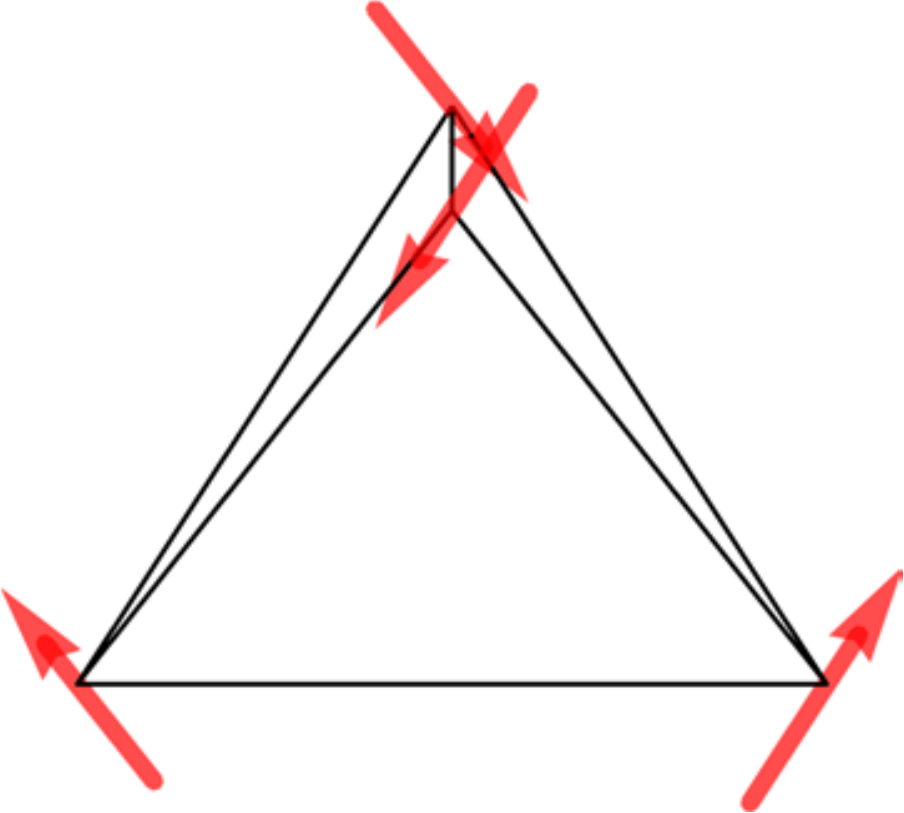}}%
\caption{
Spin configuration in the 4-sublattice coplanar antiferromagnet, $\Psi_3$, 
selected by fluctuations from the one--dimensional manifold of states 
transforming with ${\sf E}$~:
(a) viewed along $[001]$ axis; 
(b) viewed slightly off the $[110]$ axis.
At the phase boundary with the non-collinear FM phase, each of the six $\Psi_3$ 
ground states can be transformed continuously into a non-collinear FM state, 
without leaving the ground--state manifold.  
\label{fig:psi3}}
\end{figure}

The six spin configurations for $\Psi_3$ states are given by Eq.~(\ref{eq:u1gs2}) with 
$\theta_{\sf E}  =\frac{(2 n+1) \pi}{6}$, $n=0,1,2\ldots 5$.    
These states are characterised by a finite value of the order parameter ${\bf m}_{\sf E}$ 
[cf Table~\ref{table:m_lambda_global}], and by $c_{\sf E} < 0$ [cf. Eq.~(\ref{eq:ctheta})].
An example of a typical spin configuration is shown in Fig.~(\ref{fig:psi3}). 


Taken together $\Psi_2$ and $\Psi_3$ form a complete basis for the 
$\sf E$ irrep of ${\sf T}_d$.


\subsection{Palmer--Chalker phase, $\Psi_4$, with ${\sf T_2}$ symmetry}
\label{palmer-chalker}

In a region bounded by 
\mbox{$a_{\sf T_2} = a_{\sf T_{1,A^{\prime}}} $} [Eq.~(\ref{eq:boundary-T2-T1})] 
and \mbox{$a_{\sf T_2} = a_{\sf E}$} [Eq.~(\ref{eq:boundary-T2-E})] 
--- cf.  Fig.~\ref{fig:phase} ---
the energy
is minimised by setting
\begin{eqnarray}
{\bf m}_{\sf T_2}^2=1
\end{eqnarray}
and
\begin{eqnarray}
m_{\sf A_2}={\bf m}_{\sf E}=
 {\bf m}_{\sf T_{1 A^{\prime}}}={\bf m}_{\sf T_{1 B^{\prime}}}=0
\end{eqnarray}


The constraints on the total length of the spin, Eq.~(\ref{eq:spinconstraints}),  
further imply that 
\begin{eqnarray}
{\bf m}_{\sf T_2}^2=1 \\
m_{\sf T_2}^y m_{\sf T_2}^z=0 \\
m_{\sf T_2}^x m_{\sf T_2}^z=0 \\
m_{\sf T_2}^x m_{\sf T_2}^y=0
\end{eqnarray}
giving us a set of 6 ground states
\begin{eqnarray}
{\bf m}_{\sf T_2}=
\begin{pmatrix}
\pm 1 \\
0 \\
0
\end{pmatrix},
\begin{pmatrix}
0 \\
\pm1 \\
0
\end{pmatrix},
\begin{pmatrix}
0 \\
0 \\
\pm 1
\end{pmatrix}.
\end{eqnarray}


\begin{figure}[!t]
\centering
\subfigure[]{%
  \includegraphics[width=.4\linewidth]{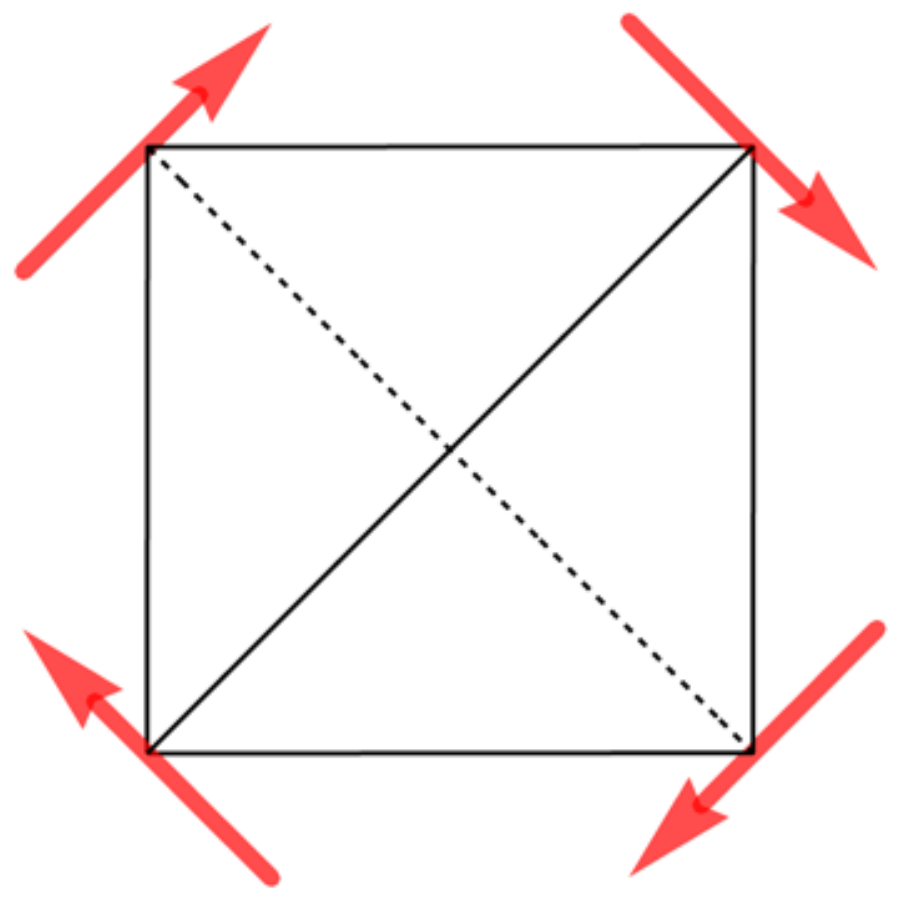}}%
    \qquad
\subfigure[]{%
  \includegraphics[width=.4\linewidth]{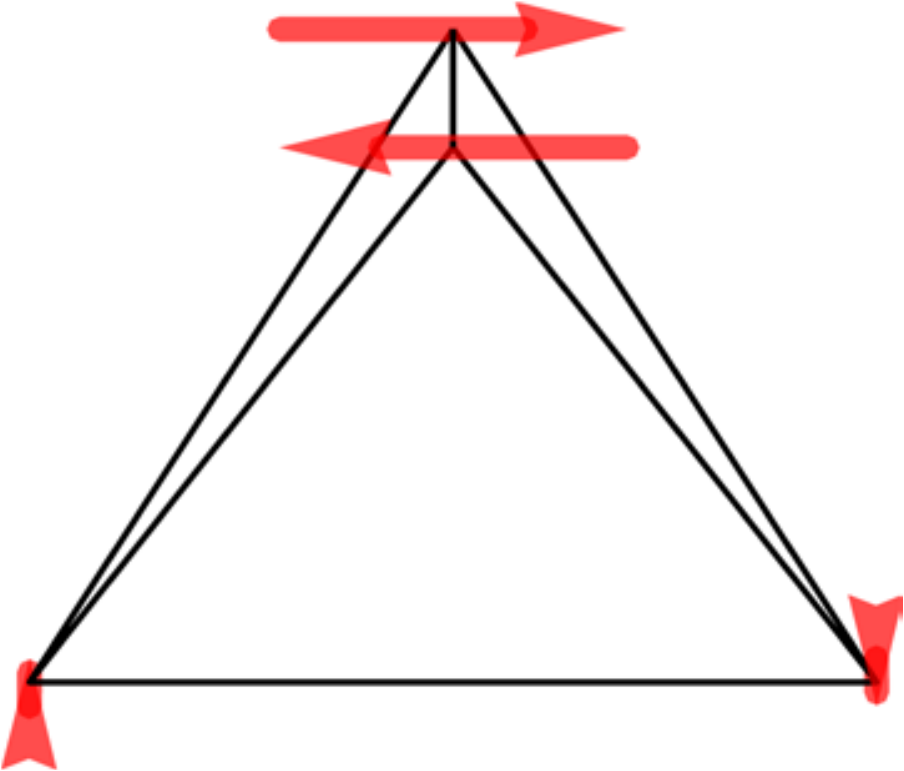}}%
\caption{
Spin configuration in the 4-sublattice Palmer--Chalker phase, $\Psi_4$, 
transforming with the ${\sf T_2}$ irrep of $T_d$~:
(a) viewed along $[001]$ axis; 
(b) viewed slightly off the $[110]$ axis.
At the phase boundary with the $\Psi_2$ phase, each of the six Palmer--Chalker 
ground states can be transformed continuously into a $\Psi_2$ state.
\label{fig:PCphase}}
\end{figure}


Within these ground states, spins are arranged in helical manner in a 
common $[100]$ plane, with a typical spin configuration given by 
(see Fig.~\ref{fig:PCphase}).
\begin{eqnarray}
\mathbf{S}_0&= & S \left(\frac{1}{\sqrt{2}},-\frac{1}{\sqrt{2}},0\right)  \nonumber \\
\mathbf{S}_1&= & S \left(-\frac{1}{\sqrt{2}},-\frac{1}{\sqrt{2}},0\right) \nonumber \\
\mathbf{S}_2&= & S \left(\frac{1}{\sqrt{2}},\frac{1}{\sqrt{2}},0\right)   \nonumber \\
\mathbf{S}_3&= & S \left(-\frac{1}{\sqrt{2}},\frac{1}{\sqrt{2}},0\right)
\label{eq:pcgs-2}
\end{eqnarray}

This phase is the ``Palmer--Chalker" phase, first
identified as the ground state of a model with antiferromagnetic
nearest neighbour Heisenberg interactions and long-range
dipolar interactions on the pyrochlore lattice \cite{palmer00}.
The 6 degenerate ground states in this phase are described 
by the basis vectors $\Psi_{4,5,6}$ [\onlinecite{poole07}], which 
are all equivalent by symmetry.
For brevity we refer to this phase as $\Psi_4$, but it should be remembered that all
 three basis vectors $\Psi_{4,5,6}$ are equivalent ground states.

The Palmer-Chalker states are superficially similar to the
$\Psi_3$ states [Section \ref{subsec:psi3}]
, being coplanar, antiferromagnet configurations with
all spins lying in a $[100]$ plane.
However, their symmetry properties are quite different (as is expressed in the fact
that they transform according to different irreps of $T_d$.)
A simple example of these different symmetry properties is their behaviour
under the 3 $C_2$ rotations around $\langle100\rangle$ axes.
The $\Psi_3$ configurations are invariant under all such rotations.
The PC states are invariant under one such rotation (the one around the
axis normal to all the spins), but reverse all spin orientations under the
other two.

\subsection{Boundary between Palmer--Chalker phase and the one--dimensional 
                   manifold of states with ${\sf E}$ symmetry}
\label{ssec:T2E}

The boundary between the Palmer--Chalker phase and the one--dimensional manifold 
of states with ${\sf E}$ symmetry occurs when $a_{\sf E} = a_{\sf T_2}$
[cf. Eq.~(\ref{eq:boundary-T2-E})].  
In this case, ${\mathcal H}_{\sf ex}^{[{\sf T_d}]}$~[Eq.~(\ref{eq:HTd})]
is minimized by setting 
\begin{eqnarray}
{\bf m}_{\sf E}^2+{\bf m}_{\sf T_2}^2=1
\end{eqnarray}
and
\begin{eqnarray}
m_{\sf A_2}={\bf m}_{\sf T_{1 A^{\prime}}}={\bf m}_{\sf T_{1 B^{\prime}}}=0 .
\end{eqnarray}
Substituting 
\begin{eqnarray}
   {\bf m}_{\sf E} = \ m_{\sf E} (\cos \theta_{\sf E},\  \sin \theta_{\sf E}) \; ,
   \label{eq:thetaEm}
\end{eqnarray}
and imposing the constraint Eq.~(\ref{eq:spinconstraints}), we find
\begin{eqnarray}
2 m_{\sf E} m_{\sf T_2}^x \sin(\theta_{\sf E})-m_{\sf T_2}^y m_{\sf T_2}^z=0 \nonumber \\
2 m_{\sf E} m_{\sf T_2}^y \sin\left(\theta_{\sf E}-\frac{2 \pi}{3}\right)
-m_{\sf T_2}^x m_{\sf T_2}^z=0 \nonumber \\
2 m_{\sf E} m_{\sf T_2}^z \sin\left(\theta_{\sf E}+\frac{2 \pi}{3}\right)-
m_{\sf T_2}^x m_{\sf T_2}^y=0.
\label{eq:U1PCconstraints}
\end{eqnarray}


\begin{figure}
\begin{centering}
\includegraphics[width=.8\columnwidth]{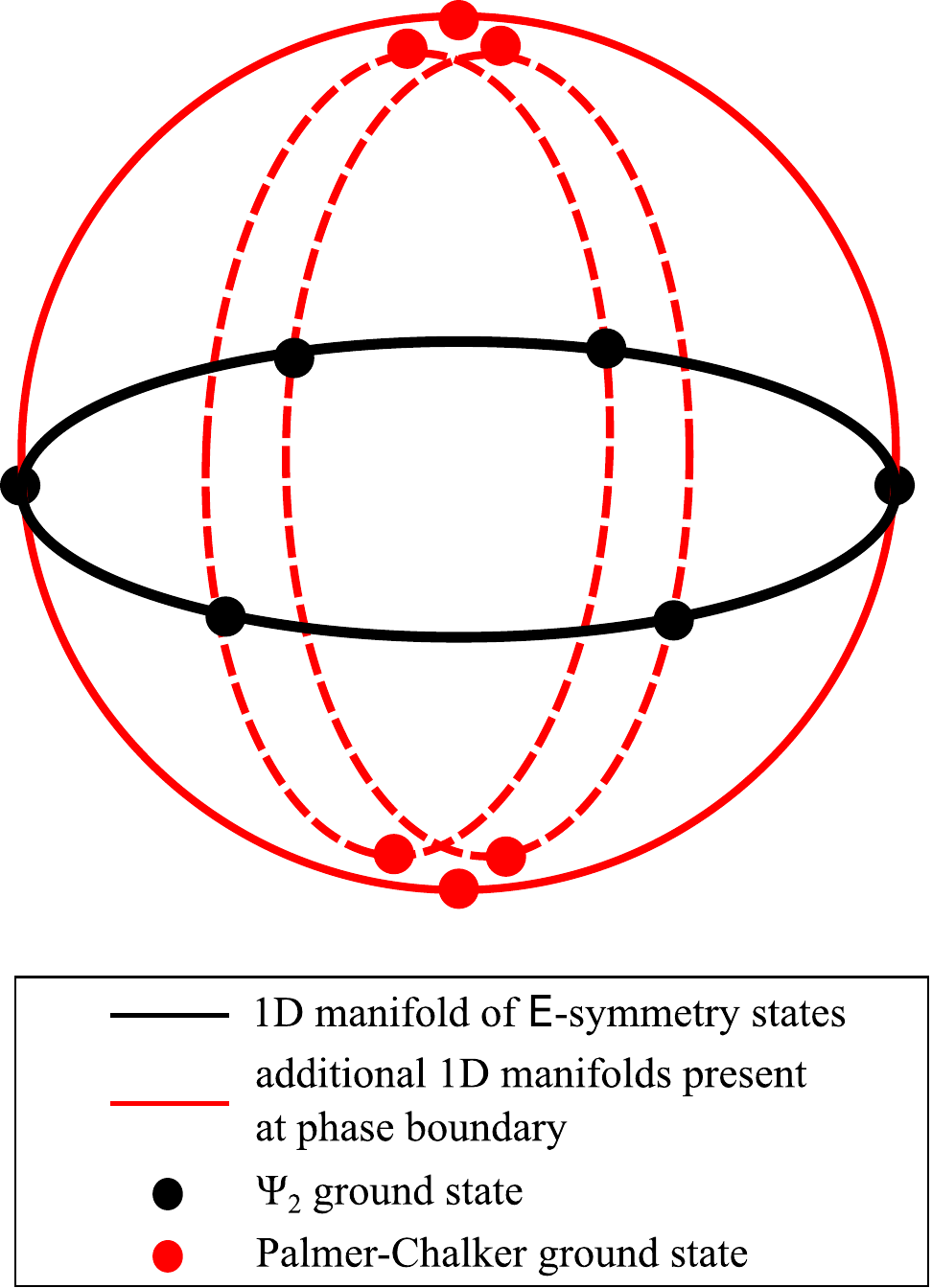}
\par\end{centering}
\caption{
Structure of the ground state manifold at the boundary between the 
Palmer--Chalker (PC) phase and the one--dimensional manifold of 
states with ${\sf E}$ symmetry. 
The black circle denotes the manifold of ${\sf E}$--symmetry ground 
states, including the six $\Psi_2$ ground states (black dots).
At the boundary with the PC phase, this manifold branches at 
the $\Psi_2$ states, to connect with three, additional, one--dimensional 
manifolds.
These manifolds in turn interpolate to the six Palmer--Chalker 
ground states with ${\sf T_2}$ symmetry (red dots).
An exactly equivalent picture holds on the boundary between the non-collinear 
ferromagnet (FM), and the one--dimensional manifold 
of states with ${\sf E}$ symmetry. 
However in this case the different manifolds intersect at the $\Psi_3$ states.
}
\label{fig:manifold-T2-E} 
\end{figure}


It is easy to show that there are no solutions to Eqs.~(\ref{eq:U1PCconstraints})
where more than one component of ${\bf m}_{\sf T_2}$ is finite.
There are, however, three distinct one--dimensional manifolds which connect pairs of 
Palmer--Chalker states to the one--dimensional manifold of $\sf E$-symmetry states:
\begin{eqnarray}
{\bf m}_{\sf E} 
   = \cos(\alpha) 
     \begin{pmatrix}
        1 \\
        0
    \end{pmatrix}, 
    \
{\bf m}_{\sf T_2}
   = \sin(\alpha) 
       \begin{pmatrix}
        1 \\
        0 \\
        0
      \end{pmatrix} 
\label{eq:spoke1}
\\
{\bf m}_{\sf E} 
   = \cos(\beta)
        \begin{pmatrix}
           -\frac{1}{2} \\
           \frac{\sqrt{3}}{2}
      \end{pmatrix}, 
      \
{\bf m}_{\sf T_2}
   = \sin(\beta) 
        \begin{pmatrix}
             0 \\
             1 \\
             0
         \end{pmatrix} 
\label{eq:spoke2}
\\
{\bf m}_{\sf E} 
    = \cos(\gamma)
         \begin{pmatrix}
            -\frac{1}{2} \\
           -\frac{\sqrt{3}}{2}
        \end{pmatrix}, 
        \
{\bf m}_{\sf T_2}
      = \sin(\gamma) \begin{pmatrix}
            0 \\
            0 \\
            1
         \end{pmatrix} 
\label{eq:spoke3}
\end{eqnarray}
where the angles $\alpha$, $\beta$ and $\gamma$ run from $0$ to $2\pi$.  


A typical spin configuration for one of the three connecting manifolds is
\begin{eqnarray}
{\bf S}_{0} & = & S \sqrt{\frac{2}{3}} \left( -\cos(\alpha),
   \:\cos \left( \alpha+\frac{\pi}{3} \right),
   \:\cos \left( \alpha-\frac{\pi}{3} \right) \right) \nonumber\\
{\bf S}_{1} & = & S \sqrt{\frac{2}{3}} \left( -\cos(\alpha),
   \:-\cos \left( \alpha+\frac{\pi}{3} \right),
   \:-\cos \left( \alpha-\frac{\pi}{3} \right) \right) \nonumber\\
{\bf S}_{2} & = & S \sqrt{\frac{2}{3}} \left( \cos(\alpha),
   \:\cos \left( \alpha-\frac{\pi}{3} \right),
   \:-\cos \left( \alpha+\frac{\pi}{3} \right) \right),\nonumber\\
{\bf S}_{3} & = & S \sqrt{\frac{2}{3}} \left( \cos(\alpha),
   \:-\cos \left( \alpha-\frac{\pi}{3} \right),
   \:\cos \left( \alpha+\frac{\pi}{3} \right) \right) .\label{con1}
\end{eqnarray}
where $\alpha = 0$, $\pi$ correspond to the $\Psi_2$ 
ground states with $\theta_{\sf E}  = 0$, $\pi$,  
and $\alpha = \pi/2$, $3\pi/2$ correspond to two
of the six Palmer--Chalker ground states.
These manifolds are illustrated in Fig.~\ref{fig:manifold-T2-E}.


By application of the ``Lego--brick'' rules given
in Section~\ref{Lego-brick-rules}, the ground state degeneracy on this
phase boundary must be at least $\mathcal{O}(2^{L^2})$.
This follows from the fact that, on the boundary, the Palmer--Chalker ground 
states  $\Psi_4$ share two spin orientations with the neighbouring $\Psi_3$ configuration.
This, in turn, is connected with the $\mathcal{O}(L^2)$ number of zero modes 
appearingin spin wave expansions around the $\Psi_2$ configurations at this boundary 
[cf. Section~\ref{section:classical-spin-wave-E}].


We note that a special case of this phase boundary is studied for pure 
XY spins (i.e. with infinite easy--plane anisotropy) in Ref.~[\onlinecite{mcclarty14}], 
finding the same structure of ground state manifolds.

\subsection{Boundary between the non-collinear ferromagnet and the 
                   one--dimensional manifold of states with ${\sf E}$ symmetry}

The boundary between the non-collinear ferromagnet and the one--dimensional manifold 
of states with ${\sf E}$ symmetry occurs when $a_{\sf E} = a_{\sf T_{1, A^\prime}}$
[cf. Eq.~(\ref{eq:boundary-E-T1})].  
In this case, ${\mathcal H}_{\sf ex}^{[{\sf T_d}]}$~[Eq.~(\ref{eq:HTd})]
is minimised by setting 
\begin{eqnarray}
{\bf m}_{\sf T_{1,A'}}^2 + {\bf m}_{\sf E}^2 = 1
\end{eqnarray}
and
\begin{eqnarray}
m_{\sf A_2}  = {\bf m}_{\sf T_{1,B'}}  = {\bf m}_{\sf T_2}  =  0.
\end{eqnarray}

Defining, for the sake of brevity, the quantities
\begin{eqnarray}
\mu(\theta_{\sf T_1}) 
    &=& (\sqrt{2} \cos(\theta_{\sf T_1})-\sin(\theta_{\sf T_1})) 
            \nonumber  \\
\nu(\theta_{\sf T_1})
     &=& (\sin(\theta_{\sf T_1})^2+\sqrt{2}\sin(2 \theta_{\sf T_1}))
     \label{eq:munu}
\end{eqnarray}
and imposing the constraint Eq.~(\ref{eq:spinconstraints}) we obtain
\begin{eqnarray}
2 m_{\sf E} m_{\sf T_{1,A'}}^x \cos(\theta_{\sf E})
    &=& - \frac{\mu(\theta_{\sf T_{1,A'}})}{\nu(\theta_{\sf T_{1,A'}})}
            m^y_{\sf {T_{1,A'}}} m^z_{\sf {T_{1,A'}}} 
            \nonumber \\
2 m_{\sf E} m_{\sf T_{1,A'}}^y \cos\left(\theta_{\sf E}-\frac{2\pi}{3}\right)
    &=& - \frac{\mu(\theta_{\sf T_{1,A'}})}{\nu(\theta_{\sf T_{1,A'}})}
           m^x_{\sf {T_{1,A'}}} m^z_{\sf {T_{1,A'}}} 
           \nonumber \\
           2 m_{\sf E} m_{\sf T_{1,A'}}^z \cos\left(\theta_{\sf E}+\frac{2\pi}{3}\right)
   &=& - \frac{\mu(\theta_{\sf T_{1,A'}})}{\nu(\theta_{\sf T_{1,A'}})} 
           m^x_{\sf {T_{1,A'}}} m^y_{\sf {T_{1,A'}}} 
           \nonumber \\
\end{eqnarray}
where $\theta_{\sf T_1}$ is the (fixed) canting angle [Eq.~(\ref{eq:FMangle})],
$\theta_{\sf E}$ is the (variable) angle within the U(1) manifold
[Eq.~(\ref{eq:thetaE})].
For the parameters considered here, the quantities $\mu(\theta_{\sf T_1}) $ 
and $\nu(\theta_{\sf T_1}) $
are always finite.
 

Arguments identical to those developed for the boundary with
the Palmer--Chalker phase, give us three further 1D manifolds
in addition to that associated with the $\sf E$ phase.
However the intersections of the manifolds are now located at 
$\theta_{\sf E}=\frac{(2n + 1) \pi}{6}$,  corresponding
to the $\Psi_3$ states.
This explains the model's general entropic preference 
for $\Psi_3$ states in the region proximate to the 
ferromagnetic phase.


A typical spin configuration for one of the three connecting manifolds,
parameterised by an angle $\eta$ is
\begin{eqnarray}
{\bf S}_{0} & = &  S \bigg( \cos(\theta_{\sf T_1}) \sin(\eta),
   \:\frac{1}{\sqrt{2}}(-\cos(\eta) +\sin(\eta) \sin(\theta_{\sf T_1}) ), \nonumber \\
&&\frac{1}{\sqrt{2}}(\cos(\eta) +\sin(\eta) \sin(\theta_{\sf T_1}) \bigg) \nonumber\\
   {\bf S}_{1} & = &  S \bigg( \cos(\theta_{\sf T_1}) \sin(\eta),
   \:\frac{1}{\sqrt{2}}(\cos(\eta) -\sin(\eta) \sin(\theta_{\sf T_1}) ), \nonumber \\
&&\frac{1}{\sqrt{2}}(-\cos(\eta) -\sin(\eta) \sin(\theta_{\sf T_1}) \bigg)\nonumber\\
   {\bf S}_{2} & = &   S \bigg( \cos(\theta_{\sf T_1}) \sin(\eta),
   \:\frac{1}{\sqrt{2}}(-\cos(\eta) -\sin(\eta) \sin(\theta_{\sf T_1}) ), \nonumber \\
&&\frac{1}{\sqrt{2}}(-\cos(\eta) +\sin(\eta) \sin(\theta_{\sf T_1}) \bigg)\nonumber\\
   {\bf S}_{3} & = & S \bigg( \cos(\theta_{\sf T_1}) \sin(\eta),
   \:\frac{1}{\sqrt{2}}(\cos(\eta) +\sin(\eta) \sin(\theta_{\sf T_1}) ), \nonumber \\
&&\frac{1}{\sqrt{2}}(\cos(\eta)-\sin(\eta) \sin(\theta_{\sf T_1}) \bigg).\label{conFMU1}
\end{eqnarray}
Here $\eta = 0$ corresponds to the $\Psi_3$ ground state with 
$\theta_{\sf E} = \pi/2$, and $\eta = \pi/2$ to one of the six FM ground states.


We note that an equivalent ground--state manifold was discussed
by Canals and coauthors~\cite{canals08}, and later by Chern \cite{chern-arXiv},
 in the context the Heisenberg antiferromagnet with 
Dzyaloshinskii--Moriya interactions on the pyrochlore lattice.
This case corresponds to a single point on the phase boundary considered here.  

\subsection{Boundary between the Palmer--Chalker phase and the non-collinear ferromagnet}
\label{FM-meets-PC}

The boundary between the Palmer--Chalker phase and the non-collinear ferromagnet 
occurs when $a_{\sf T_2} = a_{\sf T_{1, A^{\prime}}}$ 
[cf. Eq.~(\ref{eq:boundary-T2-T1})].  
In this case, ${\mathcal H}_{\sf ex}^{[{\sf T_d}]}$~[Eq.~(\ref{eq:HTd})]
is minimised by setting 
\begin{eqnarray}
&&{\bf m}_{\sf T_2}^2+{\bf m}_{\sf T_{1, A^{\prime}}}^2=1 \nonumber \\
\end{eqnarray}
and
\begin{eqnarray}
m_{\sf A_2}={\bf m}_{\sf E}={\bf m}_{\sf T_{1, B'}}=0.
\end{eqnarray}


Imposing the constraint Eq.~(\ref{eq:spinconstraints}) we obtain
\begin{eqnarray}
&&-m_{\sf T_2}^y m_{\sf T_2}^z + 
(\sin( \theta_{\sf T_1})^2
+\sqrt{2}\sin(2 \theta_{\sf T_1})) m_{\sf T_{1 A'}}^y  m_{\sf T_{1 A'}}^z  \nonumber \\
&& \ \ \ + (\sqrt{2} \cos( \theta_{\sf T_1})
-\sin( \theta_{\sf T_1})) ({\bf m_{\sf T_{1 A'}}} \times {\bf m_{\sf T_2}})_x
=0 \nonumber \\
&&-m_{\sf T_2}^x m_{\sf T_2}^z + 
(\sin( \theta_{\sf T_1})^2
+\sqrt{2}\sin(2  \theta_{\sf T_1})) m_{\sf T_{1 A'}}^x  m_{\sf T_{1 A'}}^z  \nonumber \\
&& \ \ \ + (\sqrt{2} \cos( \theta_{\sf T_1})-\sin( \theta_{\sf T_1})) 
({\bf m_{\sf T_{1 A'}}} \times {\bf m_{\sf T_2}})_y
=0 \nonumber \\
&&-m_{\sf T_2}^x m_{\sf T_2}^y + 
(\sin( \theta_{\sf T_1})^2
+\sqrt{2}\sin(2  \theta_{\sf T_1})) m_{\sf T_{1 A'}}^x  m_{\sf T_{1 A'}}^y  \nonumber \\
&& \ \ \ + (\sqrt{2} \cos( \theta_{\sf T_1})-\sin( \theta_{\sf T_1})) 
({\bf m_{\sf T_{1 A'}}} \times {\bf m_{\sf T_2}})_z
=0 \nonumber \\
\label{eq:FMPCconstraints}
\end{eqnarray}
where $\theta_{\sf T_1}$ is defined in Eq.~(\ref{eq:FMangle}).


In general, the ground state manifold on the boundary of the Palmer--Chalker phase 
is locally two--dimensional.
To establish this, we consider small deviations from a given solution 
\begin{eqnarray}
{\bf m}_{\sf T_2} &=& {\bf m}_{\sf T_2}^0 + {\bf \delta m}_{\sf T_2} \nonumber \\
{\bf m}_{\sf T_{1 A^{\prime}}} &=& {\bf m}_{\sf T_{1 A^{\prime}}}^0 + {\bf \delta m}_{\sf T_{1 A^{\prime}}}
\end{eqnarray}
and expand the constraint Eq.~(\ref{eq:FMPCconstraints}) to linear order 
in ${\bf \delta m}$.  
Generally, we find two linearly--independent solutions for 
$({\bf \delta m}_{\sf T_2},{\bf \delta m}_{\sf T_{1 A^{\prime}}})$, 
and the manifold in the vicinity of 
$({\bf m}_{\sf T_2}^0, {\bf m}_{\sf T_{1 A^{\prime}}}^0)$
is  two--dimensional.


However if we expand around a state 
$(\tilde{\bf m}_{\sf T_2}^0, \tilde{\bf m}_{\sf T_{1 A^{\prime}}}^0)$
where {\it both} order parameters are aligned with the same cubic axis, e.g.
\begin{eqnarray}
\tilde{m}_{\sf T_2}^{0y} 
= \tilde{m}_{\sf T_2}^{0z} 
= \tilde{m}_{\sf T_{1 A^{\prime}}}^{0y}
= \tilde{m}_{\sf T_{1 A^{\prime}}}^{0z}
= 0
\end{eqnarray}
one of the Eqs.~(\ref{eq:FMPCconstraints}) is satisfied trivially, 
leaving only three constraints on six variables.
It follows that the manifold is locally three-dimensional in the vicinity of 
$(\tilde{\bf m}_{\sf T_2}^0, \tilde{\bf m}_{\sf T_{1 A^{\prime}}}^0)$.

This set of ground states on the tetrahedron includes multiple states 
where one of the spins has the same direction. 
Applying the ``Lego--brick'' rules, described in Section~\ref{Lego-brick-rules},
this means that neighbouring kagome planes can be effectively decoupled in 
the ground state and there is a ground state degeneracy on the lattice
of at least $\mathcal{O}(2^L)$.

\subsection{All--in, all--out order with ${\sf A_2}$ symmetry}
\label{A2}

In the Section~\ref{ferromagnet}---\ref{palmer-chalker} we have explicitly 
discussed the different types of ordered, classical ground state 
which occur for $J_3<0, J_4=0$.
For a more general choice of parameters, with $J_4>0$ or $J_3>0$,
it is  possible to find situations where the lowest parameter 
in ${\mathcal H}_{\sf ex}^{\sf tet}$ [Eq.~(\ref{eq:HTd})] is  $a_{\sf A_2}$.   
In this case, the ground state will have 
4-sublattice order with a finite value of the order parameter $m_{\sf A_2}$
[Table~\ref{table:m_lambda_global}].  
As can be seen from the definition of the order parameter, this type 
of order is particularly simple, with all spins aligned along the 
local [111] axes, and all spins pointing either into, or out of, 
tetrahedra on the A-sublattice.  
This type of order is commonly referred to as ``all--in, all--out'' 
and is observed in some pyrochlore magnets, including 
Nd$_2$Zr$_2$O$_7$ [\onlinecite{lhotel15}].


Since this type of order does not occur for \mbox{$J_3<0, J_4=0$}, we will
not discuss it further in here.  
However we note that the order parameter $m_{\sf A_2}$ is a scalar, 
and that finite--temperature phase  transitions are therefore expected 
to fall into the Ising universality class, in the absence of a first order phase transition.

\section{Theory of classical and quantum spin wave excitations}
\label{section:spin-wave-theory} 

In order to complete the classical phase diagram described in 
Section~\ref{section:classical-ground-states}, it is necessary to understand how quantum 
and/or classical fluctuations select between the one--dimensional manifold of states 
described by ${\bf m}_{\sf E}$.  
At low temperatures, this can be accomplished by exploring the way in which 
spin-wave excitations contribute to the free energy. 
Knowledge of the spin-wave excitations also makes it possible to make
predictions for inelastic neutron scattering, discussed below, and to benchmark the 
results of the classical Monte Carlo simulations described in Section~\ref{section:finite-temperature}.


In what follows, we describe a general theory of classical, and quantum
spin-wave excitations about the different ordered states described in 
Section~\ref{section:classical-ground-states}.
In Section~\ref{section:classical-spin-wave} we establish a classical, low-temperature 
spin-wave expansion, which makes it possible to 
determine the boundary between the  $\Psi_2$ and $\Psi_3$ ground states 
for classical spins in the limit $T \to 0$ [cf. Fig.~\ref{fig:classical-phase-diagram}].
In Section~\ref{subsection:quantum-spin-wave}, we develop an equivalent quantum theory,
within the linear spin-wave approximation, which allows us to estimate the boundary between the  $\Psi_2$ and $\Psi_3$ ground states 
for quantum spins in the limit $T \to 0$ [cf. Fig.~\ref{fig:quantum-phase-diagram}].
We find that the high degeneracies at classical phase boundaries, described in 
Section~\ref{section:classical-ground-states}, strongly enhance quantum fluctuations, 
and in some cases eliminate the ordered moment entirely.
In Section~\ref{subsection:neutron-scattering} we show how both classical
and quantum spin-wave theories can be used to make predictions for
inelastic neutron scattering.
Readers interested in the relationship between classical and quantum spin--wave
theories are referred to the discussion in Ref.~[\onlinecite{seabra16}]. 

\subsection{Classical spin-wave expansion}
\label{section:classical-spin-wave}


To obtain the low-energy excitations around the ordered ground states of 
$\mathcal{H}_{\sf ex}$ [Eq. \ref{eq:Hex}] we use a description in terms
of classical spin waves, analogous to that described in Ref. [\onlinecite{shannon10}].
We begin by defining a local coordinate system by introducing a set of orthogonal 
unit vectors $\{ {\bf u}_i,  {\bf v}_i,  {\bf w}_i \}$ for each of the four
sublattices $i=0,1,2,3$ [cf. Fig.~\ref{fig:tetrahedron}].
The local ``$z$-axis'', ${\bf w}_i$, is chosen to be aligned with the
spins in a given four-sublattice ground state
\begin{eqnarray}
{\bf S}_i= S {\bf w}_i \quad \forall \ i
\end{eqnarray}
The remaining unit vectors, ${\bf u}_i$ and  ${\bf v}_i$, are only determined
up to a rotation about ${\bf w}_i$, and any convenient choice can be made.


Using this basis, the fluctuations of the spin $\textbf{S}_{ik}$ on sublattice $i$ of tetrahedron $k$ can be 
parameterized as
\begin{eqnarray}
{\bf S}_{ik}&=& 
\begin{pmatrix}
\sqrt{S} \delta u_{ik} \\
\sqrt{S} \delta v_{ik} \\
\sqrt{S^2- S \delta u_{ik}^2-S \delta v_{ik}^2}
\end{pmatrix} \nonumber \\
&\approx&
\begin{pmatrix}
\sqrt{S} \delta u_{ik} \\
\sqrt{S} \delta v_{\vec{ik}} \\
S - \frac{1}{2}  \delta u_{ik}^2- \frac{1}{2}  \delta v_{ik}^2
\end{pmatrix}.
\label{eq:fluc}
\end{eqnarray}
Substituting Eq.~(\ref{eq:fluc}) into $\mathcal{H}_{\sf ex}$ [Eq.~(\ref{eq:Hex})]
we obtain
\begin{eqnarray}
\mathcal{H}_{\sf ex}  
   &=& \sum_{{\sf tet} \  k} \sum_{i < j} {\bf S}_{ik} \cdot {\bf J}_{ij}\cdot {\bf S}_{jk} \nonumber \\
   &=& {\mathcal E}_0 +  \mathcal{H}_{\sf ex}^{\rm CSW} + \ldots
\label{eq:spin-wave-expansion}
 \end{eqnarray}
where
\begin{eqnarray}
{\mathcal E}_0  &=& \frac{N S^2}{4} \sum_{i,j=0}^{3} {\bf w}_i \cdot {\bf J}_{ij} \cdot {\bf w}_j 
\label{eq:E0}
\end{eqnarray}
is the classical ground-state energy of the chosen \mbox{4-sublattice} state, 
and %
\begin{eqnarray}
\mathcal{H}_{\sf ex}^{\sf CSW}
 &=&  \frac{S}{2} \sum_k \sum_{i, j=0}^3 \nonumber \\
&&   \bigg[
 - \frac{1}{2} (\delta u_{ik}^2 +\delta u_{jk}^2 +\delta v_{ik}^2 +\delta v_{jk}^2)
  \left( {\bf w}_i \cdot {\bf J}_{ij} \cdot {\bf w}_j\right) \nonumber \\
 &+& \ \delta u_{i k} \delta u_{j k} \left( {\bf u}_i \cdot {\bf J}_{ij} \cdot {\bf u}_j \right)
 +  \delta v_{i k} \delta v_{j k} \left( {\bf v}_i \cdot {\bf J}_{ij} \cdot {\bf v}_j \right) \nonumber \\
 &+&  \delta u_{i k} \delta v_{j k} \left( {\bf u}_i \cdot {\bf J}_{ij} \cdot {\bf v}_j \right) 
 +  \delta v_{i k} \delta u_{j k} \left( {\bf v}_i \cdot {\bf J}_{ij} \cdot {\bf u}_j \right)
  \bigg] \nonumber \\
  \label{eq:Hfluc}
\end{eqnarray}
describes the leading effect of (classical) fluctuations about this state.  
Performing a Fourier transformation, we find 
\begin{eqnarray}
\mathcal{H}_{\sf ex}
   &=& \frac{N S^2}{4} \sum_{i,j=0}^{3} 
         {\bf w}_i \cdot {\bf J}_{ij} \cdot {\bf w}_j 
\nonumber \\
                  &+& \frac{1}{2}  \sum_{{\bf q}} \tilde{u}(-{\bf q})^T \cdot {\bf M}({\bf q})
                 \cdot \tilde{u}({\bf q})
\end{eqnarray}
Here $ \tilde{u}({\bf q})$ is the vector 
\begin{eqnarray}
\tilde{u}({\bf q})=\bigg( 
\delta u_0 ({\bf q}), 
\delta u_1 ({\bf q}),
\delta u_2 ({\bf q}),
\delta u_3 ({\bf q}), \nonumber \\
\delta v_0 ({\bf q}),
\delta v_1 ({\bf q}),
\delta v_2 ({\bf q}),
\delta v_3 ({\bf q}) \bigg)^T,
\end{eqnarray}
and ${\bf M}({\bf q})$ the $8 \times 8$ matrix 
\begin{eqnarray}
{\bf M}({\bf q})=
2 S \begin{pmatrix}
{\bf M}^{11}({\bf q})
& {\bf M}^{12}({\bf q}) \\
{\bf M}^{21}({\bf q})
& {\bf M}^{22}({\bf q})  \\
  \end{pmatrix}  
  \label{eq:M}
\end{eqnarray}
built from $4 \times 4$ blocks 
\begin{eqnarray}
{\bf M}^{11}_{ij}({\bf q})
   &=&\cos({\bf q} \cdot {\bf r}_{ij} )  \nonumber \\
&&\bigg( {\bf u}_i \cdot {\bf J}_{ij} \cdot {\bf u}_j 
 - \delta_{ij}  \sum_{l} \left(  {\bf w}_l \cdot {\bf J}_{lj} \cdot {\bf w}_j   \right)  
\bigg) \nonumber \\
\\
{\bf M}^{12}_{ij}({\bf q})
   &=&{\bf M}^{21}_{ji}({\bf q})=\cos({\bf q} \cdot {\bf r}_{ij} ) 
\bigg( {\bf v}_i \cdot {\bf J}_{ij} \cdot {\bf u}_j  
\bigg) \\
{\bf M}^{22}_{ij}({\bf q})
   &=&\cos({\bf q} \cdot {\bf r}_{ij} ) \nonumber \\
&& \bigg( {\bf v}_i \cdot {\bf J}_{ij} \cdot {\bf v}_j 
 - \delta_{ij}  \sum_{l} \left(  {\bf w}_l \cdot {\bf J}_{lj} \cdot {\bf w}_j   \right)  
\bigg) \nonumber \\
\end{eqnarray}
where \mbox{$i, j  \in \{0, 1, 2,  3\}$} and \mbox{${\bf r}_{ij} = {\bf r}_j - {\bf r}_i$} 
[cf. Eq.~(\ref{eq:r})].


The matrix ${\bf M}({\bf q})$ [Eq.~(\ref{eq:M})] can be diagonalized 
by a suitable orthogonal transformation, \mbox{${\bf U}=({\bf U}^T)^{-1}$} to give
\begin{equation}
{\mathcal H}_{\sf ex}^{\sf CSW}
           =  \frac{1}{2}  \sum_{{\bf q}} \sum_{\nu=1}^{8}
           \kappa_{\nu {\bf q}} \upsilon_{\nu {\bf q}}  
           \upsilon_{\nu -{\bf q}}
           \label{eq:HlowT}
\end{equation}
where the eight normal modes of the system are given by 
\begin{eqnarray}
\upsilon({\bf q})={\bf U} \cdot \tilde{u}({\bf q})
\label{eq:orthogonaltransformation}
\end{eqnarray}
with associated eigenvalues 
$\kappa_{\nu} ({\bf q})$.
Since ${\mathcal H}_{\sf ex}^{\sf CSW}$ [Eq.~(\ref{eq:HlowT})] is quadratic in 
$\upsilon_{\nu {\bf q}}$, the associated partition function can be calculated
exactly
\begin{eqnarray}
\mathcal{Z}^{\sf CSW}_{\sf ex} &=& \left( \frac{1}{\sqrt{2 \pi}} \right)^{2N} \exp{\left( \frac{-\mathcal{E}_0}{T}  \right)} \int \left[\prod_{\nu=1}^{8} \prod_{{\bf q}}d\upsilon_{\nu {\bf q}}\right]\nonumber \\
&&\exp{\left( -\frac{1}{2}
  \frac{\sum_{\nu=1}^{8} \sum_{{\bf q}} \kappa_{\nu {\bf q}} \upsilon_{\nu {\bf q}} 
   \upsilon_{\nu -{\bf q}} }{T} 
 \right)} \nonumber \\
 & =&  \exp{\left( \frac{-\mathcal{E}_0}{T} \right)}  \prod_{\nu=1}^{8} \prod_{{\bf q}} \left( \sqrt{\frac{T}{\kappa_{\nu {\bf q}}}}  \right). 
  \label{eq:partitionfunction}
\end{eqnarray}
It follows that, for $T \to 0$, the free energy of the system is given by
\begin{eqnarray}
\mathcal{F}_{\sf ex}^{\sf low-T} &=& 
   {\mathcal E}_0 
   + \frac{T}{2} \sum_{\nu {\bf q}}  \ln \kappa_{\nu {\bf q}} 
   - N T \ln T 
   + {\mathcal O}(T^2).   \nonumber \\
   \label{eq:FlowT}
\end{eqnarray}
Where the ${\mathcal O}(T^2)$ corrections arise from the higher order, spin wave
interaction, terms neglected in Eq.~(\ref{eq:spin-wave-expansion}).


Within this classical, low-T expansion, the eigenvalues $\kappa_{\nu} ({\bf q})$ 
correspond to independent, low--energy modes, which determine the 
physical properties of the states, and have the interpretation of a classical 
spin-wave spectrum.
However the classical spectrum $\kappa_{\nu} ({\bf q})$ should {\it not} be confused 
with the quantum spin-wave dispersion $\omega_{\nu} ({\bf q})$, measured in inelastic 
neutron scattering experiments and discussed in Section~\ref{subsection:quantum-spin-wave}.
%


\begin{figure*}
\includegraphics[width=0.95\textwidth]{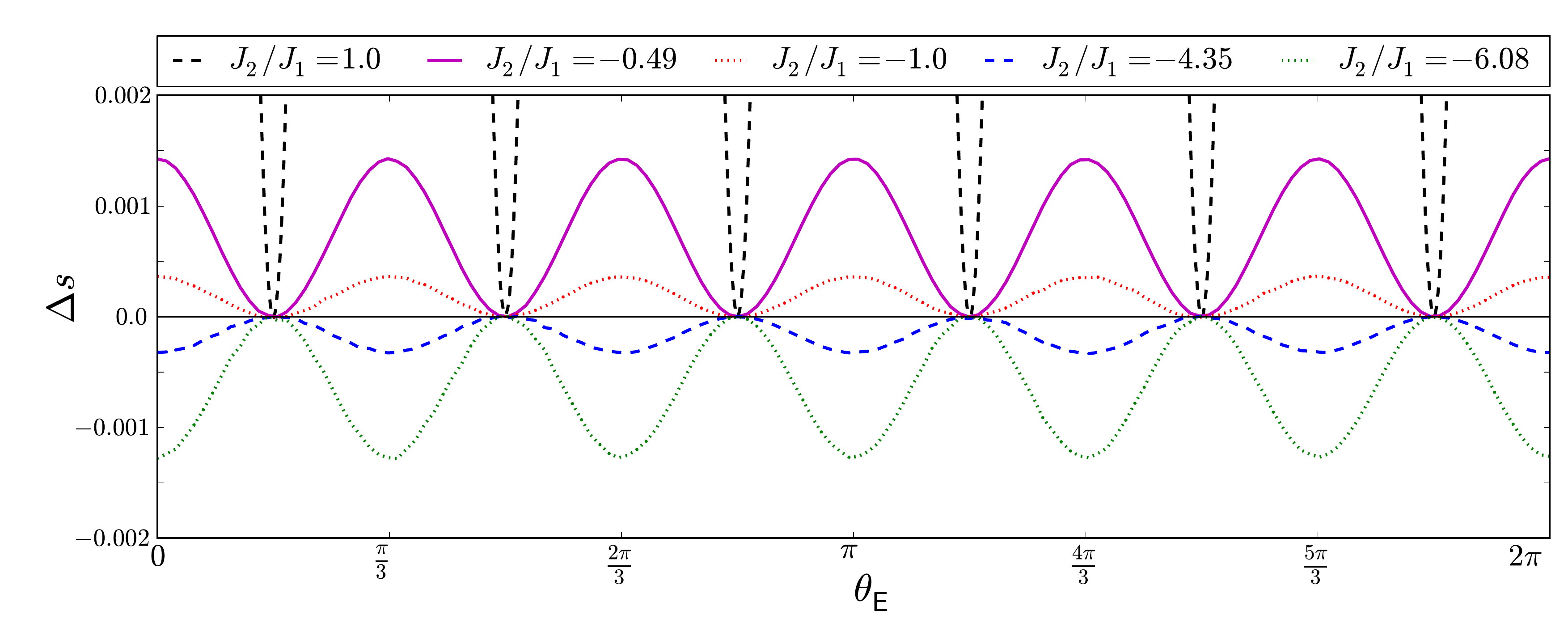}
\caption{
Variation of entropy per spin within the one--dimensional manifold of 
states with symmetry ${\sf E}$. 
Entropy ${\mathcal S}(\theta_{\sf E})$ has been estimated using the low-temperature 
expansion [Eq.~(\ref{eq:S})], for a range of values of $J_2$, with the entropy 
of the $\Psi_3$ state subtracted as a reference, i.e. 
$\Delta s_{\theta_{\sf E}} = [{\mathcal S}(\theta_{\sf E}) - {\mathcal S}(\pi/6)]/ N$.
The parameters $J_1=0.115 \text{meV}$ and $J_3=-0.099 \text{meV}$ were fixed 
at values appropriate to Er$_2$Ti$_2$O$_7$ [\onlinecite{savary12-PRL109}], 
setting $J_4 \equiv 0$.
In all cases, $\Delta s_{\theta_{\sf E}}$ repeats with period $\pi/3$.  
For a choice of $J_2$ appropriate to Er$_2$Ti$_2$O$_7$  
[$J_2/J_1=-0.49$ --- solid purple line],  entropy takes on its maximum value for 
$\theta_{\sf E}=\frac{n \pi}{3}$, with $n=0,1,2,3,4,5$, corresponding to the six 
$\Psi_2$ ground states.  
The extreme variation in entropy at the boundary of the Palmer--Chalker phase 
[$J_1=J_2$ --- dashed black line], reflects the presence of an $\mathcal{O}(L^2)$ 
set of zero modes in the spectrum of $\Psi_2$ ground state.
None the less, the entropy difference between $\Psi_2$ and $\Psi_3$,  
$\Delta s_{\pi/3} \approx 0.18$ remains finite.
For sufficiently negative $J_2$ (dashed blue line, dotted green line) 
$\Delta  s_{\pi/3} < 0$, and fluctuations select the $\Psi_3$ state.
All results have been calculated from Eq.~(\ref{eq:S}), 
with the sum evaluated numerically by a Monte Carlo method.
Statistical errors are smaller than the point size.
}
\label{fig:entropy}
\end{figure*}

\subsection{Ground-state selection within the one--dimensional manifold 
                   of states with ${\sf E}$ symmetry}
\label{section:classical-spin-wave-E}

Energy alone does not select between the one--dimensional manifold of states 
with ${\sf E}$ symmetry~\cite{champion03,champion04}.
However quantum fluctuations~\cite{zhitomirsky12,savary12-PRL109,wong13},
thermal fluctuations at low temperature~\cite{champion03,champion04},
thermal fluctuations near the ordering temperature \cite{oitmaa13, zhitomirsky14},
structural disorder~\cite{maryasin14,andreanov15} 
and structural distortion~\cite{maryasin14} {\it are} effective in selecting an ordered ground state.
In what follows, we use knowledge of the free energy within a classical spin-wave theory, 
$\mathcal{F}_{\sf ex}^{\sf low-T}$ [Eq.~(\ref{eq:FlowT})], to determine which of the possible 
${\sf E}$-symmetry ground states is selected by thermal fluctuations in the limit $T \to 0$.
A parallel treatment of the quantum problem is given in Ref.~[\onlinecite{wong13}]
and a classical analysis applied to the limiting case of pure XY spins (i.e. with
infinite easy plane anisotropy) is given in Ref.~[\onlinecite{mcclarty14}].


As a first step, it is helpful to write down a minimal, symmetry allowed,
form for the free energy in terms of the components of
\begin{eqnarray}
{\bf m}_{\sf E}=
\begin{pmatrix}
m_{\sf E} \cos(\theta_{\sf E}) \\
m_{\sf E} \sin(\theta_{\sf E}) 
\end{pmatrix}
\end{eqnarray}
(cf.~[\onlinecite{javanparast15}]).
Keeping only those terms which respect the lattice
symmetries, and going up to 6th order in $m_{\sf E}$
\begin{eqnarray}
{\mathcal F}_{\sf E} 
   &=& {\mathcal F}_0
    +  \frac{1}{2} \; a \; m_{\sf E}^2 
    + \frac{1}{4} \; b \; m_{\sf E}^4 
    + \frac{1}{6} \; c \; m_{\sf E}^6 \nonumber\\   
  &&  + \frac{1}{6} \; d \; m_{\sf E}^6 \; \cos(6\,\theta_{\sf E})
    + {\mathcal O} (m_{\sf E}^8)
\label{eq:Landau}
\end{eqnarray}
where ${\mathcal F}_0$ is an unimportant constant.

Note that Eq.~(\ref{eq:Landau}) does not contain the symmetry
allowed coupling to $m_{\sf A_2}$:
$$
m_{\sf A_2} m_{\sf E}^3 \cos(3 \theta_{\sf E})
$$
which appears in [\onlinecite{javanparast15}].
This is for two reasons:
(i) we are considering the $T\to0^{+}$ limit where a finite
value of $m_{\sf A_2}$ is energetically unfavourable and will
be very small;
(ii) $m_{\sf A_2}$ can, in any case, be integrated out
to arrive at Eq. (\ref{eq:Landau}) with renormalized coefficients
for the 6th order terms.

It follows from Eq. (\ref{eq:Landau}) that i) a suitable  order parameter for 
symmetry breaking within this manifold is $c_{\sf E} = \cos 6 \theta_{\sf E}$ 
[\onlinecite{chern-arXiv, zhitomirsky14}], cf.~Eq.~(\ref{eq:ctheta}), and that 
ii) the two states spanning ${\bf m}_{\sf E}$, $\Psi_2$ and  $\Psi_3$, 
are distinguished only at sixth-order (and higher) in
$m_{\sf E}$ [\onlinecite{mcclarty09}].
The quantity $c_E$ is a secondary order parameter, in the sense that 
a finite expectation value of $c_{\sf E}$ is induced by coupling to the
primary order parameter ${\bf m}_{\sf E}$.
These facts have important consequences for the finite temperature
phase transition into the paramagnet, as discussed below.


For $T \to 0$, we can parameterise ${\mathcal F}_{\sf E}$ [Eq.~(\ref{eq:Landau})]
from ${\mathcal F}_{\sf ex}^{\sf low-T}$ [Eq.~(\ref{eq:FlowT})].  
Since ${\mathcal H}_{\sf ex}^{[{\sf T_d}]}$ ~[Eq.~(\ref{eq:HTd})] is
quadratic in ${\bf m}_{\sf E}$, all other terms in the free energy 
must be of purely entropic origin.
Moreover, 
symmetry requires that
the entropy 
associated with the ${\sf E}$--symmetry states will vary as 
\begin{eqnarray}
{\mathcal S}_{\sf E} (\theta_{\sf E})  
   &=& N \sum_{n=0, 1, 2\ldots} s_n \cos (6n \theta_{\sf E})
\end{eqnarray}
The sign of the coefficients $s_n$ then determines the ground state 
selected by fluctuations.
Taking the derivative
of Eq.~(\ref{eq:FlowT})
with respect to $T$ allows us to
explicitly calculate  ${\mathcal S}_{\sf E} (\theta_{\sf E}) $ 
\begin{eqnarray}
\frac{{\mathcal S}_{\sf E} (\theta_{\sf E})}{N} &=&
-\frac{1}{N}\frac{\partial {\mathcal F}_{\sf ex}^{\sf low-T}}{\partial T}
 \nonumber \\
    &=& \ln{T} + 1 - \frac{1}{2 N} \sum_{{\bf q}} 
\ln{\left( \det({\bf M}_{\sf \theta_{\sf E}} ({\bf q})) \right)} 
\label{eq:S}
\end{eqnarray}
[cf.~Ref.~[\onlinecite{shannon10}]], where ${\bf M}_{\sf \theta_{\sf E}}({\bf q})$ is the $8 \times 8$ matrix 
defined in Eq.~(\ref{eq:M}), calculated by expanding around a state with a particular value
of $\sf \theta_{\sf E}$.
These results are illustrated in Fig.~\ref{fig:entropy}.
Equivalent calculations, carried out numerically for all parameters associated
with ${\sf E}$--symmetry ground states, lead to the phase boundary between 
$\Psi_2$ and  $\Psi_3$ shown in Fig.~\ref{fig:classical-phase-diagram}.
For parameters appropriate to Er$_2$Ti$_2$O$_7$ [\onlinecite{savary12-PRL109}], 
we find that fluctuations select a $\Psi_2$ ground state, in keeping with 
published work~\cite{champion04,mcclarty09,savary12-PRL109,zhitomirsky12}.


We can now learn more about how ground state selection works by realising
that, for some choices of parameters, the 
operation connecting different ${\sf E}$--symmetry ground states 
becomes an {\it exact} symmetry of the Hamiltonian.
This is most easily seen in a coordinate frame tied to 
the local [111] axis, as described in Section~\ref{section:local-frame}.
Considering ${\mathcal H}^{\sf local}_{\sf ex}$~[Eq.~(\ref{eq:Hross})], 
for the simple choice of parameters
$$
(J_{zz},\ J_{\pm},\ J_{\pm \pm},\ J_{z \pm}) 
    = (0,\  J,\ 0,\ 0) 
    \quad \quad J > 0
$$
the ground state belongs to ${\sf E}$ and the Hamiltonian reduces
to that of an ``XY'' ferromagnet.  
In this case the entire one--dimensional manifold of 
${\sf E}$--symmetry states are connected by an explicit symmetry 
of the Hamiltonian (rotation around the local $\langle 111 \rangle $ axes).
It follows that order-by-disorder is ineffective, and 
the ground state retains its U(1) symmetry 
--- for a related discussion, see [\onlinecite{wong13}].


To gain insight into the phase diagram for \mbox{$J_3 < 0,\ J_4 \equiv 0$}  
[cf.~Fig.~\ref{fig:classical-phase-diagram}], we expand about a point in parameter space
\begin{eqnarray}
(J_{zz},\ J_{\pm},\ J_{\pm \pm},\ J_{z \pm}) 
   & =& (-2J,\ J,\ 0,\ 0)
    \quad \quad J > 0 \nonumber \\
\implies (J_1, J_2, J_3, J_4)  &=& (2J, -2J, 0, 0).\nonumber
\end{eqnarray}
At this point the ground state manifold is formed from linear
combinations of $E$ and $A_2$ symmetry states and 
the entire ground state manifold is connected by an exact symmetry
of the Hamiltonian, so once again there is no order by disorder.
For $J_3<0$, states with a finite value of $m_{\sf A_2}$
are removed from the ground state manifold and fluctuations select
a ground state from amongst the ${\sf E}$ states.
It follows that, for \mbox{$J_3 \to 0^{-},\ J_4 \equiv 0$},  the phase boundary 
between the $\Psi_2$ and $\Psi_3$ states should tend to the 
line \mbox{$J_2/|J_3| = - J_1/|J_3|$} [cf. Fig.~\ref{fig:classical-phase-diagram}].  


To see which phase is preferred for finite $J_3$, we expand the 
difference in entropy $\mathcal{S}_{\sf E} (\theta_{\sf E})$ 
between the $\Psi_2$ and $\Psi_3$ ground states
\begin{eqnarray}
\Delta s_{\pi/3}
   = \frac{\mathcal{S}_{\sf E}(\pi/3) 
     - \mathcal{S}_{\sf E} (\pi/6)}{N} 
\end{eqnarray}
in powers of $J_{\pm \pm}$ and $J_{z \pm}$.  
We do this by writing the matrix ${\bf M}({\bf q})$ [Eq.~\ref{eq:M}] as
\begin{eqnarray}
{\bf M}({\bf q}) = {\bf M}_0({\bf q}) + \epsilon {\bf X}({\bf q})
\end{eqnarray}
where ${\bf M}_0({\bf q})$ is the matrix associated with the high-symmetry point,
and ${\bf X}({\bf q})$ that associated with the perturbation, and noting that 
\begin{eqnarray}
 &&\ln( \det( {\bf M}_0 + \epsilon {\bf X})) 
   =  \ln( \det({\bf M}_0) ) \nonumber \\
&&
+ \sum_{n=1}^{\infty}
 (-1)^{(n+1)} \frac{\epsilon^n}{n}  
 \Tr \bigg[ \left(  {\bf X} \cdot {\bf M}_0^{-1} \right)^n \bigg].  
 \end{eqnarray}
We then expand in powers of $J_{\pm\pm}$ and $J_{z\pm}$.


We find that the leading correction to $\Delta s_{\pi/3}$ is
\begin{eqnarray}
\Delta s_{\pi/3}
    \approx a \left( \frac{J_{\pm\pm}}{J_{\pm}} \right)^3
\end{eqnarray}
where $a=0.0045$.
It follows that, for sufficiently small $J_3$, the phase boundary between 
$\Psi_2$ and $\Psi_3$ should tend to the line $J_{\pm\pm} = 0$, with the 
$\Psi_2$ phase favoured for $J_{\pm\pm}>0$ and $\Psi_3$ favoured 
for $J_{\pm\pm}<0$.
Numerical evaluation of Eq.~(\ref{eq:S}), in the limit $J_3 \to 0$, 
yields results in agreement with these arguments [cf.~Fig.~ \ref{fig:classical-phase-diagram}].


On the line $J_{\pm\pm} = 0$ itself, we find that the leading correction 
to the difference in entropy is 
\begin{eqnarray}
\Delta s_{\pi/3} \approx b \left( \frac{J_{z\pm}}{J_{\pm}} \right)^6
\end{eqnarray}
with $b=-5.3 \times 10^{-5}$. 
Hence the $\Psi_3$ state is weakly preferred, and the phase boundary will bend 
towards positive $J_2/|J_3|$, as observed in Fig.~\ref{fig:classical-phase-diagram} of the main text.
Since $J_{z\pm}$ is a term which drives ferromagnetic out of plane fluctuations, a negative sign 
for $b$ is consistent with the argument that $\Psi_3$ is better connected to the 
ferromagnetic phase, and hence has a softer spectrum for ferromagnetic out-of-plane fluctuations. 
On the other hand, symmetries of the $\Psi_{2}$ states have been shown to allow for small antiferromagnetic out-of-plane ordering, of the ``all in, all out'' type~\cite{javanparast15}.


In the limit $|J_3| \gtrsim (|J_1|, |J_2|)$, numerical evaluation of Eq.~(\ref{eq:S}) 
yields the more complex, reentrant behaviour, as seen in Fig.~\ref{fig:classical-phase-diagram}. 
This behaviour occurs over a very narrow region of parameter space, and is discussed 
in detail in Ref. [\onlinecite{wong13}] for the case of quantum, as opposed to thermal, 
order by disorder.

\subsection{Quantum spin-wave theory}
\label{subsection:quantum-spin-wave}

Quantum spin-wave theories for ${\bf q}=0$, 4-sublattice classical ground states of 
$\mathcal{H}_{\sf ex}$~[Eq.~(\ref{eq:Hex1})] have been discussed by a number 
of authors, with attention focused on comparison with inelastic neutron scattering
in applied magnetic field~\cite{ross11-PRX1,savary12-PRL109}, and the 
way in which quantum fluctuations select between the one--dimensional
manifold of states with ${\sf E}$ symmetry~\cite{savary12-PRL109,zhitomirsky12,wong13}.
To date, all calculations have been carried out in the linear spin-wave approximation
\begin{eqnarray}
{\mathcal H}_{\sf ex} 
    &\approx& {\mathcal E}_0 \left(1 + \frac{1}{S} \right)
            \nonumber \\
      && \;  + \sum_{{\bf q}} 
             \sum_{\nu=0}^{3} \omega_{\nu}({\bf q})
             \left(
                  b_{\nu}^{\dagger}({\bf q})
                  b^{\phantom \dagger}_{\nu}({\bf q}) +\frac{1}{2} 
             \right) 
             + \ldots  
             \nonumber \\
\label{eq:Hlsw}
\end{eqnarray}
where ${\mathcal E}_0$ is the classical ground state energy defined in Eq.~(\ref{eq:E0}), 
$\omega_{\nu}({\bf q})$ the spin-wave dispersion, and $b^{\phantom \dagger}_{\nu}$ 
a set of non-interacting bosons describing spin-wave modes with band index $\nu=0,1,2,3$.   


\begin{figure*}
\centering
\includegraphics[width=1.8\columnwidth]{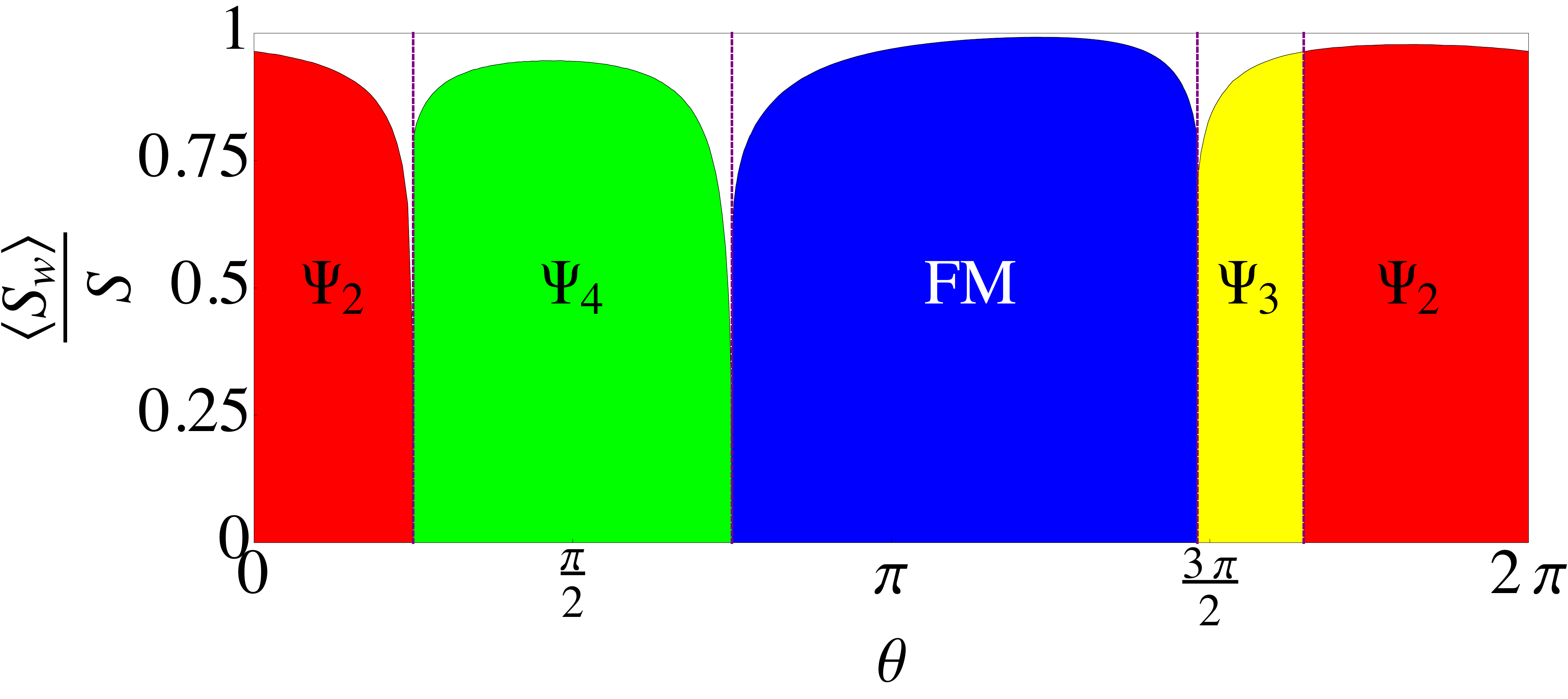}
\caption{
Fraction of the classical moment achieved in ordered phases
of a pyrochlore magnet with anisotropic exchange interactions, 
within linear spin-wave theory.
Away from the phase boundaries, the ordered moment is close to 
its full classical value.
All results are obtained within a linear spin--wave 
analysis of $\mathcal{H}_{\sf ex}$~[Eq.~(\ref{eq:Hex1})] 
--- cf. Appendix~\ref{appendix:linear-spin-wave} --- with $J_1 / |J_3| = 3 \cos\theta$, 
$J_2 / |J_3| = 3 \cos\theta$, $J_3 < 0$, $J_4 \equiv 0$, corresponding to the white circle in 
Fig.~\ref{fig:classical-phase-diagram} 
and Fig.~\ref{fig:quantum-phase-diagram}.
}
\label{fig:ordered-moment}
\end{figure*}    
  

Since the anisotropic exchange model $\mathcal{H}_{\sf ex}$~[Eq.~(\ref{eq:Hex1})] 
does not, in general, posses any continuous symmetry,  spin-wave excitations 
about a  4-sublattice classical ground states will generally be gapped, and quantum
effects are small.
However the enlarged ground-state manifolds occurring where different 
symmetry ground states meet can lead to ``accidental'' degeneracies in 
the spin-wave spectrum, and large quantum fluctuations about the ordered
state.
The effect of these fluctuations on classical order may be estimated by 
calculating the correction to the ordered moment on sublattice $i$, $\langle S^{w}_i \rangle$
defined in Eq.~\ref{eq:HPsw}. Details of calculations are given in Appendix~\ref{appendix:linear-spin-wave}.
In all of the 4-sublattice phases described in this text, 
$\langle S^{w}_i \rangle$ is the same for all sublattices $i=0,1,2,3$.


In Fig.~\ref{fig:ordered-moment} we show the effect of 
quantum fluctuations on the classical, zero-temperature 
ground states of $\mathcal{H}_{\sf ex}$~[Eq.~(\ref{eq:Hex1})].    
For parameters which are deep within the ordered phases,
the ordered moment approaches
its full classical value.
However the enlarged ground-state degeneracies on classical
phase boundaries lead to additional zero-modes in the 
spin-wave spectrum, and correspondingly larger corrections.
Corrections to the ordered moment diverge logarithmically 
approaching the boundary with the Palmer--Chalker 
phase from $\Psi_2$, where there are entire planes of zero modes, 
and approaching the boundary with the ferromagnetic phases from 
the Palmer--Chalker phase.


In Fig.~\ref{fig:quantum-phase-diagram} [Section~\ref{section:introduction}], 
we show the quantum phase diagram of $\mathcal{H}_{\sf ex}$~[Eq.~(\ref{eq:Hex1})], 
within linear spin-wave theory.   
Regions where quantum fluctuations eliminate the ordered moment entirely, 
are shaded white.    
The effect is strongest where the degeneracy of the classical ground state 
is highest, i.e. approaching the Heisenberg line
$$
J_1/|J_3| = J_2/|J_3|  \to \infty \; ,
$$
and in the vicinity of the special point~\cite{benton16} 
$$
J_1/|J_3| \; ,  \; J_2/|J_3|  \to 0 \; .
$$
The absence of an ordered moment within linear spin-wave 
theory can indicate a region where conventional magnetic order breaks
down entirely, and typically underestimates the extent of any 
unconventional order~\cite{chandra88,schulz96,shannon04,shannon06}.   
It therefore seems reasonable to suggest that the vicinity of these phase
boundaries will be favorable places to find novel quantum ground states and quantum spin liquids.
We will return to this point in our discussion of Er$_2$Sn$_2$O$_7$, 
in Section~\ref{section:Er2Sn2O7}.   

\subsection{Cross section in neutron scattering}
\label{subsection:neutron-scattering}

Inelastic neutron scattering experiments measure the dynamical structure factor
\begin{eqnarray}
&&S(\mathbf{q}, \omega) = \nonumber \\
&& \sum_{\alpha, \beta=1}^{3} \sum_{i, j=0}^{3} 
   \left( \delta_{\alpha \beta}- \frac{q_{\alpha} {q_\beta}}{q^2} \right)
   \langle 
      m_{\alpha}^{i}(-{\bf q}, -\omega)  m_{\beta}^{j}({\bf q}, \omega) 
   \rangle \nonumber\\
   \label{eq:defSq-inelastic}
\end{eqnarray}
where the projection operator 
\begin{eqnarray}
\left( \delta_{\alpha \beta} - \frac{q_{\alpha} {q_\beta}}{q^2} \right) \nonumber
\end{eqnarray}
reflects the fact the neutron interacts with the components of the
spin transverse to the momentum transfer~${\bf q}$, and 
\begin{eqnarray}
&& m^{i}_{\alpha}({\bf q}) \nonumber \\
&&   =\frac{1}{\sqrt{2\pi}} \sqrt{\frac{4}{N}} \sum_{\beta=1}^3
         g_i^{\alpha \beta}  \int dt
         \left( 
            \sum_{{\bf R}_i} e^{-i\omega t}e^{i {\bf q}.\bf{R}_i} S_i^{\beta}({\bf R}_i, t) 
         \right) \nonumber \\
\end{eqnarray}
is the Fourier transform of the magnetic moment associated with the rare-earth ions,
for a given sublattice $i=0,1,2,3$.   
The associated g-tensor $g_i^{\alpha \beta}$ is defined in 
Appendix \ref{appendix:local-coordinate-frame}.


The equal-time structure factor measured in energy-integrated, 
quasi-elastic neutron scattering, is given by
\begin{eqnarray}
&&S(\mathbf{q}, t=0)=\frac{1}{\sqrt{2\pi}}\int_{-\infty}^{\infty} d{\omega} \ S(\mathbf{q}, \omega)
\nonumber \\
&&=
 \sum_{\alpha, \beta=1}^{3} \sum_{i, j=0}^{3} 
   \left( \delta_{\alpha \beta}- \frac{q_{\alpha} {q_\beta}}{q^2} \right) \nonumber \\
&&\qquad \qquad \times  \langle 
      m_{\alpha}^{i}(-{\bf q},t=0)  m_{\beta}^{j}({\bf q}, t=0) 
   \rangle.
   \label{eq:defSq}
\end{eqnarray}
For many purposes it is also convenient to resolve the equal-time 
structure factor $S({\bf q}) = S({\bf q}, t=0)$ into spin-flip ({\sf SF}) and non-spin flip ({\sf NSF}) components, 
or comparison with experiments carried out using polarised neutrons.
For neutrons with polarisation along a direction \mbox{${\bf \hat{n}} \perp {\bf q}$}, 
these are given by 
\begin{eqnarray}
S^{\sf NSF}({\bf q})
   &=& \sum_{i, j=0}^{3} 
          \langle 
               ( {\bf m}^{i}(-{\bf q}) \cdot {\bf \hat{n}} )  
               ({\bf m}^{j}({\bf q})  \cdot {\bf \hat{n}} )
          \rangle \nonumber\\
          \label{eq:sqnsf}\\
S^{\sf SF}({\bf q})
   &=& \sum_{i, j=0}^{3} 
           \frac{1}{q^2}
                    \langle 
                          ( {\bf m}^{i}(-{\bf q} ) 
                          \cdot 
                          \left( 
                               {\bf \hat{n}} \times {\bf q}) 
                           \right) 
                   \nonumber \\
        &&   \qquad  \qquad \qquad \times 
                    \left( 
                        {\bf m}^{j}({\bf q})  \cdot ({\bf \hat{n}} \times {\bf q})
                   \right)
                   \rangle
                   \nonumber \\
                    \label{eq:sqsf}
\end{eqnarray}
In this article, where we quote results for {\sf SF} and {\sf NSF} components of $S({\bf q})$, 
we follow the conventions of Fennell {\it et al.}~[\onlinecite{fennell09}] and 
consider \mbox{${\bf \hat{n}}=(1, -1, 0)/\sqrt{2}$}.


Connection to theory is made by using simulation, or spin-wave theory to 
evaluate the correlations of the magnetic moments $m_{\alpha}^{i}({\bf q}, t)$.   
Equal-time correlations $S({\bf q})$ [Eqs. (\ref{eq:defSq}-\ref{eq:sqsf})] can be calculated  
directly using classical Monte Carlo simulation described in Section~\ref{section:finite-temperature}, or using the classical spin wave theory described in Section~\ref{section:classical-spin-wave}.
The quantum spin-wave theory described in Section~\ref{subsection:quantum-spin-wave} and 
Appendix~\ref{appendix:linear-spin-wave}, gives access to the full dynamical 
structure factor $S({\bf q},\omega)$ [Eq.~(\ref{eq:defSq-inelastic})], as measured by 
inelastic neutron scattering.


In the case of the classical spin-wave theory discussed in Section~\ref{section:classical-spin-wave}, 
equal time correlations $\langle m_{\alpha}^{i}(-{\bf q})  m_{\beta}^{j}({\bf q}) \rangle$ 
are expressed in terms of the correlations of the spin wave modes $\upsilon_{\nu} (\mathbf{q})$
using Eqs. (\ref{eq:fluc}) and (\ref{eq:orthogonaltransformation}).
The required correlation functions $\langle \upsilon_{\nu {\bf q}} \upsilon_{\lambda -{\bf q}} \rangle $
can then be obtained directly from Eq.~(\ref{eq:partitionfunction}) 
\begin{eqnarray}
\langle \upsilon_{\nu {\bf q}} \upsilon_{\lambda -{\bf q}} \rangle 
    = \delta_{\nu \lambda} \frac{T}{\kappa_{\nu {\bf q}}}.
\end{eqnarray}


In Figs.~\ref{fig:SqYTO},~\ref{fig:ETOsq},~\ref{fig:J1=0sq}, explicit comparison is made 
between the equal-time structure factor $S({\bf q})$ calculated within 
classical Monte Carlo simulation, and from a low-temperature classical spin wave theory.
We find excellent, quantitative agreement between the two approaches. 
This confirms classical spin-wave theory as a useful link between the exact analytical 
zero-temperature theory developed in Sections~\ref{section:model} 
and~\ref{section:classical-ground-states} and the finite-temperature 
simulations presented in Section~\ref{section:finite-temperature}.\\


In the case of the quantum spin-wave theory described in Section~\ref{subsection:quantum-spin-wave} 
and Appendix~\ref{appendix:linear-spin-wave}, it is necessary to re-express the spin correlation functions
$
\langle S_i^{\alpha} S_j^{\beta} \rangle
$
in terms of the spin-wave operators $b^{\dagger}_{\nu}, b_{\nu}$ [cf. Eq.~(\ref{eq:Hlsw})]
This can be accomplished using the Bogoliubov
transformations described in Appendix~\ref{appendix:linear-spin-wave}.
Within linear spin-wave theory, the correlations of $b_{\nu}$ take on a simple form
\begin{eqnarray}
&&\langle b^{\dagger}_{\nu}(\mathbf{q}, \omega) b^{\dagger}_{\nu'}(-\mathbf{q}, -\omega) \rangle=
\langle b_{\nu}(\mathbf{q}, \omega) b_{\nu'}(-\mathbf{q}, -\omega) \rangle=0 \nonumber \\
\\
&&\langle b_{\nu}(\mathbf{q}, \omega) b_{\nu'}^{\dagger}(\mathbf{q}, \omega) \rangle
\nonumber \\
&& \qquad \qquad=
\delta_{\nu \nu'} \delta(\omega-\omega_{\nu}(\mathbf{q}))
   +  \langle b_{\nu}^{\dagger}(\mathbf{q}, \omega) b_{\nu'}(\mathbf{q}, \omega) \rangle \nonumber \\
&&\qquad \qquad =\delta_{\nu \nu'} \delta(\omega-\omega_{\nu}(\mathbf{q})) (1+n_{B}(\omega_{\nu}(\mathbf{q})))
\end{eqnarray}
where $\delta(\omega-\omega_{\nu}(\mathbf{q}))$ is the Dirac delta function enforcing 
conservation of energy and $n_{B}(\omega)$ is the Bose-Einstein distribution
\begin{eqnarray}
n_B(\omega)=\frac{1}{\exp\left( \frac{\omega}{T}\right)-1} \; .
\end{eqnarray}


\section{Finite-temperature properties}
\label{section:finite-temperature} 

The symmetry analysis and ``Lego--brick'' rules described in Section~\ref{section:model}, 
analysis of ground-state energy described in Section~\ref{section:classical-ground-states},  
and spin-wave theory described in Section~\ref{section:spin-wave-theory}, 
together make it possible to determine the ordered ground-states 
of the anisotropic exchange model $\mathcal{H}_{\sf ex}$~[Eq.~(\ref{eq:Hex1})], in the limit $T\to 0$.  
The resulting classical ground-state phase diagram is shown in  
phase diagram Fig.~\ref{fig:classical-phase-diagram}.
%
%
However the interesting and unusual properties of rare-earth pyrochlore oxides
all come from experiments carried out at finite temperature, and often relate to 
paramagnetic, rather than ordered phases.
We have therefore used classical Monte Carlo simulations to explore
the physics of $\mathcal{H}_{\sf ex}$ at finite temperature.
The main conclusions of these simulations are summarized in the finite-temperature 
phase diagram Fig.~\ref{fig:finite-temperature-phase-diagram}.


We note that, although the systems we wish to describe are fundamentally
quantum in nature, the classical simulations can be expected to give a qualitatively 
correct description of the physics in the high temperature, paramagnetic phase,
where thermal fluctuations destroy quantum coherence.
The classical description should also work, at least in some
respects, at low temperature as long as the ground state
of the quantum system is in a classically ordered phase.
Ultimately, the best justification for our approach will come
{\it a posteriori} from agreement with neutron scattering experiments [cf. Fig. {\ref{fig:Sq}}].


In what follows we document the nature of phase transitions from the paramagnet 
into each of the ordered phases [Sections~\ref{subsection:PM-to-PC}---\ref{subsection:PM-to-Psi3}], 
and explore how the enlarged ground-state degeneracies at classical phase boundaries 
manifest themselves at finite temperature [Section~\ref{section:living-on-the-edge}].
Technical details of simulations are given in Appendix~\ref{appendix:classical-MC}.  


\begin{figure}[t]
\centering\includegraphics[width=0.93\columnwidth]{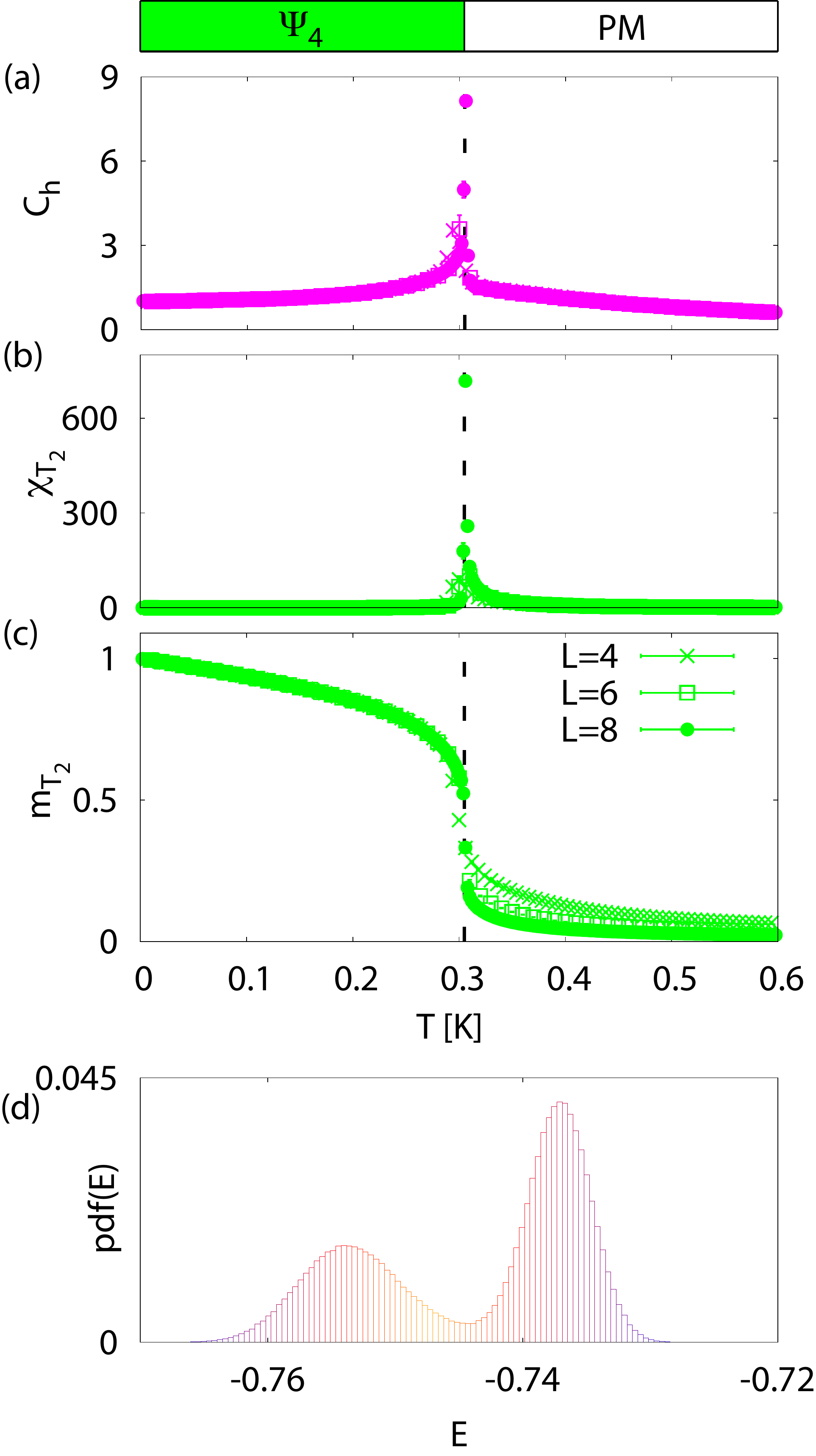}
\caption{Finite-temperature phase transition from the paramagnet into the 
Palmer--Chalker phase [$\Psi_4$], as determined by classical Monte Carlo 
simulation of $\mathcal{H}_{\sf ex}$~[Eq.~(\ref{eq:Hex1})], for parameters 
\mbox{$J_1 = 0$\ meV}, 
\mbox{$J_2 = 0.3$\ meV},
\mbox{$J_3 = -0.1$\ meV},
\mbox{$J_4 = 0$\ meV}. 
a) Temperature dependence of the specific heat $c_h(T)$.
b) Temperature dependence of the order-parameter susceptibility, $\chi_{\sf T_2}(T)$.
c) Temperature dependence of the order parameter, $|{\bf m}_{\sf T_2}(T)|$.    
d) Probability distribution of the energy $E$ evaluated at the transition temperature for a cluster of size $L=12$.
The black dashed line in (a)-(c) indicates a first-order phase transition  
at $T_{\sf T_2} = 305 \pm 5$ mK. 
Simulations were performed for clusters of $N=16L^3$ spins, 
with $L=4,6,8,12$.  
\label{fig:PMtoPC}}
\end{figure}

\subsection{Finite-temperature transition from the paramagnet into the Palmer--Chalker phase}
\label{subsection:PM-to-PC}

The most revealing feature of any broken-symmetry state is usually 
its finite-temperature phase transition.
In Fig.~\ref{fig:PMtoPC} we show simulation results for the finite-temperature
phase transition from the paramagnet into Palmer--Chalker phase with 
${\sf T_2}$ symmetry.   
Simulations were carried out for parameters 
$$ (J_1,\  J_2,\ J_3,\ J_4) = (0, 0.3, -0.1,  0)\quad \text{meV} $$ 
deep within the Palmer--Chalker phase.
Clear evidence for a phase transition can be found in the anomalies in both the 
specific heat $c_h(T)$ [Fig.~\ref{fig:PMtoPC}(a)], and the order-parameter 
susceptibility $\chi_{\sf T_2}(T)$ [Fig.~\ref{fig:PMtoPC}(b)] at \mbox{$T_{\sf T_2} = 305  \pm 5$\ mK}.


Symmetry permits a continuous phase transition between the paramagnet
and the Palmer--Chalker phase.
However for this parameter set, fluctuations drive the transition first order, 
as is evident from the discontinuity in the value of the order parameter 
${\bf m}_{\sf T_2}$ for $T=T_{\sf T_2}$ [Fig.~\ref{fig:PMtoPC}(c)],
and the double peak in the probability distribution for the energy [Fig.~\ref{fig:PMtoPC}(d)].

\subsection{Transition from the paramagnet into the non-colinear ferromagnetic phase}
\label{subsection:PM-to-FM}

\begin{figure}
\centering\includegraphics[width=0.93\columnwidth]{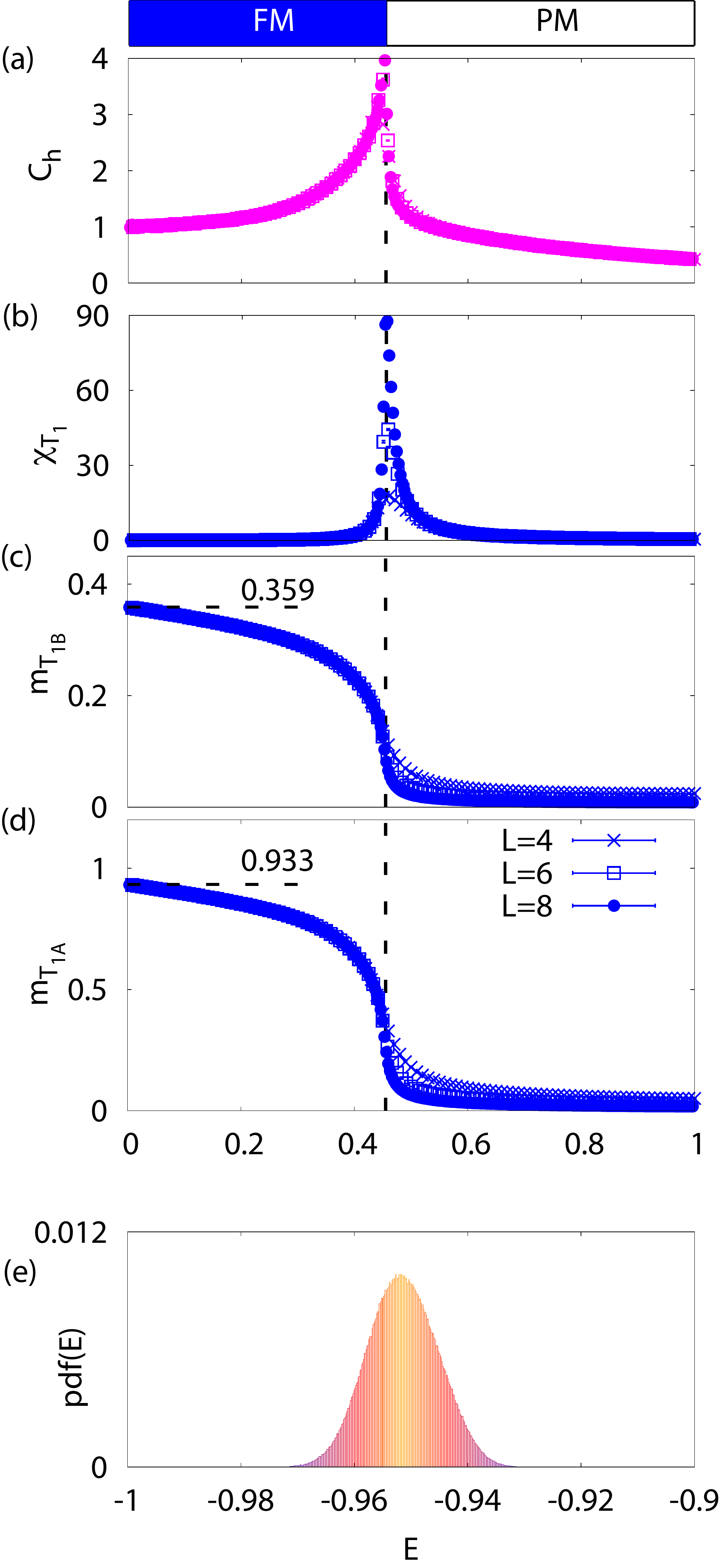}
\caption{
Finite-temperature phase transition from the paramagnet into the 
non-collinear ferromagnet (FM), as determined by classical Monte Carlo 
simulation of $\mathcal{H}_{\sf ex}$~[Eq.~(\ref{eq:Hex1})], for parameters 
appropriate to Yb$_2$Ti$_2$O$_7$ \cite{ross11-PRX1}, i.e.
\mbox{$J_1 = -0.09$\ meV}, 
\mbox{$J_2 = -0.22$\ meV},
\mbox{$J_3 = -0.29$\ meV}
setting
\mbox{$J_4 = 0$\ meV}. 
a)  Temperature dependence of the specific heat $c_h(T)$.
b) Temperature dependence of the order-parameter susceptibility, $\chi_{\sf T_1}(T)$.
c) Temperature dependence of the order parameter, $|{\bf m}_{\sf T_{1, B}}(T)|$.
d) Temperature dependence of the order parameter, $|{\bf m}_{\sf T_{1, A}}(T)|$.
e) Probability distribution of the energy $E$ evaluated at the transition 
    for a system of size $L=12$.
The black dashed line in (a)--(d) indicates a phase transition 
at \mbox{$T_{\sf T_1} = 455 \pm 5\ \text{mK}$}.  
Simulations were performed for clusters of $N=16L^3$ spins, 
with $L=4,6,8,12$. 
}
\label{fig:PMtoFM}
\end{figure}

In Fig.~\ref{fig:PMtoFM} we show simulation results for the finite-temperature
phase transition from the paramagnet into the non-colinear ferromagnet, 
for parameters appropriate to Yb$_2$Ti$_2$O$_7$ [\onlinecite{ross11-PRX1}], 
setting $J_4 =0$  
$$ (J_1,\  J_2,\ J_3,\ J_4) = (-0.09, -0.22, -0.29,  0)\quad \text{meV} $$ 
Anomalies in both the specific heat $c_h(T)$ [Fig.~\ref{fig:PMtoFM}(a)] 
and order-parameter susceptibility $\chi_{\sf T_1}(T)$ [Fig.~\ref{fig:PMtoFM}(b)]
at \mbox{$T_{\sf T_1} = 455  \pm 5\ \text{mK}$}, provide clear evidence of a phase transition.


At low temperatures, the temperature--dependence of the order parameters
${\bf m}_{\sf T_{1,A}}$ and ${\bf m}_{\sf T_{1,B}}$ [Fig.~\ref{fig:PMtoFM}(c)--(d)] 
converges on the values expected from the zero--temperature analysis of Section~\ref{ferromagnet},
and with a slope predicted by a low-temperature expansion about the FM ground state (not shown).


The single peak in the probability distribution for the energy [Fig.~\ref{fig:PMtoFM}(e)]
suggests that for these parameters, the thermal phase transition from paramagnet to 
non-collinear FM in a classical model is probably continuous and at most very weakly first order.


\subsection{Transition from the paramagnet into the $\Psi_2$ phase}
\label{subsection:PM-to-Psi2}

\begin{figure}[h!]
\centering\includegraphics[width=0.93\columnwidth]{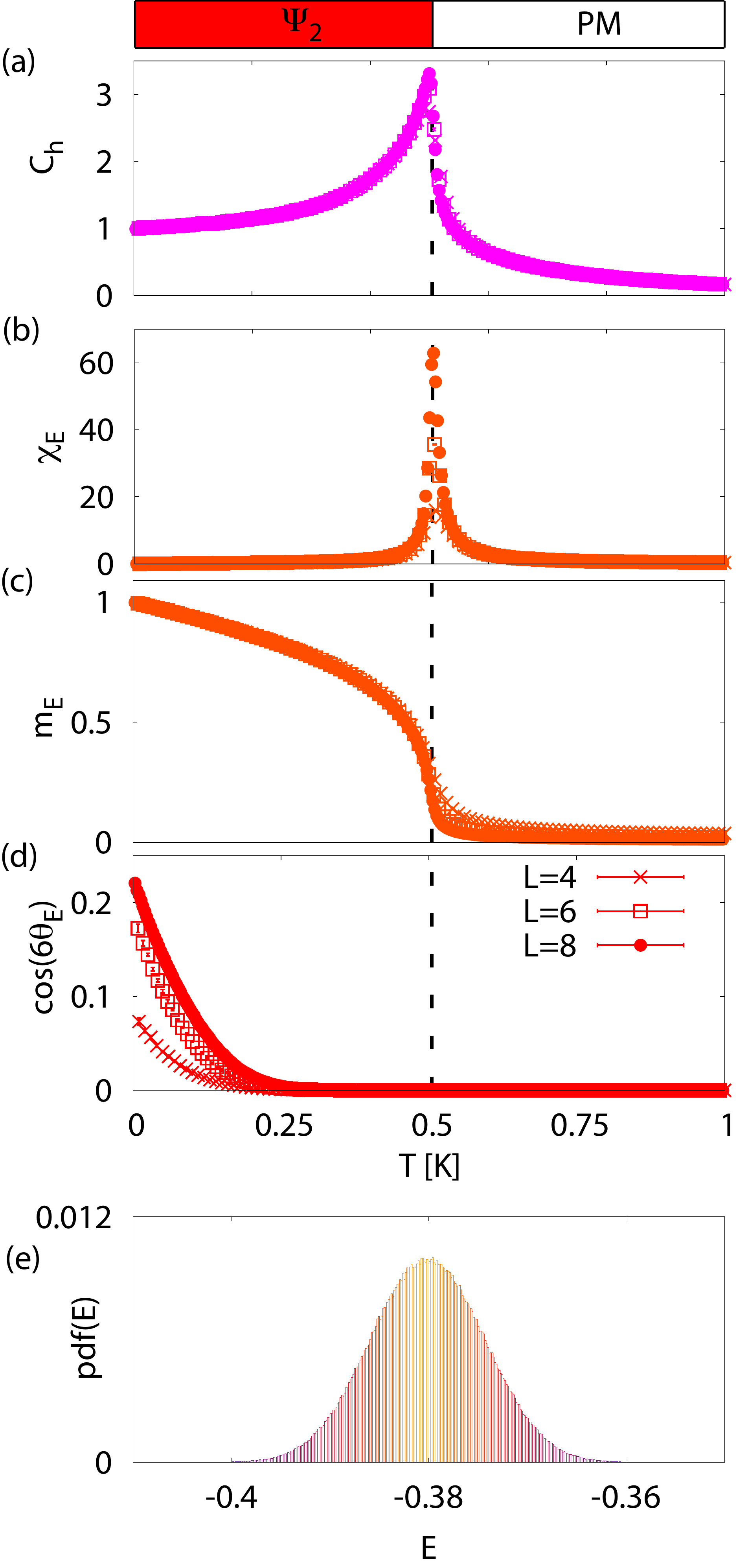}
\caption{
Finite-temperature phase transition from the paramagnet into the non-coplanar 
antiferromagnet $\Psi_2$, as determined by classical Monte Carlo 
simulation of $\mathcal{H}_{\sf ex}$~[Eq.~(\ref{eq:Hex1})], for parameters 
appropriate to Er$_2$Ti$_2$O$_7$ \cite{savary12-PRL109}, i.e.
\mbox{$J_1 = 0.11$\ meV}, 
\mbox{$J_2 = -0.06$\ meV},
\mbox{$J_3 = -0.1$\ meV}
setting
\mbox{$J_4 = 0$\ meV}. 
a) Temperature dependence of the specific heat $c_h(T)$.
b) Temperature dependence of the order-parameter susceptibility, $\chi_{\sf E}(T)$.
c) Temperature dependence of the order parameter, $|{\bf m}_{\sf E}(T)|$.
d) Temperature dependence of the secondary order parameter, $\cos 6\theta_{\sf E}$.
e) Probability distribution of the energy $E$ evaluated at the transition 
    temperature 
    for a system of size $L=12$.
The black dashed line indicates a continuous phase transition 
at \mbox{$T_N=505 \pm 5$\ mK}.  
Simulations were performed for clusters of $N=16L^3$ spins, 
with $L=4,6,8,12$. 
}
\label{fig:PMtoPsi2}
\end{figure}

In Fig.~\ref{fig:PMtoPsi2} we show simulation results for the finite-temperature
phase transition from the paramagnet into the $\Psi_2$ phase, 
for parameters appropriate to Er$_2$Ti$_2$O$_7$~[\onlinecite{savary12-PRL109}], 
setting $J_4 =0$  
$$ (J_1,\  J_2,\ J_3,\ J_4) = (0.11, -0.06, -0.1,  0)\quad \text{meV} $$ 
This shows a number of interesting features.


Anomalies in both the specific heat $c_h(T)$ [Fig.~\ref{fig:PMtoPsi2}(a)] 
and order-parameter susceptibility $\chi_{\sf E}(T)$ [Fig.~\ref{fig:PMtoPsi2}(b)]
 at \mbox{$T_{\sf E} = 505  \pm 5\ \text{mK}$} offer clear evidence of a phase transition. 


Both the smooth evolution of the primary order parameter, 
${\bf m}_{\sf E}$~[Fig.~\ref{fig:PMtoPsi2}(c)], and the 
single peak in the probability distribution for the energy [Fig.~\ref{fig:PMtoPsi2}(e)]
suggests that the phase transition seen in simulation is at most weakly first-order.
For the clusters simulated, we confirm that it is possible to obtain a fairly good collapse of 
data for $\chi_{\sf E}(T)$ [Fig.~\ref{fig:PMtoPsi2}(b)] using 3D XY exponents~\cite{zhitomirsky14}.


However there are only a discrete number of $\Psi_2$ ground states, and 
a finite value of  $|{\bf m}_{\sf E}|$ alone does not imply $\Psi_2$ order.
Evidence for the $\Psi_2$ ground state comes from the secondary 
order parameter $c_{\sf E} = \cos 6 \theta_{\sf E} > 0$ [Fig.~\ref{fig:PMtoPsi2}(d)].
Here simulation results are strongly size-dependent, but suggest a slow 
crossover into the $\Psi_2$ state, occurring at a $T^* \ll T_{\sf E}$, without any 
accompanying feature in $c_h(T)$ [Fig.~\ref{fig:PMtoPsi2}(a)].


On the basis of  Landau theory 
we anticipate that {\it any} finite value of $m_{\sf E}  = |{\bf m}_{\sf E}|$ will induce 
degeneracy breaking in $\theta_{\sf E}$, and that the entropic selection within the
${\sf E}$ manifold should 
therefore occur concurrently with the onset of magnetic order.
Depending on the sign of the relevant coupling, 
\begin{eqnarray}
\delta {\mathcal F}_{\sf E} = \frac{1}{6}\ d\  m_{\sf E}^6\ \cos 6 \theta_{\sf E}
\end{eqnarray}
the system will then enter either a $\Psi_2$ or a $\Psi_3$ ground state.


However, the free-energy barrier separating the $\Psi_2$ and $\Psi_3$ 
ground states is very small, and this in turn sets a very large  
length-scale for the selection of the $\Psi_2$ ground state.
Based on the low-temperature expansion $\mathcal{F}_{\sf ex}^{\sf low-T}$ 
[Eq.~(\ref{eq:FlowT})], we estimate that clusters of $N \sim 10^9$ sites
should be able to clearly resolve which state is selected at the magnetic ordering
temperature.


\begin{figure}[h!]
\centering\includegraphics[width=0.95\columnwidth]{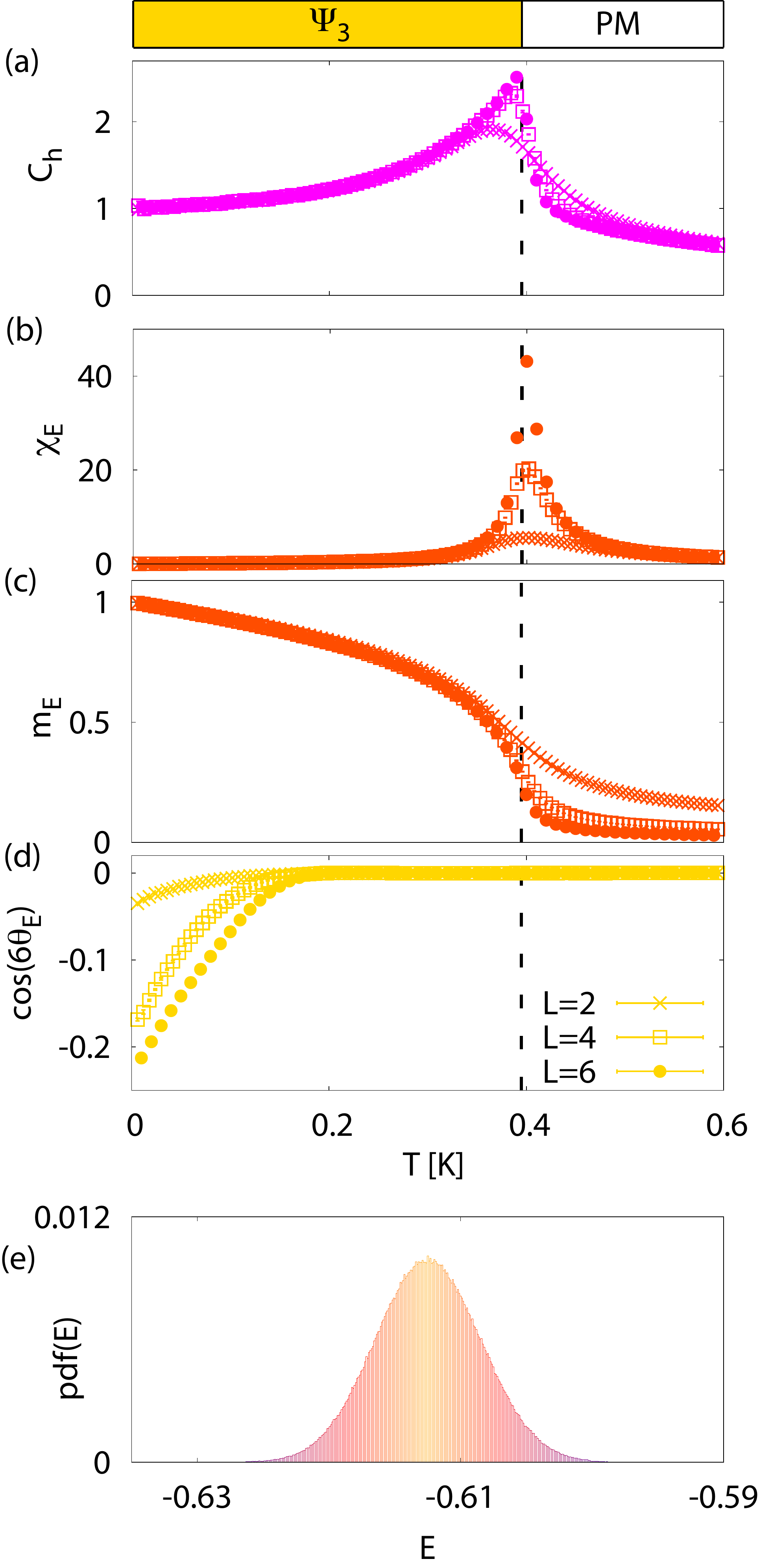}
\caption{
Finite-temperature phase transition from the paramagnet into the coplanar 
antiferromagnet $\Psi_3$, as determined by classical Monte Carlo 
simulation of $\mathcal{H}_{\sf ex}$~[Eq.~(\ref{eq:Hex1})], for parameters 
\mbox{$J_1 = 0$\ meV}, 
\mbox{$J_2 = -0.3$\ meV},
\mbox{$J_3 = -0.1$\ meV},
\mbox{$J_4 = 0$\ meV}. 
a) Temperature dependence of the specific heat $c_h(T)$.
b) Temperature dependence of the order-parameter susceptibility, $\chi_{\sf E}(T)$.
c) Temperature dependence of the order parameter, $|{\bf m}_{\sf E}(T)|$.
d) Temperature dependence of the secondary order parameter, $\cos 6\theta_{\sf E}$.
e) Probability distribution of the energy $E$ evaluated at the transition 
    temperature for a system of size $L=12$.
The black dashed line indicates a continuous phase transition 
at \mbox{$T_{\sf E}=395 \pm 5$\ mK}.  
Simulations were performed for clusters of $N=16L^3$ spins, 
with $L=2,4,6,12$. 
}
\label{fig:PMtoPsi3}
\end{figure}


\begin{figure*}
\includegraphics[width=\textwidth]{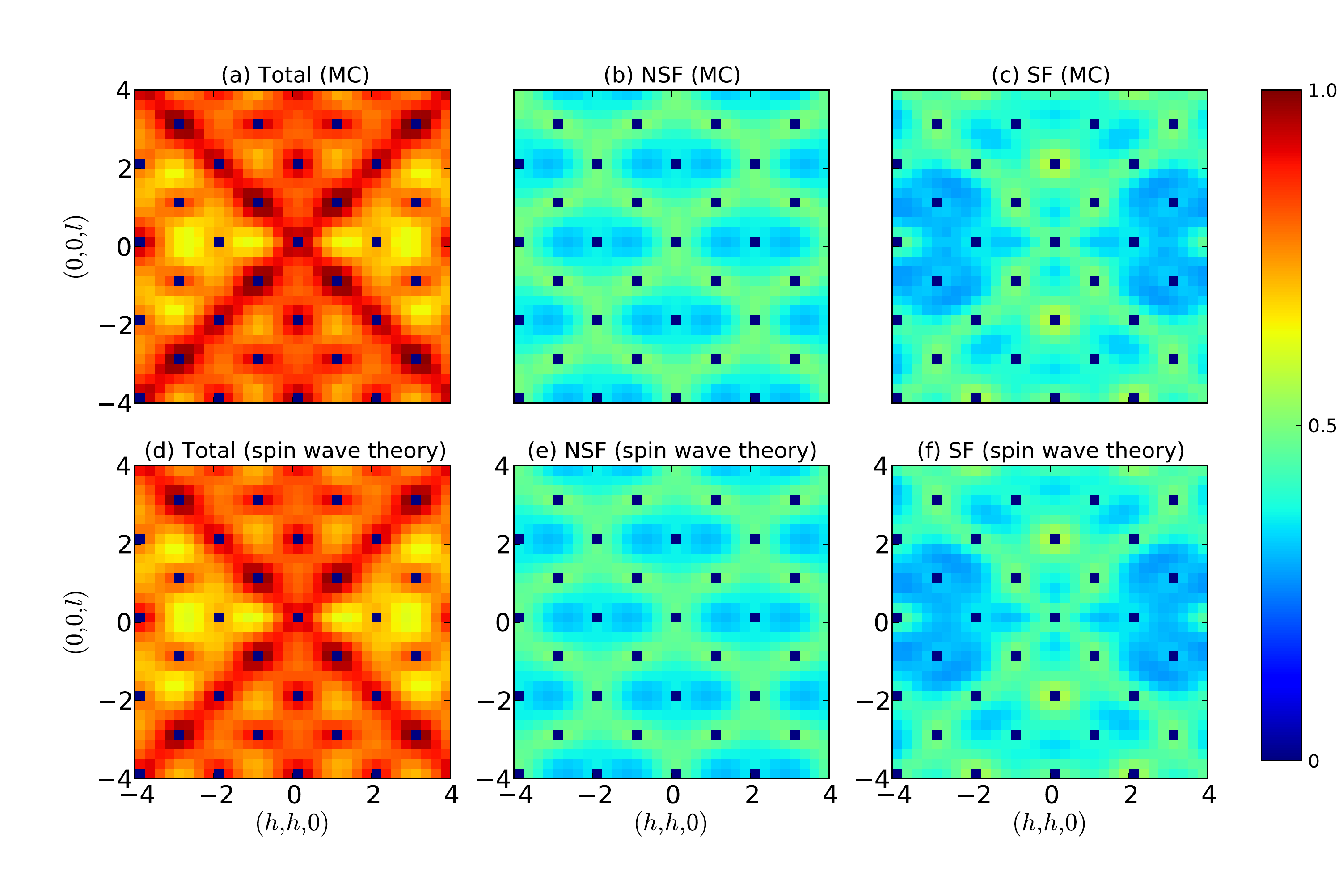}
\caption
{Comparison between results for equal-time structure factor $S({\bf q})$ obtained
in classical Monte Carlo (MC) simulation and classical low-temperature expansion
(spin wave theory) for parameters appropriate to Yb$_2$Ti$_2$O$_7$ \cite{ross11-PRX1}.   
a) Total scattering in the $(h, h, l)$ plane within MC simulation.
b) Associated scattering in the non spin-flip ({\sf NSF}) channel.
c) Associated scattering in the spin-flip ({\sf SF}) channel.
d) Total scattering in the $(h, h, l)$ plane within a spin-wave expansion
about the ferromagnetic ground state.
e) Associated scattering in the {\sf NSF} channel.
f) Associated scattering in the {\sf SF} channel.
Rods of scattering in the $(h, h, h)$ direction, associated with a low-energy 
spin-wave excitation [cf. Fig.~\ref{fig:dispersionsYTO}], are visible in both {\sf SF} and {\sf NSF} channels.
All results were obtained at $T=0.05$\ K, for exchange parameters 
\mbox{$J_1=-0.09 \text{meV}$},
\mbox{$J_2=-0.22 \text{meV}$}, 
\mbox{$J_3=-0.29 \text{meV}$}, 
setting \mbox{$J_4=0$}.
{\sf SF} and {\sf NSF} channels are defined with respect to a neutron with polarisation 
in the $(1, -1, 0)$ direction, as in Ref. \onlinecite{fennell09}. 
$S({\bf q})$ has been calculated using the experimentally measured g-tensor 
for Yb$_2$Ti$_2$O$_7$ \cite{hodges01, ross11-PRX1}, with $g_z=1.77, g_{xy}=4.18$.
In order to avoid saturating the colour scale, the intensity associated 
with Bragg peaks at reciprocal lattice vectors has been subtracted.
\label{fig:SqYTO}}
\end{figure*}

\subsection{Transition from the paramagnet into the $\Psi_3$ phase}
\label{subsection:PM-to-Psi3}

In Fig.~\ref{fig:PMtoPsi3} we show simulation results for the finite-temperature
phase transition from the paramagnet into the $\Psi_3$ phase, 
for parameters 
$$ (J_1,\  J_2,\ J_3,\ J_4) = (0, -0.3, -0.1,  0) \quad \text{meV} $$ 
close to the border with the non-collinear ferromagnet. 
Anomalies in both the specific heat $c_h(T)$ [Fig.~\ref{fig:PMtoPsi3}(a)] 
and order-parameter susceptibility $\chi_{\sf E}(T)$ [Fig.~\ref{fig:PMtoPsi3}(b)]
at \mbox{$T_{\sf E} = 395  \pm 5 \text{mK}$} offer clear evidence of a phase transition. 
Both the smooth evolution of the primary order parameter, 
${\bf m}_{\sf E}$~[Fig.~\ref{fig:PMtoPsi3}(c)], and the 
single peak in the probability distribution for the energy [Fig.~\ref{fig:PMtoPsi3}(e)]
suggest that this phase transition is continuous.


Evidence for the $\Psi_3$ ground state comes from the finite value of the secondary 
order parameter \mbox{$c_{\sf E} = \cos 6 \theta_{\sf E} < 0$} [Fig.~\ref{fig:PMtoPsi3}(d)].
This secondary order parameter shows only a slow onset, consistent with a crossover
into the $\Psi_3$ state, and is {\it very} strongly size-dependent.
As with the $\Psi_2$ state considered above, we infer that, with increasing system 
size, the temperature associated with this crossover scales towards $T = T_N$, 
and that in the thermodynamic limit, a single phase transition takes place from 
the paramagnet into the $\Psi_3$ state.


\begin{figure*}
\centering
\includegraphics[width=0.8\textwidth]{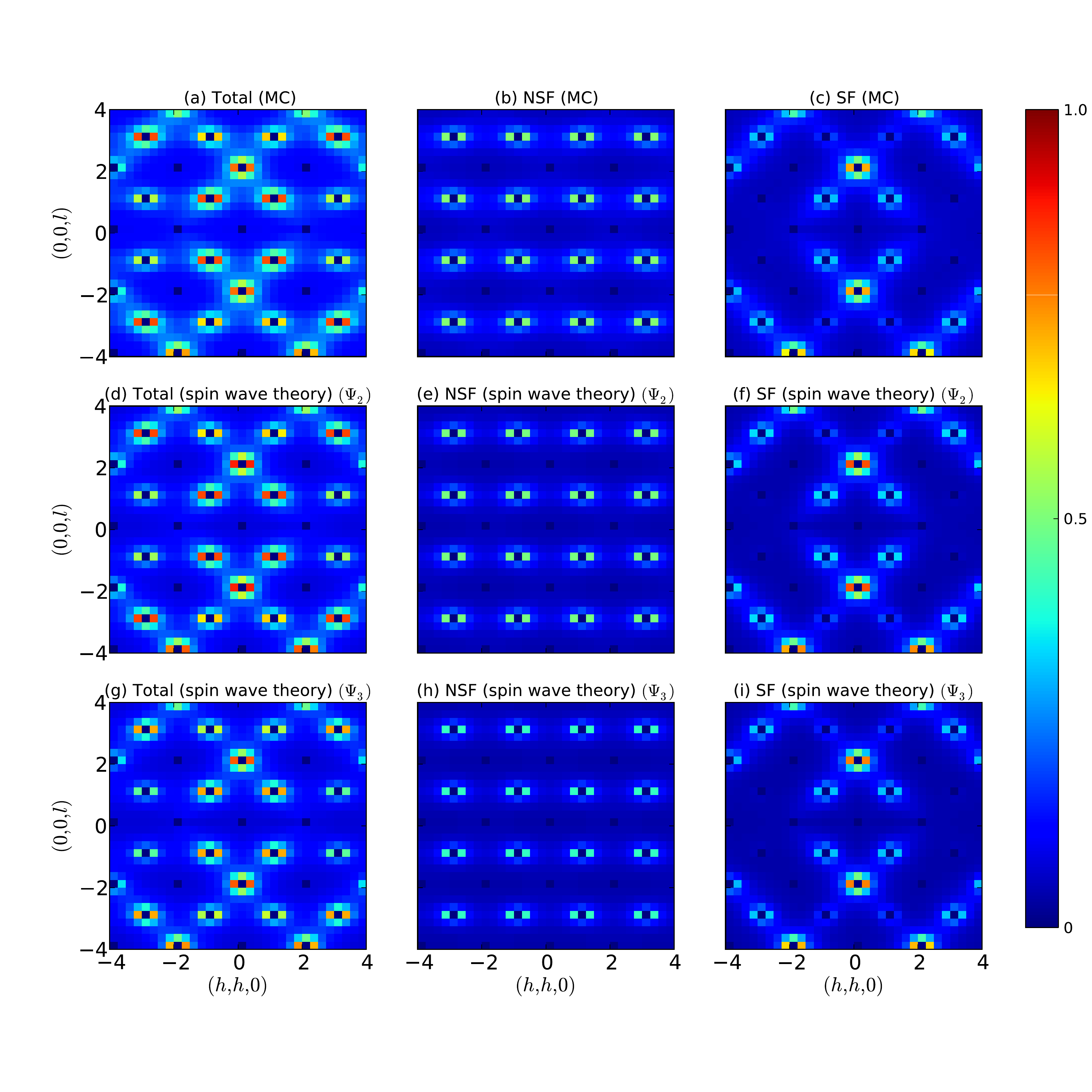}
\caption{
Comparison between results for equal-time structure factor $S({\bf q})$ obtained
in classical Monte Carlo (MC) simulation and low-temperature expansion
(classical spin wave theory) for parameters appropriate to Er$_2$Ti$_2$O$_7$.   
a) Total scattering in the $(h, h, l)$ plane within MC simulation.
b) Associated scattering in the non spin-flip ({\sf NSF}) channel.
c) Associated scattering in the spin-flip ({\sf SF}) channel.
d) Total scattering in the $(h, h, l)$ plane within a spin-wave expansion 
about a $\Psi_2$ ground state.
e) Associated scattering in the {\sf NSF} channel.
f) Associated scattering in the {\sf SF} channel.
g) Total scattering in the $(h, h, l)$ plane within a spin-wave expansion 
about a $\Psi_3$ ground state.
h) Associated scattering in the {\sf NSF} channel.
i) Associated scattering in the {\sf SF} channel.
Careful comparison of the distribution of scattering in the vicinity of the $(1, 1, 1)$, 
$(3, 3, 3)$ and $(1, 1, 3)$ reciprocal lattice vectors supports the conclusion 
that the $\Psi_2$ state is preferred for these exchange parameters, in agreement 
with experiment and the calculations described in the text.
All results were obtained at $T=0.36$\ K, for exchange parameters 
\mbox{$J_1=0.11\ \text{meV}$},
\mbox{$J_2=-0.06\ \text{meV}$},
\mbox{$J_3=-0.10\ \text{meV}$}, 
setting  \mbox{$J_4 \equiv 0$}
and
$g$-tensor parameters $g_z=2.45$ and $g_{xy}=5.97$
 [\onlinecite{savary12-PRL109}].
For clarity, intensity associated with Bragg peaks at reciprocal 
lattice vectors has been subtracted.
\label{fig:ETOsq}
}
\end{figure*}

\subsection{Comparison between Monte Carlo simulation and classical
spin wave theory}

Here, to demonstrate the validity of our theory,
we compare the structure factors, as calculated from
the classical spin wave theory ${\mathcal H}_{\sf ex}^{\sf CSW}$ [Eq.~(\ref{eq:HlowT})] 
and Monte Carlo simulation, for three different parameter sets:
the parameters of Yb$_2$Ti$_2$O$_7$ as found in Ref.~[\onlinecite{ross11-PRX1}]
where the classical ground state is ferromagnetic,
the parameters of Er$_2$Ti$_2$O$_7$ as found in Ref.~[\onlinecite{savary12-PRL109}]
where we expect the order by disorder mechanism to favour the $\Psi_2$
states
and one set of parameters where the order by disorder mechanism favours
the $\Psi_3$ states.
We find excellent, quantitative agreement between the two methods.

\begin{figure*}
\includegraphics[width=0.8\textwidth]{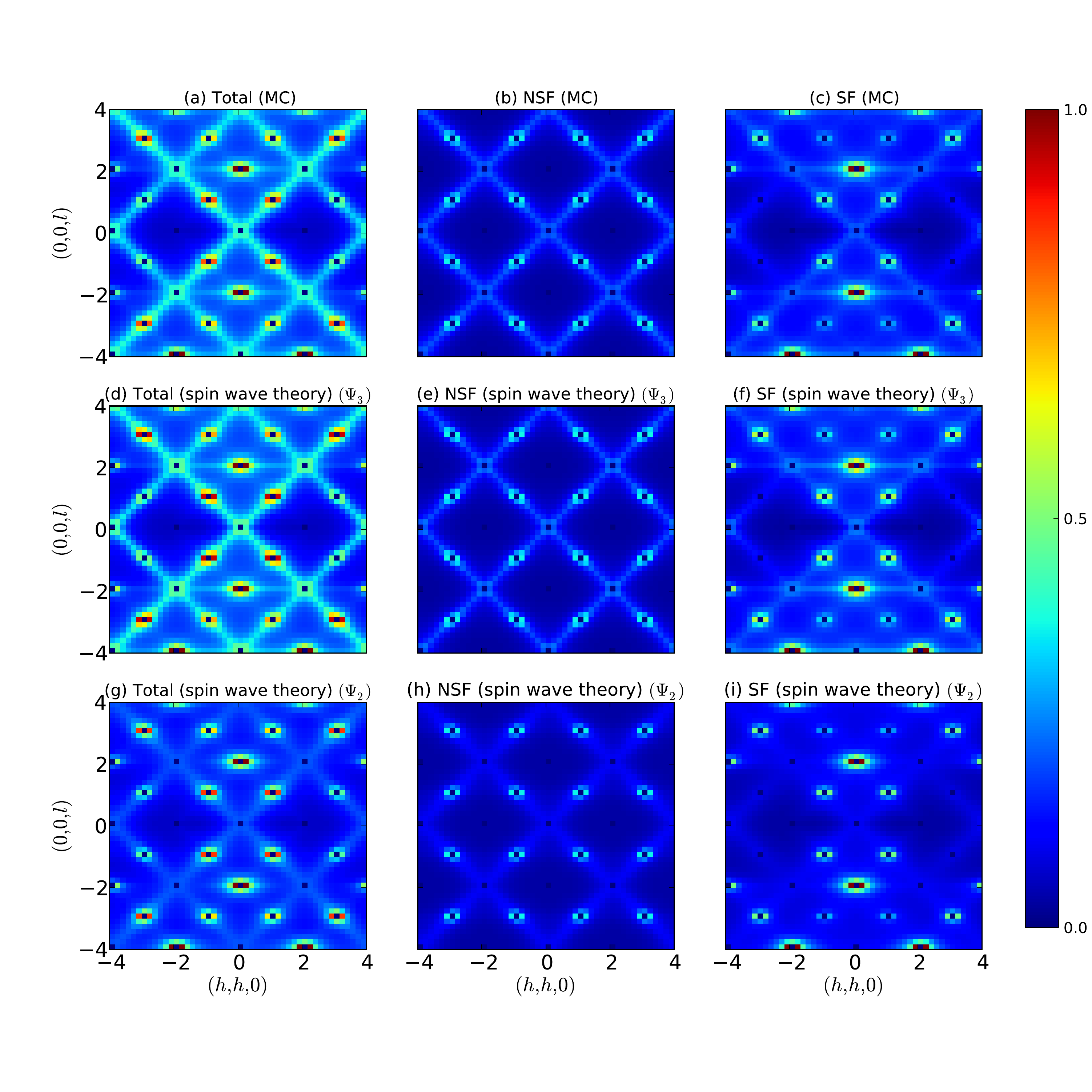}
\caption{
Comparison between results for equal-time structure factor $S({\bf q})$ obtained
in classical Monte Carlo (MC) simulation and low-temperature expansion
(classical spin wave theory) in the ordered phase for parameters 
\mbox{$J_1 = 0$},
\mbox{$J_2=-1.0\ \text{meV}$}, 
\mbox{$J_3 = -0.10\ \text{meV}$} , 
\mbox{$J_4 \equiv 0$},
approaching the non-collinear ferromagnet from within the ${\sf E}$--symmetry phase.  
a) Total scattering in the $(h, h, l)$ plane within MC simulation showing strong rod-like 
features in $[111]$ directions.
b) Associated scattering in the non spin-flip ({\sf NSF}) channel.
c) Associated scattering in the spin-flip ({\sf SF}) channel.
d) Total scattering in the $(h, h, l)$ plane within a classical spin-wave expansion 
about a $\Psi_3$ ground state, showing strong rod-like features in $[111]$ directions.
e) Associated scattering in the {\sf NSF} channel.
f) Associated scattering in the {\sf SF} channel.
g) Total scattering in the $(h, h, l)$ plane within a classical spin-wave expansion 
about a $\Psi_2$ ground state.
h) Associated scattering in the {\sf NSF} channel.
i) Associated scattering in the {\sf SF} channel.
Comparison of the scattering supports the conclusion that the $\Psi_3$
ground state is found in simulation, in agreement with the results 
of the low-T expansion [cf. Fig.~\ref{fig:entropy}].
An isotropic g-tensor $g_z=1, g_{xy}=1$ has been assumed.
For clarity, intensity associated with Bragg peaks at reciprocal 
lattice vectors has been subtracted.
}
\label{fig:J1=0sq}
\end{figure*}


In Fig.~\ref{fig:SqYTO} we show the structure factor $S({\bf q})$ calculated both
from classical spin wave theory and from Monte Carlo simulation at $T=0.05 $K, in the
{\sf NSF}, {\sf SF} and total scattering channels (see Eqs. (\ref{eq:defSq}), ({\ref{eq:sqnsf}}) 
and ({\ref{eq:sqsf}})).
We have used the experimentally determined parameters for the g-tensor \cite{hodges01}
$g_z=1.77$, $g_{xy}=4.18$  and exchange integrals \cite{ross11-PRX1} 
$J_1=-0.09 \text{meV}$, $J_2=-0.22 \text{meV}$ and $J_3=-0.29 \text{meV}$, 
setting $J_4 \equiv 0$.
The excellent, quantitative agreement between spin wave theory and simulation demonstrates
the excellent equilibration of the simulations down to $0.05$K for the parameters
of Yb$_2$Ti$_2$O$_7$.


$S({\bf q})$ is also useful for studying the entropic ground state selection 
within the one--dimensional manifold of state with ${\sf E}$ symmetry.
For a given set of parameters we may compare the diffuse scattering 
calculated in spin wave theory in expansions around the $\Psi_3$ and $\Psi_2$ 
phases with the diffuse scattering calculated in simulations.


Such a comparison is shown in Fig.~\ref{fig:ETOsq}, 
for exchange parameters appropriate to Er$_2$Ti$_2$O$_7$ 
(\mbox{$J_1=0.11\ \text{meV}$},
\mbox{$J_2=-0.06\ \text{meV}$}
and \mbox{$J_3=-0.10\ \text{meV}$}, 
setting  \mbox{$J_4 \equiv 0$})  
and temperature \mbox{$T=0.36 \text{K}$}.
From the entropy calculations shown in Fig.~\ref{fig:entropy} 
we expect the $\Psi_2$ state to be preferred for these
values of the exchange parameters.
Comparison of of the distribution of weight in the vicinity of
the $(1, 1, 1)$, $(3, 3, 3)$ and $(1, 1, 3)$ reciprocal 
lattice vectors between the Monte Carlo data
and the spin wave expansions around the $\Psi_2$ and $\Psi_3$ 
phases supports this conclusion.


Similarly, in Fig.~\ref{fig:J1=0sq} we show a comparison of the diffuse scattering between 
Monte Carlo simulations and spin wave expansions around the $\Psi_2$ and $\Psi_3$ phases
for exchange parameters approaching the non-collinear ferromagnetic 
phase ($J_1=0$, $J_2 = -1.0\ \text{meV}$ and $J_3 = -0.1\ \text{meV}$, $J_4=0$), at $T=0.4$K.
Calculations of the entropy within spin wave theory show that the $\Psi_3$ state should be preferred by
fluctuations for these parameters, and this is confirmed by the comparison of the structure factors,
in particular by the presence of bright rods along the $\langle111\rangle$ directions.

\section{Living on the edge : the influence of ground state manifolds on 
             finite--temperature phase transitions}
\label{section:living-on-the-edge}

The major assertion of this article is that many of the interesting properties of pyrochlore 
magnets  --- for example the rods of scattering observed in 
Yb$_2$Ti$_2$O$_7$, and the order-by-disorder selection of a $\Psi_2$ ground state in 
Er$_2$Ti$_2$O$_7$, see Sections~\ref{section:Er2Ti2O7} and~\ref{section:Yb2Ti2O7}
 --- are the direct consequence of competition between different ordered phases, and 
in particular, of the high ground--state degeneracy where phases with 
different symmetry meet.
While the arguments for enlarged ground state manifolds at $T=0$ are easy
to understand, it is far less obvious that this degeneracy should make itself
felt at finite temperature, especially where it is not protected by symmetry.


We can test the internal consistency of these ideas by using the probability 
distribution of the order parameter 
$${\bf m}_{\sf E} = m_{\sf E}\ (\cos\theta_{\sf E},\ \sin\theta_{\sf E})$$
[cf. Eq.~(\ref{eq:thetaE})] to deconstruct the order-by-disorder selection 
of $\Psi_2$ and $\Psi_3$ ground states in finite-temperatures simulations of 
$\mathcal{H}_{\sf ex}$ [Eq.~(\protect\ref{eq:Hex})].
The probability density function $P({\bf m}_{\sf E})$ is sensitive both to the 
formation of a one--dimensional manifold of states with ${\sf E}$ symmetry
--- which manifests itself as a ring in $P({\bf m}_{\sf E})$ --- and to
the selection of an ordered ground state within this manifold --- which will
appear as six degenerate maxima within the ring.


$P({\bf m}_{\sf E})$ also enables us to study the evolution of the ground state
manifolds at the boundaries between phases with competing symmetry --- in this case 
either with ${\sf T_2}$ or with ${\sf T_{1,A'}}$.
At these phase boundaries, ${\bf m}_{\sf E}$ takes on a new, constrained 
set of values, characteristic of the way in which different manifolds connect.
For example Eqs.~(\ref{eq:spoke1}--\ref{eq:spoke3}) predicts that, on the boundary
with the Palmer--Chalker phase, the one--dimensional manifold of states with 
$|{\bf m}_{\sf E}| = 1$ acquires ``spokes'' in the directions  
$$\theta_{\sf E} = \Big\{ 0,\ \frac{\pi}{3},\ \frac{2\pi}{3},\ \pi,\ \frac{4\pi}{3},\ \frac{5\pi}{3} \Big\}$$
connecting ${\bf m}_{\sf E} = 0$ with the six $\Psi_2$ ground states.
Observation of such a ``spoked wheel'' pattern in $P({\bf m}_{\sf E})$ at finite 
temperature would therefore confirm that the zero-temperature degeneracies
were still operative.

In an exactly parallel manner, we find that on the boundary with the
ferromagnetic phase it is possible to deform the ground state continuously from  
$\Psi_3$ to a corresponding ferromagnetic state.
For this reason $\Psi_3$ should be favoured approaching the boundary with the ferromagnet 
\mbox{[cf. Fig.~\ref{fig:classical-phase-diagram}]}. 

\begin{figure*}[h!]
\centering\includegraphics[width=17cm]{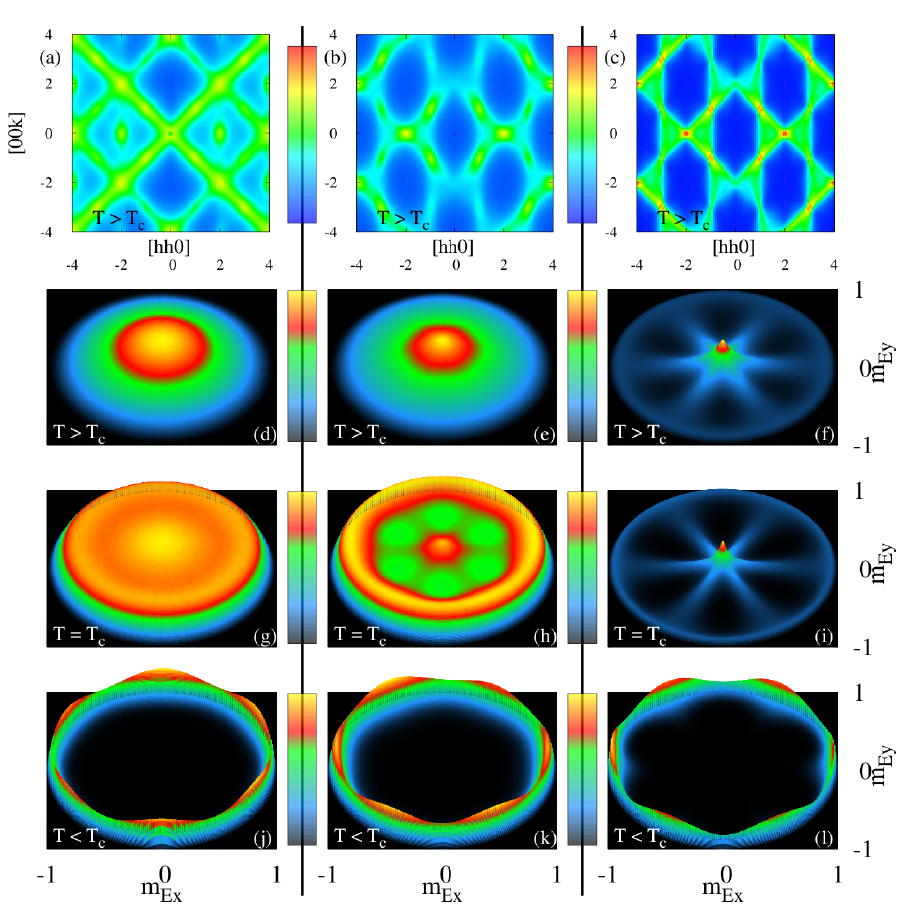}
\caption{
Influence of ground-state degeneracy on finite-temperature phase transitions, 
as revealed by the probability distribution of the order parameter 
${\bf m}_{\sf E} = m_{\sf E}\ (\cos\theta_{\sf E},\ \sin\theta_{\sf E})$~[Eq.~(\ref{eq:thetaE})].
Results are taken from simulation of $\mathcal{H}_{\sf ex}$ [Eq.~(\protect\ref{eq:Hex})],
with three different sets of exchange parameters, which we label as parameter sets (A), (B) and (C).
Parameter set (A), used to calculate panels (a, \ d, \ g, \ j), corresponds to a $\Psi_3$ ground state, 
approaching the non-collinear FM with $T_c=0.39$ K;
Parameter set (B), used to calculate panels  (b,\ e,\ h,\ k), corresponds to a $\Psi_2$ ground state, 
with $T_c=0.26$ K;
Parameter set (C), used to calculate panels (c, \ f, \ i, \ l), corresponds to a $\Psi_2$ ground state, 
on the border of the Palmer--Chalker phase with $T_c=0.065$ K.  
(a)-(c) quasi-elastic scattering $S({\bf q})$ in the paramagnetic phase $T > T_c$.
(d)-(f) corresponding results for the probability density function, ${\sf P}({\bf m}_{\sf E})$.
(g)-(i) ${\sf P}({\bf m}_{\sf E})$ at the transition temperature $T = T_c$.
(j)-(l) ${\sf P}({\bf m}_{\sf E})$ in the ordered phase $T < T_c$.
Parameter set (A): For a finite-size system, the onset of $\Psi_3$ occurs progressively, 
through (g) the emergence of a one--dimensional manifold of states with 
finite $|{\bf m}_{\sf E}|$, and then (j) the entropic selection of $\theta_{\sf E}$ 
corresponding to one of six distinct $\Psi_3$ ground states.
(a) The connection with the non-collinear FM is evident in $S({\bf q})$, with 
rods of scattering strongly reminiscent of those seen in Yb$_2$Ti$_2$O$_7$.
Parameter set (B): The same process occurs
, but in this case ${\sf P}({\bf m}_{\sf E})$ shows that $\Psi_2$ ground states are favoured 
at low temperatures (k) and even at the transition (h).
Parameter set (C):  On the boundary of the Palmer--Chalker phase, the ground state manifold includes 
additional manifolds of states which mix ${\bf m}_{\sf E}$ and ${\bf m}_{{\sf T}_2}$.
These are evident (f,i) in the ``spoked wheel'' seen in ${\sf P}({\bf m}_{\sf E})$ at 
$T \geq T_c$, and drive the entropic selection of the $\Psi_2$ ground state.
(c) The high degeneracy at this phase boundary is also evident in the ``bow-tie''
structure in $S({\bf q})$.
Further details of simulations and the parameters corresponding to 
(A), (B) and (C) are given in the text.
}
\label{fig:figure-of-doom}
\end{figure*}

In Fig.~\ref{fig:figure-of-doom} we present results for $P({\bf m}_{\sf E})$ and $S({\bf q})$ 
taken from simulations of $\mathcal{H}_{\sf ex}$ [Eq.~(\protect\ref{eq:Hex})] for three sets of 
parameters
$$(\textrm{A}) \quad (J_1,\ J_2,\ J_3,\ J_4) =  (0, -0.3, -0.1, 0) \quad \text{meV} $$
where we expect a $\Psi_3$ ground state, while approaching the border with the 
non-collinear FM [Fig.~\ref{fig:figure-of-doom} \mbox{(a,\ d,\ g,\ j)}];
$$(\textrm{B}) \quad (J_1,\ J_2,\ J_3,\ J_4) =  (0.11, 0.06, -0.1, 0) \quad \text{meV} $$
where we expect a $\Psi_2$ ground state, approaching the border with the 
Palmer--Chalker phase [Fig.~\ref{fig:figure-of-doom} \mbox{(b,\ e,\ h,\ k)}]; and
$$(\textrm{C}) \quad (J_1,\ J_2,\ J_3,\ J_4) =  (0.11, 0.11, -0.1,0) \quad \text{meV} $$
exactly on the $T=0$ border of the Palmer--Chalker phase 
[Fig.~\ref{fig:figure-of-doom} \mbox{(c,\ f,\ i,\ l)}].

\begin{figure*}[htpb] 
\includegraphics[width=18cm]{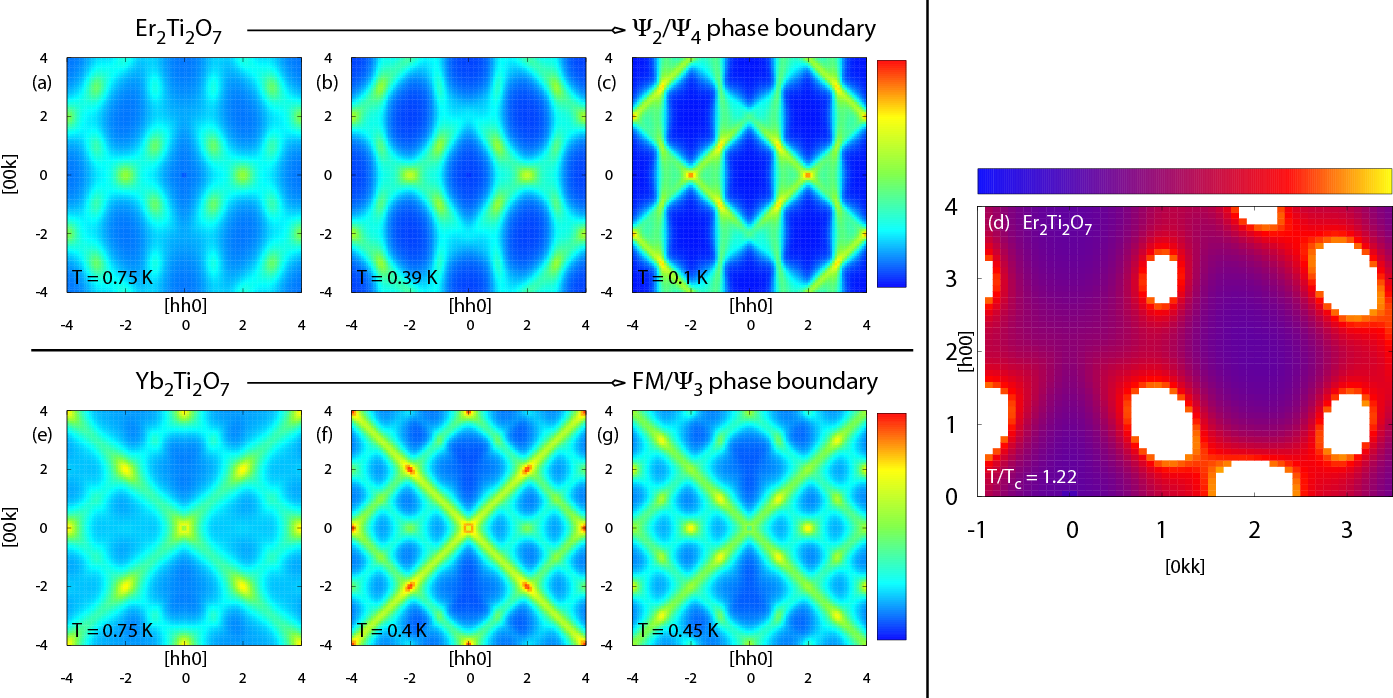}
\caption{
Correlations in the high-temperature paramagnetic phase, as revealed by the quasi-elastic 
structure factor $S({\bf q})$.
\mbox{(a-c)} results for parameters interpolating from (a) Er$_2$Ti$_2$O$_7$ 
[\onlinecite{savary12-PRL109}]
to (c) the boundary of the Palmer--Chalker phase ($\Psi_4$).
The diffuse scattering characteristic of the $\Psi_2$ phase
evolve into sharp features reminiscent of pinch points
when bordering the $\Psi_4$ phase.
Results are taken from classical Monte Carlo simulations carried out 
for 
(a) $J_2=-0.06$~\textrm{meV}, $ T=750$~\textrm{mK}; 
(b) $J_2=0.06$~\textrm{meV}, $ T=390$~\textrm{mK}; 
(c) $J_2=0.11$~\textrm{meV}, $ T=100$~\textrm{mK}.  
In all cases, 
$J_1=-0.11$~\textrm{meV}, 
$J_3=-0.1$~\textrm{meV},  
$J_4\equiv 0$, 
and $S({\bf q})$ has been calculated using g-tensor parameters appropriate 
to Er$_2$Ti$_2$O$_7$~[\onlinecite{savary12-PRL109}].
(d) detail of S(q) for parameters appropriate to Er$_2$Ti$_2$O$_7$ at $ T=616$~\textrm{mK} , 
plotted with a colour scale chosen 
to match Fig. 14 of [\onlinecite{dalmas12}] and with the same temperature ratio $T/T_{c}=1.22$
($T_{c} = 505$ mK in simulations).   
\mbox{(e-g)} results for parameters interpolating from   
Yb$_2$Ti$_2$O$_7$ [cf. Ref.~(\onlinecite{ross11-PRX1})],
to the border of the $\Psi_3$ phase.
The rods of scattering along $[111]$ directions, interpreted    
as evidence of dimensional reduction in Yb$_2$Ti$_2$O$_7$ 
[\onlinecite{ross11-PRB84}], evolve into weakly-dispersing, 
low-energy excitations in the neighbouring $\Psi_3$ phase.
Results are taken from classical Monte Carlo simulations of 
${\mathcal H}_{\sf ex}$ [Eq.~(\ref{eq:Hex1})] for 
(e) $J_1=-0.09$~\textrm{meV}, $ T=750$~\textrm{mK}; 
(f) $J_1=-0.04$~\textrm{meV}, $ T=400$~\textrm{mK}; 
(g) $J_1=-0.0288$~\textrm{meV}, $ T=450$~\textrm{mK}.  
In all cases, 
$J_2=-0.22$~\textrm{meV}, 
$J_3=-0.29$~\textrm{meV}, 
$J_4\equiv 0$, 
and $S({\bf q})$ has been calculated using g-tensor parameters appropriate 
to Yb$_2$Ti$_2$O$_7$~[\onlinecite{hodges01}].
}
\label{fig:Sq} 
\end{figure*}

The results for $S({\bf q})$ shown in Fig.~\ref{fig:figure-of-doom}(a-c), demonstrate 
the diffuse structure expected in the paramagnet in each case~: 
(A) Fig.~\ref{fig:figure-of-doom}(a) --- rods of scattering, 
reminiscent of those observed in Yb$_2$Ti$_2$O$_7$ 
[\onlinecite{bonville04,ross11-PRB84,ross09,thompson11,chang12}]; 
(B) Fig.~\ref{fig:figure-of-doom}(b) --- a diffuse web of rings, 
reminiscent to that observed in experiments on Er$_2$Ti$_2$O$_7$ 
[see Fig.~\ref{fig:Sq}(a-d) for more details], also ordering in $\Psi_2$;
(C) Fig.~\ref{fig:figure-of-doom}(c) --- ``bow-tie'' patterns reminiscent of the 
pinch points observed in the Heisenberg antiferromagnet on a pyrochlore lattice 
[\onlinecite{moessner98-PRB58}]. Indeed, the Heisenberg 
antiferromagnet corresponds to the parameters of figure~\ref{fig:figure-of-doom}(c) with $J_{3}=0$.


The results for $P({\bf m}_{\sf E})$ strongly validate our understanding of the
problem in terms of degenerate ground state manifolds, even at finite temperatures.
On the border of the Palmer--Chalker phase (C) 
$P({\bf m}_{\sf E})$ shows a diffuse spoked wheel both above $T_c$ [Fig.~\ref{fig:figure-of-doom}(f)] 
and at $T_c$ [Fig.~\ref{fig:figure-of-doom}(i)] , 
confirming that the connection  
implied by the $T=0$ ground state manifold survives even for temperatures above the 
phase transition. 
The remnant of this spoked wheel pattern is even seen at $T_c$
when we tune away from the phase boundary 
(parameter set (B)) [Fig. 23(h)].
This spoked wheel pattern in $P({\bf m}_{\sf E})$ is strong evidence that
the manifolds continue to be operative at $T_c$
and play a role in ordered state selection.
This perspective is consistent with the Ginzburg--Landau
theory presented in Ref.~\onlinecite{javanparast15}.
And it gives new insight into {\it how} the terms in the Ginzburg--Landau 
theory which lift the $U(1)$ degeneracy are generated --- namely 
by fluctuations into these low---energy manifolds.


The reason why we see the spoked wheel pattern at $T_c=0.26$ K for parameter set (B)
and not at $T_c=0.39$ K for parameter set (A) is probably a consequence of the strong 
finite size dependence of
the entropic selection between $\Psi_2$ and $\Psi_3$.


To conclude, we should add that where these two phase boundaries approach one 
another, the soft modes associated with the two different sets of manifolds compete.  
This leads to the complicated, reentrant behaviour seen in Fig.~\ref{fig:classical-phase-diagram}, 
and studied for quantum spins in [\onlinecite{wong13}].   


\section{Application to \protect{E\lowercase{r}$_2$T\lowercase{i}$_2$O$_7$}}
\label{section:Er2Ti2O7} 

Early heat capacity measurement of Er$_2$Ti$_2$O$_7$ revealed 
a phase transition at $T_c=1.25$ K, releasing an entropy 
$\Delta s \approx 0.97 k_B \ln 2$ per spin, consistent with the ordering of the 
ground state doublet of Er [\onlinecite{bloete69}].
Later, neutron scattering studies revealed the nature of the 
low temperature order, finding it to correspond to the $\Psi_2$
configurations illustrated in Fig.~\ref{fig:psi2} [\onlinecite{champion03, 
poole07, savary12-PRL109}].


The selection of the $\Psi_2$ states in Er$_2$Ti$_2$O$_7$
out of the 1D manifold of states transforming with {\sf E} symmetry has
been identified as a textbook example of order--by--disorder \cite{champion03},
with quantum zero--point fluctuations \cite{savary12-PRL109, zhitomirsky12,wong13},
low--temperature classical thermal fluctuations \cite{mcclarty14}
and thermal flucuations near the ordering temperature \cite{oitmaa13,zhitomirsky14} 
all favouring $\Psi_2$ order.
A corollary of this conclusion is that there should be a small, fluctuation
induced, gap  at ${\bf q}=0$ in the spin wave spectrum.
And, such a gap has now been observed in inelastic
neutron scattering \cite{ross14}.
We note that, an alternative scenario has been
proposed in which the selection of $\Psi_2$ comes instead from
virtual fluctuations into higher crystal field levels \cite{mcclarty09,petit14,rau-arXiv}.
Thus it may in fact be that the selection of $\Psi_2$ has multiple contributions-
both from harmonic fluctuations of the ground state order and from
virtual crystal field fluctuations.


Here, using the exchange parameters for Er$_2$Ti$_2$O$_7$ 
taken from [\onlinecite{savary12-PRL109}], we 
confirm that thermal fluctuations of classical Heisenberg spins select a 
$\Psi_2$ phase at finite temperature. 
Estimates from our Monte Carlo simulations give $T_c \approx 500 \text{mK}$ 
[Fig.~\ref{fig:PMtoPsi2}], somewhat 
lower than both experiments and those obtained in a high temperature 
series expansion of the quantum spin model \cite{oitmaa13}. 
But within the paramagnetic phase, 
our simulations of the spin structure factor $S({\bf q})$
are in excellent agreement with neutron scattering 
measurements~\cite{dalmas12} showing the build-up of long-range order,
 as can be seen from the comparison between Fig.~\ref{fig:Sq}(d) of
this work and Fig. 14 of Ref. [\onlinecite{dalmas12}].


\begin{figure*}
\includegraphics[width=0.62\textwidth]{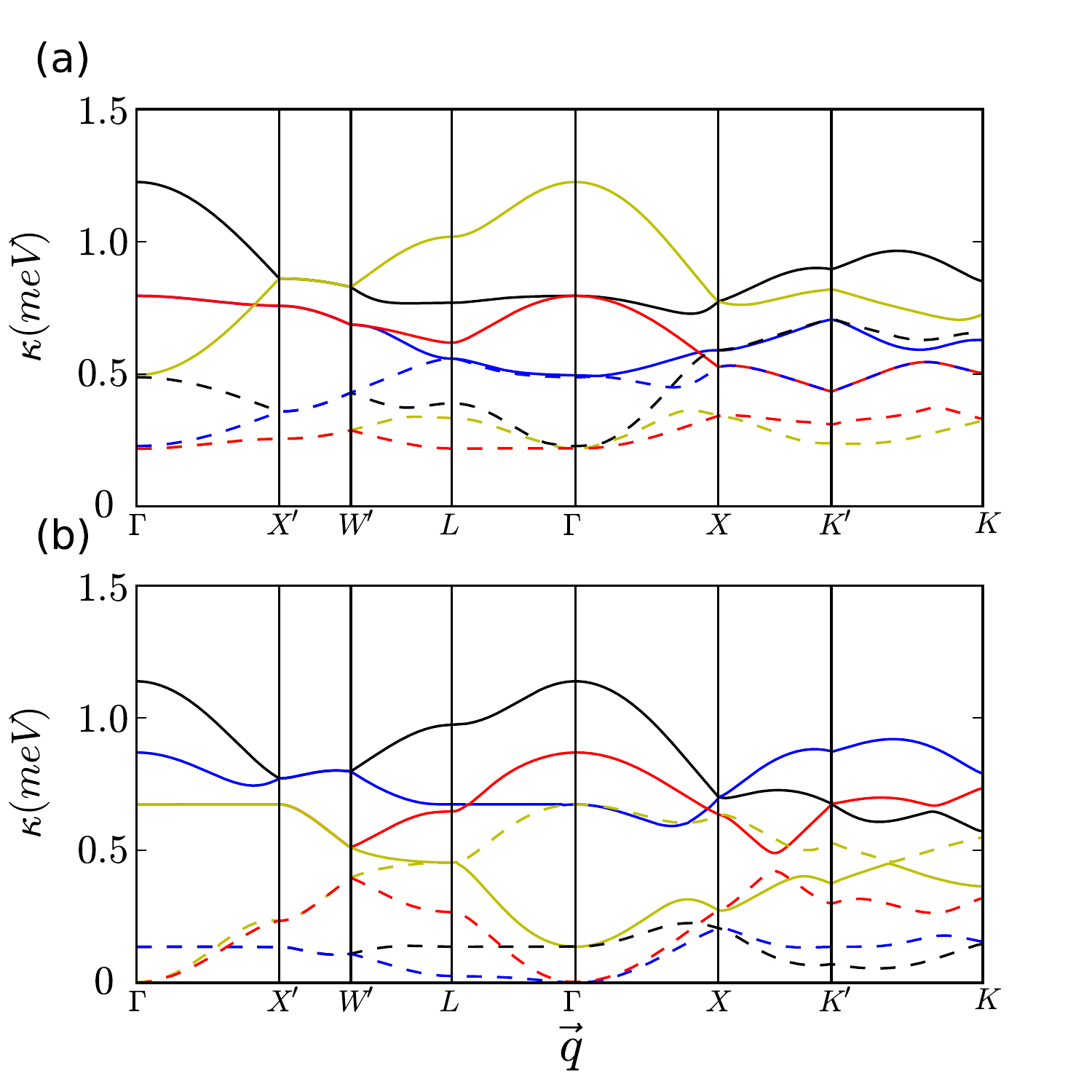}
\caption{
Spin-wave dispersion calculated within a classical, low-temperature 
expansion, showing dimensional reduction of a subset of excitations.
(a) Excitations of the FM ground state, 
for exchange parameters appropriate to Yb$_2$Ti$_2$O$_7$, 
i.e. 
\mbox{$J_1=-0.09 \text{meV}$}, 
\mbox{$J_2=-0.22 \text{meV}$}, 
\mbox{$J_3=-0.29 \text{meV}$},
setting 
$J_4=0$.  
The ferromagnet possesses a flat band in the $(h, h, h)$ ($\Gamma \to L$) 
direction at energy $\Delta\approx0.22$ meV, which gives rise to rods in the equal 
time structure factor (cf. Fig.~\ref{fig:SqYTO}).
(b) Excitations of the $\Psi_3$ ground state, for exchange parameters 
on the boundary between the $\Psi_3$ and FM phases,  
i.e. 
\mbox{$J_1=-0.029 \text{meV}$}, 
\mbox{$J_2=-0.22 \text{meV}$}, 
\mbox{$J_3=-0.29 \text{meV}$}
with
$J_4=0$.  
The $\Psi_3$ phase, on the phase boundary also possesses 
a quasi-flat band along $(h, h, h)$, which in this case is gapless 
at the $\Gamma$ point of the Brillouin zone.
This leads us to suggest that the low--energy rod--like 
features observed in the paramagnetic
phase of Yb$_2$Ti$_2$O$_7$ arise from its proximity
in parameter space to the $\Psi_3$ phase
and the low--energy modes 
which are present on the phase boundary (see discussion in Section~\ref{section:Yb2Ti2O7}).
\label{fig:dispersionsYTO}
}
\end{figure*}


A question which remains is {\it why} fluctuations should favour $\Psi_2$
in the case of Er$_2$Ti$_2$O$_7$ and more generally why they should favour
either $\Psi_2$ or $\Psi_3$ for a given set of exchange parameters $\{ J_i \}$.
Our work provides the answer to this question and, in so doing,
underlines how the properties of a frustrated magnet are strongly
influenced by competing phases.


The mechanism by which the $\Psi_2$ states are selected is inherited
from the phase boundary with the neighbouring Palmer--Chalker phase.
At this boundary three additional continuous sets of ground states apppear
connecting the 6 Palmer--Chalker ground states, to the 1D manifold of 
{\sf E} symmetry states [Fig.~\ref{fig:manifold-T2-E}].
The points in configuration space at which these sets of ground states 
meet are none other than the $\Psi_2$ configurations. 
Due to their favoured position at the junctions of the ground state manifold
the $\Psi_2$ states gain additional soft modes and are selected 
by fluctuations in the region approaching the boundary with the Palmer--Chalker
phase.
The consequences of these connected manifolds are visible
even in finite temperature simulations as shown in Fig.~\ref{fig:figure-of-doom}.


It is worth noting that an exactly parallel mechanism selects the $\Psi_3$
states for parameters proximate to the ferromagnetic 
phase.
In the region proximate to both Palmer--Chalker and Ferromagnetic phases, 
a complicated re-entrant behaviour is observed\cite{wong13} 
[Fig.~\ref{fig:classical-phase-diagram}].


\begin{figure*}
\includegraphics[width=\textwidth]{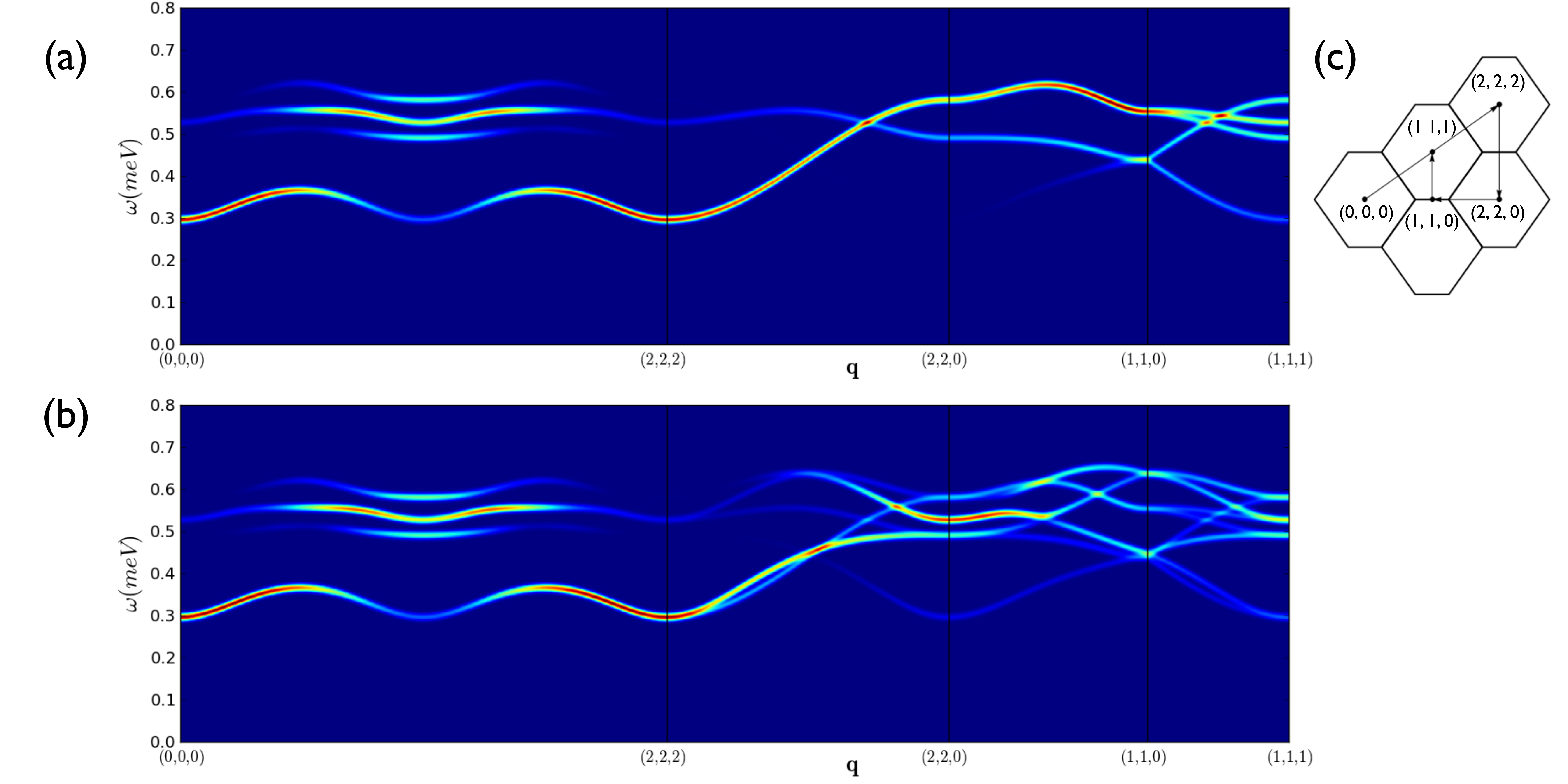}
\caption{
Prediction for the spin--wave excitations of the ferromagnetically--ordered 
ground state of Yb$_2$Ti$_2$O$_7$, calculated within linear spin--wave 
theory for the parameters given by Ross~{\it et al.} [\onlinecite{ross11-PRX1}].
(a) Spin--wave dispersion for a single ferromagnetic domain with magnetisation 
parallel to $[100]$, showing a minimum gap to excitations 
$\Delta \sim 0.3 \text{~meV}$, occurring in the zone centre.   
(b) Spin--wave dispersion averaged over the 6 possible ferromagnetic domains.   
(c) Path in reciprocal space used in making plots.   
Calculations were carried out for the anisotropic exchange model 
$\mathcal{H}_{\sf ex}$~[Eq.~(\ref{eq:Hex1})], with results 
convoluted with a Gaussian of full width at half maximum 
$0.014 \text{~meV}$ to mimic finite experimental resolution. 
The relative intensity of scattering is shown in false color.  
Details of calculations are given in 
Section~\ref{subsection:quantum-spin-wave} 
and Appendix~\ref{appendix:linear-spin-wave}.
}
\label{fig:LSW-Yb2Ti2O7}
\end{figure*}


Our work shows that the preference of fluctuations for $\Psi_2$ ordering in
Er$_2$Ti$_2$O$_7$ is a property inherited from a nearby phase boundary
where the $\Psi_2$ states sit at the junctions of a connected ground state
manifold.
In the context of this result it is interesting to ask how the spin correlations
evolve as the exchange parameters are tuned from those appropriate to Er$_2$Ti$_2$O$_7$ to the
boundary of the Palmer--Chalker phase.
This is illustrated using Monte Carlo simulations of the spin structure factor $S(\mathbf{q})$ 
in  Fig.~\ref{fig:Sq}(a-c).
For the parameters appropriate to Er$_2$Ti$_2$O$_7$, our simulations reproduce the
smooth features observed in experiment.
As the phase boundary
is approached these smooth features 
evolve into sharp, pinch-point like
features, associated with the large ground state manifold
on the phase boundary.
In the limit $J_3\to0$ these pinch-point like
features become the pinch points associated with the
Coulomb phase of the $O(3)$ Heisenberg model on
the pyrochlore lattice \cite{moessner98-PRB58}.


Our study of the ground state selection in Er$_2$Ti$_2$O$_7$ 
emphasises
that the properties of frustrated magnets can be strongly
influenced by the soft modes appearing on nearby boundaries.
This same essential insight also manifests itself -- in rather different ways --
in the study of Yb$_2$Ti$_2$O$_7$ and Er$_2$Sn$_2$O$_7$, to which we now
turn.


\section{Application to Y\lowercase{b}$_2$T\lowercase{i}$_2$O$_7$}
\label{section:Yb2Ti2O7} 

Like its sister compound Er$_2$Ti$_2$O$_7$, Yb$_2$Ti$_2$O$_7$
was first identified as undergoing a finite temperature ordering
transition in the heat capacity study by Bl{\"o}te {\it et al.}, nearly 50 years ago
\cite{bloete69}.
That study revealed a sharp anomaly in the heat capacity at $T_c=0.214$K, 
with a corresponding release of entropy of $\Delta s=0.97 k_B \ln 2$ per spin.
Since then, the presence of this phase transition in 
stoichiometric Yb$_2$Ti$_2$O$_7$ has been
debated in the literature, with different groups, with
different samples, reporting differing results 
for the presence or absence of magnetic order.
Nevertheless, it is now widely accepted that stoichiometric Yb$_2$Ti$_2$O$_7$ undergoes a thermodynamic phase transition into a state with finite magnetisation at a temperature $T_c \sim 0.2$K [\onlinecite{hodgesPRL01, yasui03, chang12, lhotel14, chang14,robert15, bhattacharjee16, gaudet16, hallas16,yaouanc16}].


For the parameters given by Ross~{\it et al.} [\onlinecite{ross11-PRX1}], 
the theory developed in Section~\ref{section:classical-ground-states} 
predicts that Yb$_2$Ti$_2$O$_7$ has a ${\bf q} = 0$ 
ground state, with non--collinear order ferromagnetic order.
This ``splayed ferromagnet'' is consistent with the interpretation 
of neutron scattering experiments given in 
[\onlinecite{chang14, robert15, gaudet16, hallas16}].
An ordered ground state of this type would normally be expected to support 
coherent, dispersing spin--wave excitations, with a finite gap coming from the 
anisotropy of exchange interactions, as illustrated in Fig.~\ref{fig:LSW-Yb2Ti2O7}.


Curiously, however, gapped, coherent spin--waves have yet to be 
observed in Yb$_2$Ti$_2$O$_7$, with a succession of experiments reporting 
a broad, gapless continuum at low 
temperatures~\cite{ross09,robert15,gaudet16,hallas16}.    
The origin of this gapless continuum remains a puzzle, although the presence 
of competing classical ordered phases must ultimately impact on 
quantum excitations \cite{jaubert15}.
It is also important to recall that the ordered ground state breaks only 
the point--group symmetries of the anisotropic exchange model 
$\mathcal{H}_{\sf ex}$~[Eq.~(\ref{eq:Hex1})], and so spin--waves cannot 
be interpreted as Goldstone modes.
It follows that interaction effects may play an important role, even at 
low orders in $1/S$.   
And it is interesting to note that the broad continuum observed in experiment 
has more in common with semi--classical simulations of the spin--excitations 
of the paramagnetic phase of Yb$_2$Ti$_2$O$_7$ 
[\onlinecite{robert15,taillefumier-unpub}], than with the linear spin--wave 
excitations of the ground state, as shown in Fig.~\ref{fig:LSW-Yb2Ti2O7}.
This point will be discussed further elsewhere~\cite{taillefumier-inprep}.


While the nature of the ground state of Yb$_2$Ti$_2$O$_7$ has proved controversial, 
and the associated excitations remain to be understood, all neutron--scattering 
experiments agree about the signature feature of its paramagnetic phase 
--- striking ``rod''--like structures along the $\langle111\rangle$ directions of reciprocal space. 
First observed more than ten years ago~\cite{bonville04}, these rods of 
scattering have since been interpreted as evidence of dimensional 
reduction~\cite{ross11-PRB84,ross09} and, in the 
context of $\mathcal{H}_{\sf ex}$~[Eq.~(\ref{eq:Hex1})], as evidence 
of significant anisotropic exchange 
interactions~\cite{ross11-PRX1,thompson11,cao09-PRL103}. 
They are a robust feature of $S({\bf q})$, as calculated from 
$\mathcal{H}_{\sf ex}$~[Eq.~(\ref{eq:Hex1})] 
within both the (semi-classical) random phase 
approximation~\cite{thompson11,chang12}, and 
classical Monte Carlo simulations~\cite{jaubert15} [Fig.~\ref{fig:Sq}]. 
However, despite their ubiquity, the origin of these rods of 
scattering remains mysterious.


To understand the origin of the rods of scattering
we must once again look to the influence of
the phase boundaries.
The classical ground states of $\mathcal{H}_{\sf ex}$ reduce to a 
set of independent kagome planes on the boundary between FM and 
Palmer--Chalker phases, and to a set of independent chains on 
the boundary between the $\Psi_2$ and Palmer--Chalker phases. 
However the rods of scattering seen in Yb$_2$Ti$_2$O$_7$ occur 
for parameters where the ground state of $\mathcal{H}_{\sf ex}$ is expected to 
be ordered and fully three-dimensional~\cite{ross11-PRX1}.
Indeed, our classical Monte Carlo simulations predict 
that Yb$_2$Ti$_2$O$_7$ orders at $450\ \text{mK}$ 
[cf.~Fig.~\ref{fig:PMtoFM}], a little higher than the 
$T_c \approx 200\ \text{mK}$ found in experiment.


Within the scenario of multiple--phase competition, rods of scattering can be traced back 
to dimensionally-reduced excitations, due to quasi-degenerate lines 
of low-lying spin wave excitations, which evolve into low-lying excitations 
of $\Psi_3$ on the boundary between the FM and the $\Psi_3$ phases 
[cf.~Fig.~\ref{fig:dispersionsYTO}]. 
This progression is also clear in the evolution of $S({\bf q})$ from 
parameters appropriate to Yb$_2$Ti$_2$O$_7$ [Fig.~\ref{fig:Sq}(e)] 
to the border of the $\Psi_3$ phase [Fig.~\ref{fig:Sq}(g)], 
also shown in Ref.~[\onlinecite{jaubert15}]. 


Seen in this light, the observation of rods of scattering in 
Yb$_2$Ti$_2$O$_7$ is a  consequence of the
proximity of competing ordered states- in this case the
$\Psi_2$ and $\Psi_3$ states.
The importance of these competing ground states in
driving the unusual physics of Yb$_2$Ti$_2$O$_7$ has
been underlined in recent work \cite{robert15, jaubert15}.
In Ref. [\onlinecite{jaubert15}] it was shown 
that both quantum and thermal fluctuations 
bring the phase boundary between ferromagnetic
and ${\sf E}$ symmetry phases closer to the parameter regime
appropriate to Yb$_2$Ti$_2$O$_7$.
In this sense, Yb$_2$Ti$_2$O$_7$ really is a material ``living on the
edge'' between differing magnetic orders, and the rods of scattering
are a manifestation of this.  

\section{Application to E\lowercase{r}$_2$S\lowercase{n}$_2$O$_7$}
\label{section:Er2Sn2O7} 

Like Er$_2$Ti$_2$O$_7$ and Yb$_2$Ti$_2$O$_7$, 
the magnetic ions in Er$_2$Sn$_2$O$_7$ have a Kramers 
doublet ground state \cite{matsuhira02, alam14}, and their interactions are believed 
to be well-described by $\mathcal{H}_{\sf ex}$~[Eq.~(\ref{eq:Hex1})] 
[\onlinecite{guitteny13}].
Correlations reminiscent of the Palmer--Chalker phase have been 
observed in neutron scattering~\cite{guitteny13}, and magnetization 
measurements show some evidence of spin-freezing at low temperatures\cite{guitteny13}. 
Nonetheless, Er$_2$Sn$_2$O$_7$ shows no evidence 
of magnetic order, in thermodynamic measurements \cite{sarte11,guitteny13}, 
$\mu$SR \cite{lago05}, or neutron scattering \cite{sarte11,guitteny13}, 
down to a temperature of $20\ \text{mK}$~[\onlinecite{lago05}]. 


\begin{figure}
\includegraphics[width=\columnwidth]{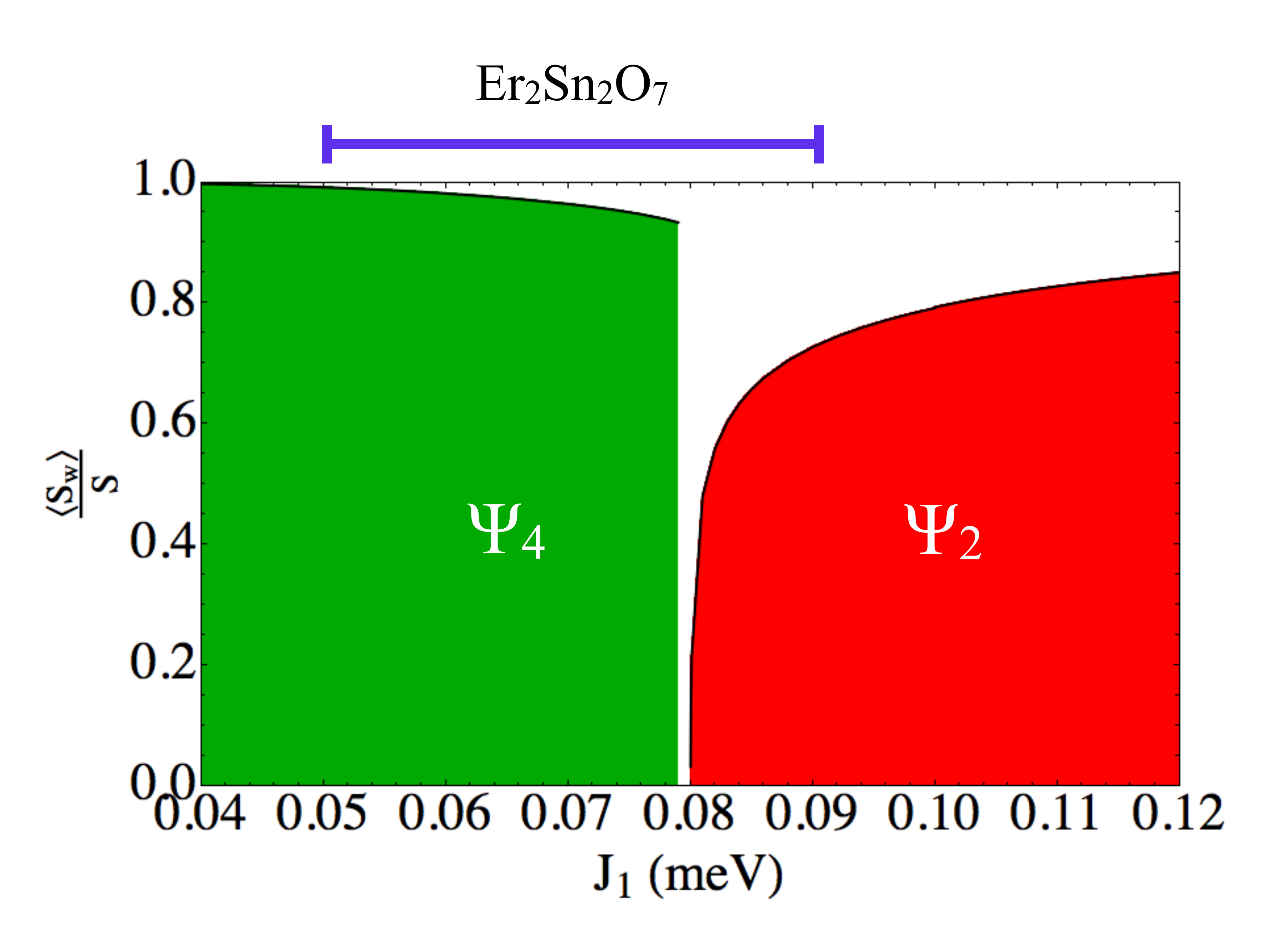}
\caption{
Ordered moment in the region of parameter space relevant to 
Er$_2$Sn$_2$O$_7$, as calculated in linear spin wave theory.
On the Palmer--Chalker ($\Psi_4$) side of the phase boundary, the 
quantum correction to the ordered moment is small.
However, the correction diverges on the approach the phase boundary 
from the non--coplanar antiferromagnet ($\Psi_2$), indicating the possibility 
of a region of quantum disorder between these two phases.
Experimental estimates of exchange parameters in 
Er$_2$Sn$_2$O$_7$ [\onlinecite{guitteny13}] place it close to this 
phase boundary, making it a good candidate for the observation of 
quantum spin--liquid physics.
Calculations were carried out for the anisotropic exchange model 
$\mathcal{H}_{\sf ex}$~[Eq.~(\ref{eq:Hex1})], as described in 
Section~\ref{subsection:quantum-spin-wave} 
and Appendix~\ref{appendix:linear-spin-wave}, with parameters 
$J_2=0.08 \ \text{meV}$, $J_3=-0.11 \ \text{meV}$ taken from 
[\onlinecite{guitteny13}], setting the Dzyaloshinskii-Moriya 
interaction $J_4=0$.
The error bars on the estimated value of $J_1$ 
are taken from Ref. [\onlinecite{guitteny13}].
}
\label{fig:eso-quantum}
\end{figure}


The exchange parameters determined for Er$_2$Sn$_2$O$_7$ 
in Ref. [\onlinecite{guitteny13}] would place it extremely close
to the phase boundary between the Palmer--Chalker and $\Psi_2$  
states [Fig.~\ref{fig:classical-phase-diagram}].
Classical Monte Carlo simulations with this parameter set
predict a phase transition into the Palmer--Chalker state 
 at \mbox{$T_c \approx 200 \text{mK}$}.
However, we can once more gain further insight by looking at the behaviour
of the model $\mathcal{H}_{\sf ex}$ approaching the phase boundary.


As the phase boundary is approached, the ground--state value of
the ordered moment, as calculated in linear spin wave theory,
is reduced by quantum fluctuations.
This is illustrated in Fig.~\ref{fig:eso-quantum}.
Approaching the boundary from the Palmer--Chalker
side, this quantum correction is small.
However, approaching the boundary from the $\Psi_2$
side, the correction is logarithmically divergent.
Since spin wave theory typically underestimates quantum effects,
this divergence is a likely indicator of a region of quantum
disorder between the Palmer--Chalker and $\Psi_2$
region of the phase diagram [cf. Fig.~\ref{fig:quantum-phase-diagram}].
The placement of Er$_2$Sn$_2$O$_7$  immediately adjacent to this
classical phase boundary thus makes it a prime candidate for
the observation of quantum spin liquid physics.


We can see therefore that the competition between 
Palmer--Chalker and $\Psi_2$
ordering in Er$_2$Sn$_2$O$_7$ enhances quantum
fluctuations in that material and may even stabilize a 
quantum disordered state.
Such a scenario would be consistent with the lack
of observed magnetic order in Er$_2$Sn$_2$O$_7$ 
and would make Er$_2$Sn$_2$O$_7$ the first example
of a pyrochlore spin liquid with dominantly XY--like interactions.
While the recent years have seen considerable theoretical
advances in the understanding of quantum spin liquid states
occurring close to the Ising (spin ice) limits of $\mathcal{H}_{ex}$
[\onlinecite{shannon12, savary12-PRL108, benton12, lee12, savary13, gingras14, hao14}],
the limit of dominant XY interactions has been much less explored
for quantum spins.
A deeper understanding of Er$_2$Sn$_2$O$_7$ calls for further work
in this direction.

\section{Other pyrochlore magnets}
\label{section:other-pyrochlores} 

The family of rare--earth pyrochlore oxides R$_2$M$_2$O$_7$
is very diverse \cite{gardner10}, and it is becoming even more so,
with high--pressure synthesis techniques allowing for many new
combinations of rare earth, R, and transition metal, M
to be realized \cite{wiebe15}.
In this Section we briefly explore the properties 
of further examples of pyrochlore magnets.  
We will mostly restrict our comments to materials 
based on the Kramers ions such as Er$^{3+}$, Yb$^{3+}$, 
where the model ${\mathcal H}_{\sf ex}$ [Eq.~(\ref{eq:Hex1})], 
with couplings of the form of ${\bf J}_{01}$ [Eq.~(\ref{eq:Jij})], 
offers a completely a completely general description of 
nearest--neighbor interactions on the pyrochlore lattice.  
That said, many of the same physical phenomena  
arise in pyrochlore magnets based on non--Kramers ions 
and, where these have a doublet ground state, 
interactions may also be described using ${\mathcal H}_{\sf ex}$ 
[\onlinecite{onoda11, lee12}].  
With this in mind, we also make a few brief comments about
materials based on the non--Kramers ions Tb$^{3+}$ and Pr$^{3+}$.


Yb$_2$Sn$_2$O$_7$ [\onlinecite{dun13, yaouanc13, lago14}]
and Yb$_2$Pt$_2$O$_7$ [\onlinecite{cai16}] 
have both been identified as having ferromagnetic ground states, 
and may therefore 
be placed in the non-collinear ferromagnet ($\sf T_1$)
region of our phase diagram. 
Meanwhile, the ground state of another Yb based 
system, Yb$_2$Ge$_2$O$_7$ has been shown in neutron scattering
experiments to belong to the manifold of ${\sf E}$ symmetry states,
although any ground state selection between $\Psi_2$ and $\Psi_3$
configurations has yet to be determined \cite{dun14, hallas16-PRB93}.
The progression, as a function of decreasing transition metal ionic 
radius Sn$\to$Pt$\to$Ti$\to$Ge for the Yb$_2{\text M}_2$O$_7$ compounds
thus tunes across the phase boundary between the ferromagnetic
and $\Psi_2/\Psi_3$ regions in Fig. \ref{fig:classical-phase-diagram}
[\onlinecite{dun14, jaubert15}].
The spin excitations above these ordered states remain a puzzle, however,
with a recent systematic study of the Sn, Ti and Ge compounds 
showing an absence of coherent
spin waves in all three materials, and a continuity of the inelastic
neutron scattering spectrum across the finite temperature
ordering transition in each case \cite{hallas16}.

Amongst Er based pyrochlores, Er$_2$Ge$_2$O$_7$ has 
been observed to order antiferromagnetically at $T_N=1.41$K
[\onlinecite{dun15}]. Neutron scattering 
experiments reveal this to belong
to the $\Psi_2/\Psi_3$ region of the phase diagram in Fig. 
\ref{fig:classical-phase-diagram}. The behaviour
of the intensity of the magnetic Bragg peaks under
external magnetic field suggests that fluctuations
select a $\Psi_3$ ground state out of the {\sf E} symmetry 
manifold for Er$_2$Ge$_2$O$_7$ [\onlinecite{dun15}]. 
Er$_2$Pt$_2$O$_7$, also orders antiferromagnetically at 
$T_N=0.3$K. 
If the variation of 
exchange constants with the size of the transition
metal ion M is montonic, then it would be expected for this material to lie
near the boundary between the $\Psi_2$ and Palmer-Chalker
regions of the phase diagram in Fig. \ref{fig:classical-phase-diagram}.
However, this assumption could fail in the case of Er$_2$Pt$_2$O$_7$
since Pt$^{4+}$ ion differs from \{Ti$^{4+}$, Sn$^{4+}$, Ge$^{4+}$ \} 
in that it possesses a partially filled d-shell~\cite{cai16,hallas16b}.

The Gd based pyrochlores Gd$_2$M$_2$O$_7$ have
attracted significant research interest over a period
of nearly two decades \cite{gardner10, raju99}. 
The physics of these pyrochlores is somewhat different 
from (e.g.) Yb and Er based systems
because the Gd$^{3+}$ ions have vanishing orbital angular
momentum L=0. The interactions of the $S=7/2$ Gd spins are
thus quite isotropic, and the nearest neighbour 
anisotropic exchange which is the focus of this article 
is a less important consideration than further neighbour
interactions- including dipole-dipole interactions.
Nevertheless, combining the effect of nearest neighbour 
antiferromagnetic exchange and the nearest neighbour 
part of the dipole-dipole interaction our theory does 
predict a Palmer-Chalker ground state
which is consistent with observations on Gd$_2$Sn$_2$O$_7$ 
[\onlinecite{wills06, quilliam07}]. 

In Gd$_2$Ti$_2$O$_7$ further neighbour interactions drive a
complex phenomenology involving multiple phase transitions 
and ``partially ordered" states \cite{champion01, bonville03,
stewart04, javanparast15-PRL114}. The precise nature of the
magnetic ground state remains a matter of discussion in 
the literature \cite{paddison15}, and is beyond the scope of the 
present study. Recently, another Gd pyrochlore, Gd$_2$Pb$_2$O$_7$ has 
been synthesised presenting an antiferromagnetic ordering  
transition at T=0.81K  into an as yet unidentified 
ground state \cite{hallas15}.

Nd based pyrochlores have attracted considerable
recent attention and provide examples of ``all-in, all out''
ordering on the pyrochlore lattice \cite{bertin15, 
ciomaga-hatnean15, xu15, lhotel15}.
In particular, it has recently come to light that
 Nd$_2$Zr$_2$O$_7$ exhbits the novel
phenomenon of ``moment fragmentation''
in which the
spin correlations simultaneously show the
 pinch points of a Coulomb phase  and 
the Bragg peaks of an ordered state \cite{brooks14, petit16, benton16-arXiv}.
However, we note that 
the ground doublet of the Nd ions is of 
the dipolar-octupolar type \cite{xu15, lhotel15}. In this case, the anisotropic 
exchange interactions would have a different form to Eq. (6)
[\onlinecite{huang14}].
Recently, a Ce based pyrochlore Ce$_2$Sn$_2$O$_7$,
also believed to belong to the group of dipolar-octupolar
pyrochlores,
has been synthesised
which appears to show
an absence of magnetic order down to $T= 0.02$K
[\onlinecite{sibille15}], suggesting it as a promising
candidate spin--liquid system.

There are also many pyrochlore systems where the
rare-earth ion $R$ is a non-Kramers ion, such as Tb$^{3+}$ or
Pr$^{3+}$ [\onlinecite{gardner10}].
Many of these systems also exhibit two--fold
degenerate crystal field ground states, 
and a pseudospin-1/2 description of the
magnetic degrees of freedom may be appropriate 
\cite{onoda11, curnoe13, mukherjee14}
(although in the Tb$^{3+}$ systems the picture is
complicated by the relatively small gap to crystal field
excitations \cite{gingras00, mirebeau07, zhang14}).
In this case the nearest neighbour bilinear exhange Hamiltonian for these
pseudospins takes the form of Eq. (\ref{eq:Hross}), with the constraint that
${\sf J}_{z\pm}=0$ [\onlinecite{onoda11, lee12}] --- cf. Table~\ref{table:J-in-local-frame}.   
The pseudospin ${\bf S}$ must also be interpreted differently in this
case with the part of the pseudospin perpendicular to the local
$[111]$ directions corresponding to a quadrupolar degree of freedom.
Thus, an ``easy plane'' order of the pseudospins (as occurs in the $\sf E$
and $\sf T_2$ regions of our phase diagram) actually corresponds to
quadrupolar order for non-Kramers ions.

Amongst the Tb pyrochlores, Tb$_2$Sn$_2$O$_7$ is known to exhibit
ferromagnetic order \cite{mirebeau05, petit12, mcclarty10-arXiv}
and therefore belongs to the ${\sf T_1}$ region of
the phase diagram.
Tb$_2$Ti$_2$O$_7$, meanwhile, has been studied for a long time
as a candidate spin--liquid \cite{gardner99, gardner01}
exhibiting power-law spin correlations
\cite{fennell12, petit12-PRB86}.
The apparent spin--liquid behaviour has been linked with quantum
spin ice physics \cite{molavian07}, but it has recently been proposed that an alternative
form of spin--liquid physics may be at work \cite{benton16}.
Recent studies have also revealed the presence
of competing ordered states, with quadrupolar order (corresponding
to a Palmer-Chalker like configuration of the pseudo-spins
\cite{taniguchi13, takatsu16})
and antiferromagnetic ${\bf q}=(1/2, 1/2, 1/2)$ order
\cite{fritsch13, kermarrec15}
 observed
depending on the sample details and the experimental cooling protocol.
Tb$_2$Ge$_2$O$_7$ has recently come to light as another system of interest
for spin liquid physics, with no long range order observed down to 20mK
and short range ferromagnetic correlations \cite{hallas14}.
 This may suggest that Tb$_2$Ge$_2$O$_7$ lives in a region
of disorder, proximate to the ferromagnetic (${\sf T_1}$) phase, such
as that proposed in Ref. [\onlinecite{benton16}].

At the same time, the Pr based pyrochlores, Pr$_2$M$_2$O$_7$ 
(M=Sn, Zr, Hf, Pb) have emerged as promising candidates for
quantum spin ice physics, exhibiting an absence of magnetic
order and dynamic, spin--ice like correlations
\cite{zhou08, kimura13, hallas15, sibille16}.

\section{Conclusion}
\label{conclusions} 

Rare-earth pyrochlore oxides offer a
veritable treasure trove of novel physical phenomena, 
ranging from classical and quantum spin liquids,
to dimensional reduction, and phases governed by 
order--by--disorder effects. 
In this Article we have established a general theory of multiple--phase competition in
materials with anisotropic exchange interactions on the 
pyrochlore lattice, and shown how it can be 
can be applied to three specific materials~: Er$_2$Ti$_2$O$_7$, 
Yb$_2$Ti$_2$O$_7$ and Er$_2$Sn$_2$O$_7$.
The recurring theme throughout this analysis is of materials  
``living on the edge'', in the sense of having properties which are dictated
by the competition between neighbouring forms of magnetic order.   


Starting from a very general model of 
interactions between 
nearest--neighbour spins on the pyrochlore lattice, 
$\mathcal{H}_{\sf ex}$~[Eq.~(\ref{eq:Hex1})], we have used an analysis 
based on point--group symmetry to establish the exact, classical, ground--state 
phase diagram [Sections~\ref{section:model} and \ref{section:classical-ground-states}].  
As a by--product, we provide a complete classification of possible four--sublattice 
ordered states, according to the way they lift the symmetry of a single tetrahedron.
Moreover, using the ``Lego--brick'' rules developed
in Section~\ref{Lego-brick-rules}, we are able to identify the 
conditions under which the classical ground state manifold 
undergoes a dimensional reduction into independent planes 
or chains of spins, opening the door to new physical phenomena.  


We have given particularly careful consideration 
to the ground--state manifolds in the limit 
where the symmetric off diagonal exchange
$J_3<0$ and the  Dzyaloshinskii-Moriya interaction
$J_4=0$  [Section~\ref{section:classical-ground-states}]. 
Based on the experimental parameterisations of exchange interactions 
for Kramers pyrochlores \cite{ross11-PRX1, savary12-PRL109, guitteny13}, 
this limit is of particular experimental relevance.
We have elucidated the nature of the 
expanded ground state manifolds
which occur at the phase boundaries of this model, and it is these
which drive much of the physics of the surrounding regions of
parameter space.


Stepping out of the ground state manifold,
we have given, in Section~\ref{section:spin-wave-theory},
calculations of the spin wave excitations in the ordered phases.
Among other things, this allows us to determine the ground state
selection by both quantum and thermal fluctuation and to
identify regions of the phase diagram where classical order
will be melted by quantum fluctuations.


We have also studied the finite--temperature properties of the anisotropic
exchange model [Eq.~(\ref{eq:Hex1})] using classical Monte Carlo 
simulations, presented in Sections~\ref{section:finite-temperature}
and \ref{section:living-on-the-edge}.
These simulations make it possible to determine the
finite temperature phase diagram [Fig. (\ref{fig:finite-temperature-phase-diagram})]
and to show how the expanded ground state manifolds on
the phase boundaries manifest themselves at finite temperature
[Fig.~\ref{fig:figure-of-doom}].


The implications of our theory for three specific pyrochlore materials ---
Er$_2$Ti$_2$O$_7$, Yb$_2$Ti$_2$O$_7$ and Er$_2$Sn$_2$O$_7$ ---
are expounded in Sections~\ref{section:Er2Ti2O7}, \ref{section:Yb2Ti2O7} 
and \ref{section:Er2Sn2O7}.
We find that the influence of nearby phase boundaries accounts for the
ground state selection by fluctuations in Er$_2$Ti$_2$O$_7$,
the apparent dimensional reduction in the paramagnetic phase of
Yb$_2$Ti$_2$O$_7$, and the suppression of magnetic order in
Er$_2$Sn$_2$O$_7$.
The unusual properties of these three materials
can be understood as ``living on the edge'' --- having properties 
controlled by the competition between different ground states.


As discussed in Section \ref{section:other-pyrochlores},
the family of rare earth pyrochlore magnets is a large one,
extending well beyond the three materials covered  
in Sections~\ref{section:Er2Ti2O7} to \ref{section:Er2Sn2O7}.
In particular, recent work has seen the synthesis of rare earth pyrochlores
${\text R}_2{\text M}_2$O$_7$ with ${\text M}=$Ge, Pt, Pb, Os, Zr, Hf
[\onlinecite{dun15, cai16, hallas15, zhao16, ciomaga15, sibille16}].
This work suggests the possibility to move around the phase diagram shown in 
Fig.~\ref{fig:classical-phase-diagram} and Fig.~\ref{fig:quantum-phase-diagram}, 
using chemical or physical pressure.
This would be particularly interesting in cases where systems related
by a change of transition metal ion $\text M$ live on opposite sides
of a classical phase boundary. 
One might then hope to tune through a region of strong quantum fluctuations
using substitution of the transition metal ion.
Such an opportunity would seem to present 
itself for Er$_2{\text M}_2$O$_7$ with $\text M=$Sn, Ti
and  Yb$_2{\text M}_2$O$_7$ with $\text M=$Ge, Ti [\onlinecite{jaubert15}].


Our analysis may also be useful in the study of related systems,
such as the rare-earth spinel CdEr$_2$Se$_4$ [\onlinecite{lago10}],
where the Er ions also form a pyrochlore lattice.
Looking further afield, a modification of our theory could be
used in the understanding of ``breathing'' pyrochlore compounds,
where the tetrahedra of the pyrochlore lattice alternate in size
\cite{okamoto13, tanaka14, kimura14, benton15}.
It is also interesting to note recent neutron scattering experiments on 
NaCaCo$_2$F$_7$, a pyrochlore material with quenched exchange
disorder \cite{ross16}.
The observed diffuse scattering in that material is rather similar to that
predicted by our Monte Carlo simulations in
Fig.~\ref{fig:Sq}(b).
This may spring from a connection between the low--energy configurations
found in the clean limit of the anisotropic exchange model, studied in this Article, 
and the low--energy configurations of the disordered system.


From a theoretical perspective, our work also highlights the importance of large,
classical,  ground--state degeneracies which are {\it not} related to the well--studied 
examples of spin ice, or of the Heisenberg antiferromagnet on a pyrochlore lattice.
These degenaracies, which emerge in a number of different limits of 
$\mathcal{H}_{\sf ex}$~[Eq.~(\ref{eq:Hex1})], could lead to 
novel forms of classical or quantum spin liquid, as well as
entirely new forms of classical and quantum order \cite{benton-thesis}.   
One such case, where fluctuations lead to a spin--liquid described
by a rank--2 tensor field with a continuous gauge symmetry, has been 
developed in Ref.~[\onlinecite{benton16}].  
However there are many other regions of parameter space where strong 
fluctuations persist to low temperature \cite{ludo-unpub}, 
and the majority of these have yet to be fully explored.
It seems that the study of rare earth pyrochlore magnets with anisotropic 
exchange interactions may have many more surprises yet in store.


\bigskip


\section*{Acknowledgments}   
The authors are pleased to acknowledge helpful conversations with 
Bruce Gaulin, 
Michel Gingras,
Alannah Hallas,
Edwin Kermarrec, 
Isabelle Mirebeau,
Sylvain Petit,
Karlo Penc,
and Kate Ross, 
and a critical reading of the manuscript by Mathieu Taillefumier.
This work was supported by the 
Theory of Quantum Matter Unit of the
Okinawa Institute of Science and Technology 
Graduate University.

\appendix

\section{g-tensor in local and global coordinate frames}
\label{appendix:local-coordinate-frame}


The local crystal-electric field (CEF), acting on a given magnetic ion, affects both 
the character of its ground state, and the nature of its exchange interactions 
with other magnetic  ions.
For this reason, it is often convenient chose a coordinate frame 
$$
\{ {\bf x}_i^{\sf \ local},  {\bf y}_i^{\sf \ local}, {\bf z}_i^{\sf \ local} \}
$$
which is tied to the local CEF on site $i$.
We can accomplish this by choosing $\mathbf{z}^{\sf local}_i$ to be parallel with the 
 $[111]$ axis on site $i$, i.e. the local axis with $C_3$-symmetry

For the tetrahedron shown in Figure~\ref{fig:tetrahedron}, the magnetic ions labelled 
$S_0$, $S_1$, $S_2$ and $S_3$ occupy positions 
\begin{eqnarray}
&{\bf r}_0 =  \frac{a}{8} \left( 1, 1, 1 \right) 
\qquad 
&{\bf r}_1 =  \frac{a}{8} \left( 1,-1,-1 \right) 
\nonumber \\
&{\bf r}_2 =  \frac{a}{8} \left( -1,1,-1 \right) 
\qquad 
&{\bf r}_3 = \frac{a}{8} \left( -1,-1,1 \right) 
\; ,
\label{eq:r}
\end{eqnarray}
relative to the centre of the tetrahedron, in units such that the cubic, 16-site unit 
cell of the pyrochlore lattice occupies a volume $V = a^3$.     
The local $[111]$ axes on these sites are given by
\begin{eqnarray}
&\mathbf{z}^{\sf local}_0 = \frac{1}{\sqrt{3}}(1,1,1)
\qquad
&\mathbf{z}^{\sf local}_1 = \frac{1}{\sqrt{3}}(1,-1,-1) 
\nonumber \\
&\mathbf{z}^{\sf local}_2 = \frac{1}{\sqrt{3}}(-1,1,-1)
\qquad
&\mathbf{z}^{\sf local}_3 = \frac{1}{\sqrt{3}}(-1,-1,1) 
\; . \nonumber
\label{eq:local-111-axis}\\
\end{eqnarray}
In defining ($\mathbf{x}^{\sf local}_i, \mathbf{y}^{\sf local}_i$) 
we follow the conventions of Ross {\it et al.} 
[\onlinecite{ross11-PRX1}], and make the convenient choice
\begin{eqnarray}
&\mathbf{x}^{\sf local}_0 = \frac{1}{\sqrt{6}}(-2,1,1)
\qquad
&\mathbf{x}^{\sf local}_1 = \frac{1}{\sqrt{6}}(-2,-1,-1)
\nonumber \\
&\mathbf{x}^{\sf local}_2 = \frac{1}{\sqrt{6}}(2,1,-1)
\qquad
&\mathbf{x}^{\sf local}_3 = \frac{1}{\sqrt{6}}(2,-1,1)
\; , \nonumber
\label{eq:local-easy-plane-x}\\
\end{eqnarray}
such that all ${\bf y}_i^{\sf \ local}$ lie in a common plane
\begin{eqnarray}
&\mathbf{y}^{\sf local}_0 = \frac{1}{\sqrt{2}}(0,-1,1)
\qquad
&\mathbf{y}^{\sf local}_1 = \frac{1}{\sqrt{2}}(0,1,-1)
\nonumber \\
&\mathbf{y}^{\sf local}_2 = \frac{1}{\sqrt{2}}(0,-1,-1)
\qquad
&\mathbf{y}^{\sf local}_3 = \frac{1}{\sqrt{2}}(0,1,1)
\; . \nonumber
\label{eq:local-easy-plane-y}\\
\end{eqnarray}


\begin{table}
\begin{tabular}{ | c | c | c | c |}
\hline
    & 
    $\quad$Yb$_2$Ti$_2$O$_7$$\quad$ & 
    $\quad$Er$_2$Ti$_2$O$_7$$\quad$ & 
    $\quad$Er$_2$Sn$_2$O$_7$$\quad$ \\
\hline
\hline
$g_{xy}$ & 4.18 & 5.97 & 7.52\\
\hline
$g_z$ & 1.77 & 2.45 & 0.05\\
\hline
\end{tabular}
\caption{
Estimates of the components of the g-tensor in the local frame 
${\bf g}_{\sf local}$~[Eq.~(\ref{eq:g-tensor-local})], taken from experiment on 
Yb$_2$Ti$_2$O$_7$~[\onlinecite{hodges01}], 
\mbox{Er$_2$Ti$_2$O$_7$~[\onlinecite{savary12-PRL109}]}, 
and Er$_2$Sn$_2$O$_7$~[\onlinecite{guitteny13}].  
}
\label{table:g-tensor}
\end{table}

In this local coordinate frame, the magnetic moment
\begin{eqnarray}
m_{i}^{\alpha} 
   &=& \sum_{\beta=1}^3 g_{\sf local}^{\alpha \beta}  \mathsf{S}_i^{\beta}
\end{eqnarray}
is connected to the (pseudo) spin-1/2 operator $\mathsf{S}_i^\alpha$~[Eq.~(\ref{eq:Slocal})], 
through a g-tensor with a diagonal simple form 
\begin{eqnarray}
{\bf g}_{\sf local} &=&
    \begin{pmatrix}
       g_{xy} & 0 & 0 \\
       0 & g_{xy} & 0\\
       0 & 0 & g_z \\
    \end{pmatrix} 
    \label{eq:g-tensor-local} \\
    \nonumber
 \end{eqnarray}
where 
$\alpha$, $\beta = \{ {\bf x}_i^{\sf \ local},  {\bf y}_i^{\sf \ local}, {\bf z}_i^{\sf \ local} \}$, 
and ${\bf g}_{\sf local}$ is independent of the site considered.   
Estimates of $g_{xy}$ and $g_z$, taken from experiment on 
Yb$_2$Ti$_2$O$_7$~[\onlinecite{hodges01}], \mbox{Er$_2$Ti$_2$O$_7$~[\onlinecite{savary12-PRL109}]}, 
and Er$_2$Sn$_2$O$_7$~[\onlinecite{guitteny13}] are shown in Table~\ref{table:g-tensor}.
For rare-earth ions with Ising character, such as Dy$^{3+}$ in Dy$_2$Ti$_2$O$_7$, 
$g_z  > g_{xy}$, while for the rare-earth ions considered in this paper with easy-plane character, $g_z  < g_{xy}$.   


The g-tensor in the coordinate frame of the crystal axes, ${\bf g}_i$ [Eq.~(\ref{eq:g-tensor-global})], 
can be found by rotating ${\bf g}_{\sf local}$ [Eq.~(\ref{eq:g-tensor-local})] 
back into the global coordinate frame \mbox{$\mu$, $\nu = \{ {\bf x},  {\bf y}, {\bf z} \}$}.
Since the required rotation depends on the lattice site, the resulting $g$-tensor is sublattice-dependent
\begin{eqnarray}
&{\bf g}_{0} =
    \begin{pmatrix}
        g_{1} & g_{2} & g_{2} \\
        g_{2} & g_{1} & g_{2}\\
        g_{2} & g_{2} & g_{1} \\
    \end{pmatrix} \quad
&{\bf g}_{1} =
   \begin{pmatrix}
       g_{1} & -g_{2} &- g_{2} \\
      -g_{2} & g_{1} & g_{2}\\
      -g_{2} & g_{2} & g_{1} \\
   \end{pmatrix} \nonumber \\
&{\bf g}_{2} =
   \begin{pmatrix}
       g_{1} & -g_{2} & g_{2} \\
      -g_{2} & g_{1} & -g_{2}\\
       g_{2} & -g_{2} & g_{1} \\
   \end{pmatrix} \quad
&{\bf g}_{3} =
   \begin{pmatrix}
      g_{1} & g_{2} & -g_{2} \\
      g_{2} & g_{1} & -g_{2}\\
     -g_{2} & -g_{2} & g_{1} \\
   \end{pmatrix} \nonumber \\
   \label{eq:g-tensor}
\end{eqnarray}
where
\begin{eqnarray}
g_1 &=& \frac{2}{3} g_{xy} +\frac{1}{3} g_z \quad
g_2 = -\frac{1}{3} g_{xy} + \frac{1}{3} g_z.
\end{eqnarray}

\section{Linear spin-wave theory for a general 4-sublattice ground state}
\label{appendix:linear-spin-wave}

A general framework for linear spin-wave theory on the pyrochlore lattice 
is set out in [\onlinecite{ross11-PRX1}], following the pattern that can be found in [\onlinecite{fazekas99}].   
For completeness here we reproduce the technical steps needed to apply
such a theory to the 4-sublattice, ${\bf q}=0$ classical ground states 
discussed in Section~\ref{section:classical-ground-states}.


As with the classical spin-wave theory developed in Section~\ref{section:classical-spin-wave}, 
it is convenient to work in a local basis, in which spins are quantised
such that their local z-axis is aligned with the classical ground state.
Following Eq.~(\ref{eq:fluc}), we label these local axes 
$$\{ {\bf u}_i, {\bf v}_i, {\bf w}_i\}$$
and quantize fluctuations about the classical ground state by introducing 
Holstein-Primakoff bosons  
\begin{eqnarray}
\label{eq:HPsw}
S^w_i 
   &=& S-a^{\dagger}_i a^{\phantom \dagger}_i 
   \\
    \label{eq:HPs+}
S^+_i 
    &=& S^u_i+iS^v_i 
    = (2 S - a^{\dagger}_i a^{\phantom \dagger}_i)^{1/2} a^{\phantom \dagger}_i 
\approx \sqrt{2S} 
       a_i^{\phantom \dagger} 
      \\
      \label{eq:HPs-}
S^-_i 
   &=& S^u_i-iS^v_i
   = a^{\dagger}_i (2 S - a^{\dagger}_i a_i^{\phantom \dagger})^{1/2} 
   \approx \sqrt{2S} a^{\dagger}_i  
\end{eqnarray}
where
\mbox{$\big[ a_i^{\phantom \dagger} , a_j^{\dagger} \big] =  \delta_{ij}$} \; .


Substituting these expressions in $\mathcal{H}_{\sf ex}$ [Eq.~(\ref{eq:Hex})] 
and Fourier transforming them, we obtain
\begin{eqnarray}
{\mathcal H}_{\sf ex} 
   &\approx& {\mathcal E}_0 + {\mathcal H}^{\sf LSW}_{\sf ex} + \ldots
\label{eq:LSWapprox}
\end{eqnarray}
where ${\mathcal E}_0$ is the classical ground state energy 
defined in Eq.~(\ref{eq:E0}), and 
\begin{eqnarray}
{\mathcal H}^{\sf LSW}_{\sf ex}   
   &=&  \frac{1}{2} \sum_{{\bf q}} 
       \tilde{A}^{\dagger}({\bf q}) \cdot {\bf X}({\bf q}) \cdot
       \tilde{A}({\bf q}) 
\label{eq:HLSW}
\end{eqnarray}
describes quantum fluctuations at the level of linear spin wave theory.
Here $\tilde{A}^{\dagger}({\bf q}), \tilde{A}({\bf q})$ 
are eight-component vectors of operators
\begin{eqnarray}
\tilde{A}^{\dagger}({\bf q}) 
   = (a_0^{\dagger}({\bf q}), 
       a_1^{\dagger}({\bf q}), 
       a_2^{\dagger}({\bf q}), 
       a_3^{\dagger}({\bf q}), 
       \nonumber \\
       a_0^{\phantom \dagger}(-{\bf q}),
       a_1^{\phantom \dagger}(-{\bf q}),
       a_2^{\phantom \dagger}(-{\bf q}), 
        a_3^{\phantom\dagger}(-{\bf q}))
\end{eqnarray}
and $X({\bf q})$ is an $8 \times 8$ matrix written in block form as
\begin{eqnarray}
{\bf X}({\bf q})
   &=& 2 S \begin{pmatrix}
            {\bf X}^{11}({\bf q})
         & {\bf X}^{12}({\bf q}) \\
            {\bf X}^{21}({\bf q})
         & {\bf X}^{22}({\bf q})  \\
             \end{pmatrix} 
           \label{eq:X}  \\
{\bf X}^{11}_{ij}({\bf q}) 
   &=&
    \cos({\bf q} \cdot {\bf r}_{ij} ) \nonumber \\
   && \bigg(
     {\bf c}_i \cdot {\bf J}^{ij} \cdot  {\bf c}_j^{\ast}
      - \delta_{ij}  \sum_{l}   {\bf w}_l \cdot {\bf J}^{lj} \cdot {\bf w}_j  
     \bigg)  \\
{\bf X}^{12}_{ij}({\bf q})
   &=& {\bf X}^{21 \ast}_{ji}=\cos({\bf q} \cdot {\bf r}_{ij} ) 
       \bigg(
         {\bf c}_i \cdot {\bf J}^{ij} \cdot {\bf c}_j  
       \bigg) \\
{\bf X}^{22}_{ij}({\bf q})
   &=& \cos({\bf q} \cdot {\bf r}_{ij} ) \nonumber \\
    && \bigg( 
         {\bf c}_i^{\ast} \cdot {\bf J}^{ij} \cdot {\bf c}_j
        - \delta_{ij}  \sum_{l}  {\bf w}_l \cdot {\bf J}^{lj} \cdot {\bf w}_j   
        \bigg)  
\end{eqnarray}
where
\begin{eqnarray}
{\bf c}_i=\frac{1}{\sqrt{2}} \left( {\bf u}_i + i {\bf v}_i \right).
\end{eqnarray}


The spin-wave Hamiltonian ${\mathcal H}^{\sf LSW}_{\sf ex}$ [Eq.~(\ref{eq:HLSW})] 
can be diagonalized by a suitable Bogoliubov transformation.
We accomplish this following the method outlined in Ref.~[\onlinecite{roger83}]
by introducing new Bose operators
\mbox{$\big[ b_i^{\phantom \dagger} , b_j^{\dagger} \big] =  \delta_{ij}$}, 
such that
\begin{eqnarray}
\label{eq:bogoliubov}
B^{\dagger}({\bf q}) 
   &=&  (b_0^{\dagger}({\bf q}), 
             b_1^{\dagger}({\bf q}), 
             b_2^{\dagger}({\bf q}), 
             b_3^{\dagger}({\bf q}), 
             \nonumber \\
     &&   \qquad 
             b_0^{\phantom \dagger}(-{\bf q}),
             b_1^{\phantom \dagger}(-{\bf q}),
             b_2^{\phantom \dagger}(-{\bf q}), 
             b_3^{\phantom\dagger}(-{\bf q})) 
             \nonumber \\
   &=& \tilde{A}^{\dagger}({\bf q}) \cdot 
           {\bf U}^{ \dagger}({\bf q}) 
\end{eqnarray}
The condition that these operators are Bosonic may be written as
\begin{eqnarray}
\left[ B_i^{\phantom \dagger} ({\bf q}) ,  B_j^{\dagger} ({\bf q'}) \right] 
     &=& \sigma_{ij} \delta_{{\bf q} {\bf q}'}
\end{eqnarray}
where
\begin{eqnarray}
\hat{\sigma} = 
\begin{pmatrix}
{\bf 1} & {\bf 0} \\
{\bf 0} & -{\bf 1}
\end{pmatrix}.
\end{eqnarray}
is an $8 \times 8$ matrix (written in block form) 
and leads to a pseudo-unitary condition  on ${\bf U}({\bf q})$
\begin{eqnarray}
{\bf U}^{-1}({\bf q})=\hat{\sigma} \cdot {\bf U}^{\dagger}({\bf q}) \cdot \hat{\sigma}.
\end{eqnarray}
Substituting in Eq.~(\ref{eq:HLSW}), we obtain
\begin{eqnarray}
\label{eq:MatrixH}
&{\mathcal H}^{\sf LSW}_{\sf ex}&    \nonumber \\
   &=& \frac{1}{2} \sum_{{\bf q}} B^{\dagger} ({\bf q}) \cdot  
           {\bf U}^{ -1 \dagger}  ({\bf q})
            \cdot \mathbf{X}({\bf q}) \cdot  {\bf U}^{ -1}  ({\bf q}) 
            \cdot B^{\phantom \dagger} ({\bf q}) \nonumber\\
   &=&  \frac{1}{2} 
           \sum_{{\bf q}} B^{\dagger} ({\bf q}) \cdot  \hat{\sigma}  \cdot 
           {\bf U}({\bf q}) \cdot
            \hat{\sigma} \cdot \mathbf{X}({\bf q}) \cdot  {\bf U}^{ -1} ({\bf q})  
            \cdot  B^{\phantom \dagger} ({\bf q}) . 
            \nonumber \\
\end{eqnarray}
The object 
\mbox{$ {\bf U}({\bf q}) \cdot  \hat{\sigma} \cdot \mathbf{X}({\bf q}) \cdot  {\bf U}^{ -1}  ({\bf q})$} 
is a similarity transformation on the matrix \mbox{$\hat{\sigma} \cdot \mathbf{X}({\bf q})$}, 
and for correctly chosen ${\bf U}({\bf q})$, will be a diagonal matrix containing the 
eigenvalues of \mbox{$\hat{\sigma} \cdot \mathbf{X}({\bf q})$}.
We then arrive at
\begin{eqnarray}
{\mathcal H}^{\sf LSW}_{\sf ex} 
   &=& \frac{1}{2} \sum_{{\bf q}} B^{\dagger} ({\bf q}) \cdot  \hat{\sigma}  \cdot 
           \begin{pmatrix}
                \omega_{\nu}(\mathbf{q}) & 0 \\
                0 & -\omega_{\nu}(\mathbf{q})
           \end{pmatrix}
           \cdot B^{\phantom \dagger} ({\bf q}). \nonumber \\
\end{eqnarray}
Collecting all terms, reordering operators and inserting into Eq.~(\ref{eq:LSWapprox})  
we obtain the result quoted
in Section~\ref{subsection:quantum-spin-wave} 
\begin{eqnarray}
{\mathcal H}_{\sf ex} 
    &\approx& {\mathcal E}_0 \left(1 + \frac{1}{S} \right)
            \nonumber \\
      && \;  + \sum_{{\bf q}} 
             \sum_{\nu=0}^{3} \omega_{\nu}({\bf q})
             \left(
                  b_{\nu}^{\dagger}({\bf q})
                  b^{\phantom \dagger}_{\nu}({\bf q}) +\frac{1}{2} 
             \right) 
             + \ldots  
             \nonumber \\
\end{eqnarray}
The dispersion $\omega_{\nu}({\bf q})$ of the four branches of spin waves can be 
found by numerical diagonalization of \mbox{$\hat{\sigma} \cdot \vec{X}({\bf q})$}.

\section{Classical Monte Carlo simulation}
\label{appendix:classical-MC}

The Monte Carlo simulations described in this paper are based on the Metropolis algorithm 
with parallel tempering~\cite{swendsen86,geyer91} and  over-relaxation~\cite{creutz87}. 
The spins are modelled as classical vectors of length $|S_{i}|=1/2$ and locally updated 
using the standard Marsaglia method~\cite{marsaglia72}.   
We consider cubic clusters of linear dimension $L$, based on the 16-site cubic unit cell 
of the pyrochlore lattice, and containing $N = 16 L^3$ sites.
A Monte Carlo step (MCs) is defined as $N$ attempts to locally update a randomly 
chosen spin, and $t_{max}$ (measured in MCs) is the total Monte Carlo time over 
which data are collected.


Equilibration is performed for each temperature in two successive steps.   
First the system is slowly cooled down from high temperature (random initial spin configuration) 
to the temperature of measurement $T$ during $t_{max}/10$ MCs.   
Then, the system is equilibrated at temperature $T$ during additional $t_{max}/10$ MCs.   
After equilibration, Monte Carlo time is set to zero and measurements start and go on 
for $t_{max} \sim 10^{5}-10^{7}$ MCs.


All thermodynamical observables have been averaged over Monte Carlo time every 10 MCs, 
except for calculations of the equal-time structure factor $S({\bf q})$, where data points were 
taken every 100 MCs for efficiency.  
The parallel tempering method implies simultaneously simulating a large number of replicas
of the system in parallel, with each replica held at a different temperature.
The program then regularly attempts to swap the spin configurations of replicas
with neighbouring temperatures, in such a way as to maintain detailed 
balance~\cite{swendsen86,geyer91}.  
Simulating $\sim 120$ replicas, with swaps attempted every 100 MCs appears to offer 
a good compromise between efficiency and decorrelation for $L=6$. 


In the case of the over-relaxation method, after each Monte Carlo step, 
two further sweeps are made of the entire lattice.
Each spin feels an effective field due to the interaction with its six nearest neighbours; any 
rotation around this axis conserves the energy and is thus an acceptable move respecting 
detailed balance. 
To avoid rotating successive neighbouring spins, we first update all spins of sublattice 0, 
then sublattice 1, 2 and finally 3.  
The first iteration of all $N$ spins is deterministic, \textit{i.e.} we rotate them by the maximum 
allowed angle; while for the second iteration, a random angle of rotation is chosen for each spin. 
The generation of so many random numbers is of course time consuming but is 
recommended for better equilibration~\cite{kanki05}.
We note that convergence of the specific heat $c_h \to 1$ for $T \rightarrow 0$ 
is a good indication of the equilibration of ordered phases at low temperatures.


The main results of Monte Carlo simulations are summarised in the 
finite-temperature phase diagram Fig.~\ref{fig:finite-temperature-phase-diagram}, 
which spans all four of the ordered phases discussed in the article.
This phase diagram was determined from simulations for 64 different parameter 
sets, equally spaced on the circle defined by 
\mbox{$\sqrt{J_{1}^{2}+J_{2}^{2}} = 3\,|J_{3}|$} 
illustrated by the white circle in Fig.~\ref{fig:classical-phase-diagram}, 
with \mbox{$J_3=-0.1\ \text{meV}$} and \mbox{$J_4=0$}.   
Transition temperatures for each phase were extracted from the relevant 
order-parameter susceptibilities, as described in Section~\ref{section:finite-temperature}.  


Simulations were performed for a cluster of $N=3456$ spins ($L=6$), and 
data averaged over 10 independent runs during $t_{max}=10^{6}$ MCs. 
Parallel tempering was used, typically with 121 replicas, at temperatures 
equally-spaced from 0 to 1.2 K.
However, close to the boundaries between phases with different symmetries, the 
large number of competing ground states makes simulations difficult to equilibrate.
Here, additional data points with better statistics were sometimes necessary, typically with 
201 temperatures on a smaller temperature window, with $t_{max}=10^{7}$ MCs 
and $N=8192$ \mbox{(\textit{i.e.} L=8)}. 




\end{document}